\documentstyle[preprint,prb,aps]{revtex}

\begin{document}
\input{psfig}
\title{Quantum Fluctuations in the Frustrated Antiferromagnet
Sr$_2$Cu$_3$O$_4$Cl$_2$}

\author{A. B. Harris}
\address{Department of Physics and Astronomy, University of Pennsylvania,
Philadelphia, PA 19104}

\author{A. Aharony, O. Entin-Wohlman, and I. Ya. Korenblit}
\address{School of Physics and Astronomy, Raymond and Beverly Sackler
Faculty of Exact Sciences, Tel Aviv University, Tel Aviv 69978, Israel}

\author{R. J. Birgeneau}
\address{Center for Materials Science and Engineering, Massachusetts 
Institute of Technology, Cambridge, MA 02139}

\author{Y. J. Kim}
\address{Division of Engineering and Applied Sciences, Harvard University, 
Cambridge, MA 02138}
\address{and Center for Materials Science and Engineering,
Massachusetts Institute of Technology, Cambridge, MA 02139}
\date{\today}
\maketitle
\begin{abstract}
Sr$_2$Cu$_3$O$_4$Cl$_2$ is an antiferromagnet consisting of weakly
coupled CuO planes which comprise two weakly
interacting antiferromagnetic subsystems, I and II, which order at
respective temperatures $T_I \approx 390$K and $T_{II} \approx 40$K.
Except asymptotically near the ordering temperature, these
systems are good representations of the two-dimensional quantum
spin 1/2 Heisenberg model.  For $T< T_{II}$ there are four
low-energy modes at zero wave vector, three of
whose energies are dominated by quantum fluctuations.  For
$T_{II} < T < T_I$ there are two low energy modes.  The mode
with lower energy is dominated by quantum fluctuations.  Our
calculations of the energies of these modes (including dispersion
for wave vectors perpendicular to the CuO planes) agree extremely
well with the experimental results of inelastic neutron
scattering (in the accompanying paper) and for modes in the
sub meV range observed by electron spin resonance.  The 
parameters needed to describe quantum fluctuations are either
calculated here or are taken from the literature.  These
results show that we have a reasonable qualitative understanding
of the band structure of the lamellar cuprates needed to
calculate the anisotropic exchange constants used here.
\end{abstract}
\pacs{76.50.+g, 75.10.Jm, 75.50.Ee}

\section{INTRODUCTION}

There has been a resurgence of interest in low-dimensional
magnetism due in part
to the desire to understand high-$T_c$ superconductivity.  The
lamellar copper oxide systems, when suitably doped give rise to a
family of superconductors with $T_c$'s in the range about 30K.\cite{TC}
In these systems the Cu ions are essentially in a 3d$^9$
configuration.  Due to a large on-site Coulomb interaction, 
the states of this system which are accessible at ambient
temperature have one hole per Cu ion, and hence the manifold
of such accessible states is described by a spin 1/2 Hamiltonian
having antiferrmagnetic interactions, which are strongest between
nearest neighboring Cu ions in the CuO$_2$ plane.  That this
system is a nearly perfect realization of the two dimensional (2D)
spin 1/2 quantum Heisenberg model has been established by a
wide variety of experiments.\cite{REV}

\begin{figure}
\centerline{\psfig{figure=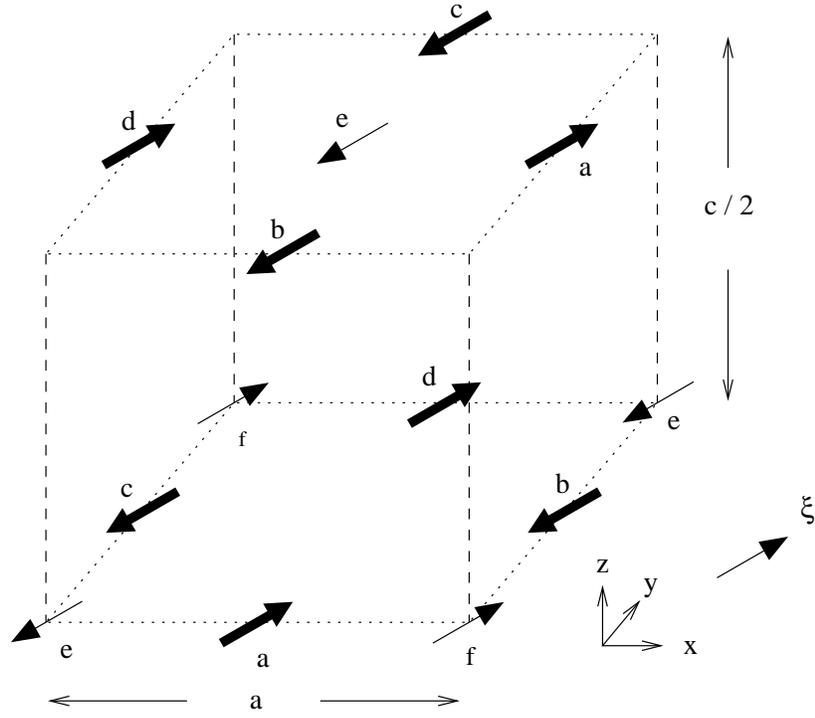}}

\vspace{0.2 in} \caption{Magnetic structure of 2342.
The CuI spins (in sublattices $\underline a$
$\underline b$ $\underline c$, and $\underline d$) are thick arrows
and the CuII spins (in sublattices $\underline e$  and $\underline f$)
are thin arrows.  The basis vectors for the magnetic unit cell are
${\bf a}_1=a (\hat x + \hat y)$, ${\bf a}_2=a (\hat x - \hat y)$, and
${\bf a}_3=\case 1/2 (a \hat x + a \hat y + c \hat z )$.
All spin directions are in the CuO ($x$-$y$) plane. 
The $\xi$ axis is defined to be collinear with the spin directions.}
\label{UCFIG}
\end{figure}

Recently, a variant of this system, Sr$_2$Cu$_3$O$_4$Cl$_2$ (2342),
has been shown to display very interesting magnetic
properties.\cite{CHOU,KASTNER,KIM}
The structure of this system\cite{STRUCT} is one
in which an additional Cu ion (which we refer to as a CuII ion)
is inserted at the center of alternate Cu plaquettes of the usual
copper lattice, whose ions we refer to as CuI's.  Although
all the Cu ions are chemically equivalent, they play very
different roles insofar as magnetism is concerned. The CuI's
order at a relatively high temperature ($T_I = 386$K) and have
properties similar to those of other lamellar cuprate
antiferromagnets.\cite{REV}  With respect to the isotropic exchange
interactions, the coupling between CuI and CuII ions is frustrated.
As a result, the CuII's order independently at a much lower
temperature, $T_{II}=39.6$K into the magnetic structure shown
in Fig. \ref{UCFIG}.  For $T_{II}<T<T_I$
a very small residual anisotropic exchange interaction
causes the CuII spins to have a small ferromagnetic moment,
the study of which\cite{KASTNER} led to the determination of
the magnetic structure which has recently been confirmed by
neutron diffraction.\cite{RJB}  The study of the statics also
led to the determination of several coupling constants
in the Hamiltonian used to model this system.

A natural continuation of this study was to investigate the
dynamics of this system, and in the accompanying paper\cite{RJB}
(which we refer to as I)
an inelastic neutron scattering study of this system is
reported.  One interesting result of these experiments was
that although the coupling between the CuI's and CuII's is
frustrated in the mean-field sense, the spin-wave spectrum
showed an incontrovertible signature of interactions between
these subsystems.\cite{KIM,RJB}  The nature of this coupling
was described by Shender in a seminal paper.\cite{EFS}  Although this
phenomenon has been identified in other materials,\cite{EFSGAP}
the effect of this coupling, caused by quantum fluctuations,
is perhaps the most dramatic in the system 2342, as described
briefly previously\cite{KIM} and in more detail in I.  As the
CuII system orders for $T<T_{II}$, the small gap
spin-wave energies are found to increase sharply.
This increase indicates that even though the CuI-CuII coupling is
frustrated in the mean-field sense, quantum fluctuations lead to
a significant interaction between sublattices.  A less obvious
type of frustration arises with respect to the in-plane anisotropy
associated with the bond anisotropy of the exchange interactions.
When the moments lies in the easy plane, the exchange tensor for
spins $i$ and $j$ in the plane has different values for directions
parallel and perpendicular to the $i$-$j$ bond.  However, within
mean field theory this anisotropy disappears when the average over
all bonds is taken.  But as before, there is a significant residual
interaction due to quantum fluctuations which gives rise to
in-plane anisotropy.  Finally, even classically frustration
can be removed by exchange anisotropy which has a form similar
to the dipolar interaction. We will refer to such exchange anisotropy
as pseudo-dipolar.

The purpose of the present paper is to
calculate the spin-wave spectrum in order to give a theoretical
interpretation to the data presented in I.  From the discussion
so far it is clear that most of these phenomena are
outside the scope of linearized spin-wave theory.  What is
required is a nonlinear spin-wave analysis, i. e. an analysis 
which includes the effects of quantum fluctuations.  In fact,
from an analysis of the magnetic structure of the cuprates\cite{SPINPRL1}
it was shown that there are several perturbations away from the
linear analysis of the isotropic Heisenberg model that one must
consider. These are the ones mentioned above, namely,
a) quantum fluctuations of otherwise frustrated interactions,
b) quantum fluctuations of the anisotropic in-plane  exchange
interactions, and c) pseudo-dipolar exchange anisotropy between
the CuI and CuII subsystems.  In a simplified way, one can
categorize these effects in the way they contribute to the
spin-wave energies, which is given by the famous formula\cite{KEFFER}
\begin{eqnarray}
\omega = \sqrt{2 H_E H_A } \ ,
\end{eqnarray}
where $H_E$ ($H_A$) is the exchange (anisotropy) field and we work
in units such that $\omega$, $H_E$, and $H_A$ are all energies,
usually given in meV (1meV/$k_B=11.6$K, 1meV/h$=241.8$Ghz.)
We will see that the out-of-plane anisotropy of the exchange
interactions gives rise to a corresponding out-of-plane
anisotropy field $H_A^{\rm out}$ which has been understood in terms
of the out-of-plane anisotropy in the exchange interactions
without reference to fluctuations.\cite{SPINPRL2,SOPR1}
In contrast, the in-plane anisotropy
of the exchange interactions, when summed over bonds, averages
to zero and therefore only contributes when fluctuations are taken into
account.\cite{SPINPRL2,SOPR1}  The mechanism studied by Shender\cite{EFS}
contributes to $H_A$ except for the Goldstone mode, whose energy becomes
nonzero only when lattice anisotropy is introduced.

One might expect that the number of coupling constants might be
so large that no useful information or test of the theory
would be possible.  As it happens, the fit to the energy of the
gaps is overdetermined and the agreement between theory and
experiment in some instances is quite remarkable, as can be seen in I.
The observation of the modes whose energy depends on the
in-plane anisotropy leads to the determination of the in-plane
anisotropy of the exchange interactions.  These quantities are
difficult to obtain experimentally.  Their values can be compared
to calculations\cite{SPINPRL2,SOPR1,SOPR2} based on the electronic
structure of the cuprates the knowledge of which may lead to a
better understanding of the high-$T_c$ superconductors.

One should recognize that at the moment inelastic neutron
scattering does not easily detect modes in the sub meV range of
energy.  As a result neutron scattering experiments have
not detected those in-plane modes whose energy depends only on
the in-plane anisotropy.  Recently, however, the modes in
the sub meV range of energy have been observed
by ESR experiments of the group at RIKEN.\cite{RIKEN1,RIKEN2}
The mere existence of these modes tends to confirm
the spin-wave calculations.  Moreover, the fact that they
are found in the predicted range of energy strongly
supports the theoretical calculations in this paper.

Briefly, this paper is organized as follows.  In Sec. II the
Hamiltonian with its various anisotropic exchange interactions
is specified.  In Sec. III we start by discussing briefly the
framework within which the calculations are to be done and we
give the Dyson--Maleev transformation\cite{D-M} to boson operators.
In Sec. IV the isotropic exchange Hamiltonian is discussed, first
within harmonic theory and then including spin-wave interactions,
which are essential to obtain a qualitatively correct spectrum.
In Sec. V the various anisotropies are included in an effective
quadratic spin-wave Hamiltonian.  In Sec. VI we give explicit
results for the spin-wave energies for the case when the transverse
wave vector is zero and show the comparison of our calculations with
the recent experiments of the MIT group.  In Sec. VII intensities of
modes are discussed, with numerical results given for zero wave
vector relative to the Bragg peaks for CuI and CuII. Our conlusions
are summarized in Sec. VIII.

\section{HAMILTONIAN}

The Hamiltonian that we intend to treat is written as
\begin{eqnarray}
{\cal H } &= & {\cal H}_1 + {\cal H}_2 \ ,
\end{eqnarray}
where ${\cal H}_1$ includes almost all the significant interactions,
namely all the intraplanar interactions and the unfrustrated
interactions between nearest neighbors in adjacent CuO planes and
${\cal H}_2$ includes small residual anisotropic interplanar
interactions involving CuII spins.  Since this latter term is totally
negligible {\it except} for extremely small wave vector and for the
lowest energy mode, it is only necessary to include contributions from
${\cal H}_2$ evaluated at zero wave vector.  Since the effects of
${\cal H}_2$ are only relevant to the extremely low frequency
spectrum, we defer consideration of ${\cal H}_2$ until Secs. V.4 and
V.5.

Thus we write ${\cal H}_1$ in tensor notation as
\begin{eqnarray}
{\cal H}_1 &=& \case 1/2 \sum_{\langle i,j \in I \rangle}
{\bf S}_i {\bf J}_I {\bf S}_j
+\sum_{\langle i \in I, j\in II\rangle} {\bf S}_i
{\bf J}_{I-II} {\bf S}_j +
\sum_{\langle i,j \in II \rangle}
{\bf S}_i {\bf J}_{II} {\bf S}_j + \sum_{i \in I} J_3 {\bf S}_i \cdot
{\bf S} _{i+\case 1/2 c\hat z} \ ,
\end{eqnarray}
where  $i \in I$($i\in II$) means that site $i$ runs over CuI (CuII)
sites and $\langle \ \ \rangle $ restricts the summation to nearest
neighbors of the indicated type in the same Cu-O plane.  The
only unfrustrated coupling between planes is that ($J_3$)
between CuI's directly above or below one another.
We will allow the couplings ${\bf J}_I$, ${\bf J}_{I-II}$,
and ${\bf J}_{II}$ to be anisotropic, whereas for simplicity
we take ${\bf J}_3$ to be isotropic.  Here and below we use a hybrid
notation for site labels in which the label $i+{\bf r}$ indicates a 
site at position ${\bf r}$ with respect to site $i$.
In ${\cal H}_2$ we include the interplanar CuI -- CuII and CuII-CuII
couplings whose isotropic parts are frustrated.

We first discuss the principal axes of the exchange tensor ${\bf J}_I$
associated with a bond between nearest neighboring CuI
spins in a CuO plane.  This bond is invariant with respect to
two mirror planes: one in the CuO plane and the other
perpendicularly  bisecting the CuI - CuI bond in question.
Accordingly, the principal axes of the CuI - CuI exchange
tensor between nearest neighbors lie along the three crystal
(1,0,0) directions, just as they would be in
the absence of the CuII's.  In that case, the exchange
tensor will have different values corresponding to the directions
i) along the bond in question, ii) perpendicular to the bond in question
but in the CuO plane, and iii) along the crystal {\underline c}
direction.  The principal axes of the other in-plane interactions
are similarly fixed by symmetry.\cite{SOPR1,RAVI}
Then the Hamiltonian ${\cal H}_1$ may be written as follows
\begin{small}
\begin{eqnarray}
&& {\cal H}_1 = \case 1/2 \sum_{i \in I} \sum_{\delta_1} \Biggl(
J_I^z S_i^z S_{i + \delta_1}^z + J_I^\parallel [ {\bf S}_i \cdot
\hat \delta_1 ] [{\bf S}_{i+\delta_1} \cdot \hat \delta_1 ]
+ J_I^\perp [{\bf S}_i \cdot \hat e_1 ] [ {\bf S}_{i+ \delta_1}
\cdot \hat e_1 ] \Biggr) \nonumber \\ &&
+ \sum_{i \in II} \sum_{\delta_{2,1}} \Biggl(
J_{I-II}^z S_i^z S_{i + \delta_{2,1}}^z + J_{I-II}^\parallel
[ {\bf S}_i \cdot \hat \delta_{2,1} ]
[{\bf S}_{i+\delta_{2,1}} \cdot \hat \delta_{2,1} ]
+ J_{I-II}^\perp [{\bf S}_i \cdot \hat e_{2,1} ]
[ {\bf S}_{i+ \delta_{2,1}} \cdot \hat e_{2,1} ] \Biggr)
\nonumber \\ &&
+ \case 1/2 \sum_{i \in II} \sum_{\delta_2} \Biggl(
J_{II}^z S_i^z S_{i + \delta_2}^z + J_{II}^\parallel [ {\bf S}_i \cdot
\hat \delta_2 ] [{\bf S}_{i+\delta_2} \cdot \hat \delta_2 ]
+ J_{II}^\perp [{\bf S}_i \cdot \hat e_2 ] [ {\bf S}_{i+ \delta_2}
\cdot \hat e_2 ] \Biggr) + J_3 \sum_{i \in I} {\bf S}_i \cdot
{\bf S}_{i + \case 1/2 c \hat z} \ ,
\end{eqnarray}
\end{small}
where $\delta_1$ ($\delta_2$) labels the nearest neighbor vectors in the
plane connecting adjacent CuI's (CuII's) and $\delta_{1,2}$
labels vectors in the CuO plane which give the
displacements of nearest neighboring CuI's relative to a CuII,
and hat indicates a unit vector.  Also $\hat e_1$, $\hat e_{2,1}$,
and $\hat e_2$ are unit vectors in the CuO plane which are perpendicular
to, respectively,  $\delta_1$, $\delta_{2,1}$, and $\delta_2$.

We separate the Hamiltonian ${\cal H}_1$ into an isotropic part,
${\cal H}_0$, and an anisotropic perturbation, ${\cal H}'$.
For that purpose we write
\begin{eqnarray}
\Delta J_1 & = & \case 1/2 (J_I^\parallel + J_I^\perp) - J_I^z \ , \ \ 
\Delta J_{12} = \case 1/2 (J_{I-II}^\parallel + J_{I-II}^\perp)
- J_{I-II}^z \ , \ \ 
\Delta J_2 = \case 1/2 (J_{II}^\parallel + J_{II}^\perp) - J_{II}^z \ , \\
\delta J_1 & = & \case 1/2 (J_I^\parallel - J_I^\perp) \ ,
 \ \ \delta J_{12} = \case 1/2 (J_{I-II}^\parallel - J_{I-II}^\perp) \ ,
 \ \ \delta J_2 = \case 1/2 (J_{II}^\parallel - J_{II}^\perp) \ , \\
\tilde J & = & \case 1/3 (J_I^\parallel + J_I^\perp + J_I^z ) \ , \ \ 
\tilde J_{12} = \case 1/3 (J_{I-II}^\parallel
+ J_{I-II}^\perp + J_{I-II}^z ) \ ,
\ \ \tilde J_2 = \case 1/3 (J_{II}^\parallel + J_{II}^\perp + J_{II}^z ) .
\end{eqnarray}
Thus the $\Delta J$'s describe the out-of-plane anisotropy (i. e. the
energy which gives rise to an easy-plane) which is responsible
for the 5 meV anisotropy gap in the spin-wave spectra of cuprates
which do not have CuII's.  Similarly, the $\delta J$'s describe
the in-plane anisotropy (i. e. the anisotropy within the easy plane)
and they i) are responsible for the weak
ferromagnetic moment\cite{CHOU,KASTNER}
induced in the CuII subsystem by the staggered moment in the CuI subsystem
and ii) contribute to the macroscopic or phenomenological
four-fold anisotropy constant $K_4$.\cite{SOPR1,CHOU,KASTNER}
(We shall see later that ${\cal H}_2$ also contributes to $K_4$.)
Note that $\delta J_{12}$ is what was called $J_{pd}$ in 
Refs. \onlinecite{CHOU} and \onlinecite{KASTNER}, but differs by
a factor of two from its definition in Refs. \onlinecite{SOPR1}
and \onlinecite{SOPR2}.  The largest coupling is
$J$ ($J_2/J \approx J_{12}/J \approx 0.1$ and $J_3/J \approx 10^{-3}$),
while the relative anisotropies, $\Delta J/J$ and $\delta J/J$ are
at most $10^{-3}$.\cite{SOPR1,SOPR2,CHOU,KASTNER}

\begin{figure}
\centerline{\psfig{figure=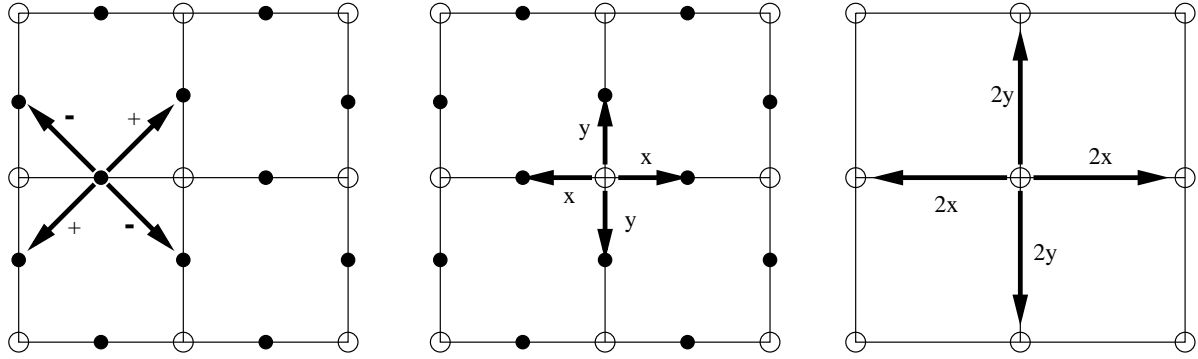}}

\vspace{0.2 in}
\caption{Nearest neighbor vectors connecting magnetic ions in a CuO
plane.  CuI spins are filled circles and CuII spins are open circles.
Left: the vectors $\delta_+$ and $\delta_-$ between nearest neighboring
CuI spins.  Center: the vectors $\delta_x$ and $\delta_y$ which
give the displacements of nearest neighboring CuI's 
relative to a CuII.  Right: the vectors $2\delta_x$ and
$2 \delta_y$ between nearest neighboring CuII spins.}
\label{DELTAFIG}
\end{figure}

With these notations the isotropic Hamiltonian is
\begin{eqnarray}
\label{HZEQ}
{\cal H}_0 &=& \case 1/2 \sum_{i \in I} \sum_{\delta_1} \tilde J
{\bf S}_i \cdot {\bf S}_{i+ \delta_1}
+ \sum_{i \in II} \sum_{\delta_{2,1}} \tilde J_{12}
{\bf S}_i \cdot {\bf S}_{i+ \delta_{2,1}}
+ \case 1/2 \sum_{i \in II} \sum_{\delta_2} \tilde J_2
{\bf S}_i \cdot {\bf S}_{i+ \delta_2}
+ \sum_{i \in I} J_3 {\bf S}_i \cdot {\bf S}_{i+ \case 1/2 c \hat z} \ ,
\end{eqnarray}
and the anisotropic perturbation is
\begin{eqnarray}
\label{HPRIMEEQ}
&& {\cal H}' =  - \case 1/2 \Delta J_1 \sum_{i \in I, \delta_1}
S_i^z S_{i+\delta_1}^z
- \Delta J_{12} \sum_{i \in II , \delta_{2,1} } S_i^z S_{i+\delta_{2,1}}^z
- \case 1/2 \Delta J_2 \sum_{i \in II, \delta_2} S_i^z S_{i+\delta_2}^z
\nonumber \\
&& + \case 1/2 \delta J_1 \sum_{i \in I, \delta_+}
\Biggl( S_i^x S_{i+\delta_+}^y + S_i^y S_{i+\delta_+}^x \Biggr)
- \case 1/2 \delta J_1 \sum_{i \in I, \delta_-}
\Biggl( S_i^x S_{i+\delta_-}^y + S_i^y S_{i+\delta_-}^x \Biggr)
\nonumber \\ && + \delta J_{12} \sum_{i \in II, \delta_x }
\Biggl( S_i^x S_{i+\delta_x}^x - S_i^y S_{i+\delta_x}^y \Biggr) 
+ \delta J_{12} \sum_{i \in II, \delta_y}
\Biggl( S_i^y S_{i+\delta_y}^y - S_i^x S_{i+\delta_y}^x \Biggr) \nonumber \\
&& + \case 1/2 \delta J_2 \sum_{i \in II} \sum_{\delta_2: j=i+\delta_2}
\Biggl[ [{\bf S}_i \cdot \hat \delta_2] [{\bf S}_j \cdot \hat \delta_2]
- [{\bf S}_i \cdot \hat e_2] [{\bf S}_j \cdot \hat e_2] \Biggr] \ ,
\end{eqnarray}
where we introduce the following sums over the $\delta$'s:
\begin{equation}
\delta_x = \pm \case 1/2 a \hat x \ , \ \ \ \ \ 
\delta_y = \pm \case 1/2 a \hat y \ , \ \ \ \ \ 
\delta_+ = \pm \case 1/2 a( \hat x + \hat y) \ , \ \ \ \ \
\delta_- = \pm \case 1/2 a( \hat x - \hat y) \ ,
\end{equation}
as shown in Fig. \ref{DELTAFIG}. In Eq. (\ref{HZEQ}),
$\tilde J= J + \case 1/3 \Delta J$ and similarly for the other
$J$'s.  Since the anisotropy in the $J$'s is so small
(at most of order $10^{-3}$), we henceforth drop the tildes.

It is convenient to express the spin components in a coordinate system
in which one axis (the $\xi$ axis) lies along the line of the staggered
magnetization.  Thus we introduce the axes $\xi$ and $\eta$ which are
obtained from $x$ and $y$ by a rotation about the $z$ axis of $\pi/4$.  Then
\begin{eqnarray}
\label{ROTEQ}
{S^x} & = &(S^\xi - S^\eta )/\sqrt 2 \ , \  \ \ \ \ \
S^y = (S^\xi + S^\eta )/\sqrt 2 \ ,
\end{eqnarray}
so that
\begin{eqnarray}
\label{HPRIMEEQ2}
&& {\cal H}' =  - \case 1/2
\Delta J_1 \sum_{i \in I, \delta_1} S_i^z S_{i+\delta_1}^z
- \Delta J_{12} \sum_{i \in I , \delta_{2,1} } S_i^z S_{i+\delta_{2,1}}^z
- \case 1/2 \Delta J_2 \sum_{i \in II, \delta_2} S_i^z S_{i+\delta_2}^z
\nonumber \\
&& + \case 1/2 \delta J_1 \sum_{i \in I, \delta_+}
\Biggl( S_i^\xi S_{i+\delta_+}^\xi - S_i^\eta S_{i+\delta_+}^\eta \Biggr)
+ \case 1/2 \delta J_1 \sum_{i \in I, \delta_-}
\Biggl( S_i^\eta S_{i+\delta_-}^\eta - S_i^\xi S_{i+\delta_-}^\xi \Biggr)
\nonumber \\
&& - \delta J_{12}  \sum_{i \in II, \delta_x }
\Biggl( S_i^\xi S_{i+\delta_x}^\eta + S_i^\eta S_{i+\delta_x}^\xi \Biggr)
+ \delta J_{12} \sum_{i \in II, \delta_y}
\Biggl( S_i^\xi S_{i+\delta_y}^\eta + S_i^\eta S_{i+\delta_y}^\xi \Biggr) 
\nonumber \\ && - \delta J_2 \sum_{i \in e}
\sum_{\delta_x: j=i+2\delta_x} 
\Biggl( S_i^\xi S_j^\eta + S_i^\eta S_j^\xi \Biggr)
+ \delta J_2 \sum_{i \in e} \sum_{\delta_x: j=i+2\delta_y} 
\Biggl( S_i^\xi S_j^\eta + S_i^\eta S_j^\xi \Biggr) \ ,
\end{eqnarray}
where, in the last line, $i \in e$ indicates that the sum is taken
over only half the CuII spins, i. e. those on the $e$ sublattice
(see Fig. \ref{UCFIG}).

\section{BOSON HAMILTONIAN}

\subsection {Overview of the Calculation}
Since the CuI - CuII interaction is frustrated,
the CuI and CuII sublattices are decoupled within mean-field theory
or within harmonic spin-wave theory at zero wave vector.  In other
words, to calculate the energy gaps at zero wave vector we will need
to include fluctuations, as first indicated by Shender.\cite{EFS}  Here, in
view of the myriad of terms in the Hamiltonian, we need to proceed
in as systematic a way as possible.  In the original work of Shender\cite{EFS}
it was found that the effective coupling between sublattices, which
depends on fluctuations beyond mean-field theory or beyond harmonic
spin-wave theory involved energies of relative order $1/S$ with respect
to energies encountered in mean-field theory.  Accordingly, here we
will calculate all relevant effects in the spin-wave spectrum due to 
anharmonic perturbations up to first order in $1/S$.  Therefore we
analyze perturbative contributions at one-loop order.  To be more specific,
we will introduce the usual Dyson-Maleev boson representation\cite{D-M}
of spin operators,
in terms of which anharmonic perturbations involving three (four) boson
operators are of relative order $1/\sqrt S$ ($1/S$).  This
means that we treat four-operator perturbations within first-order
perturbation theory and three-operator perturbations within second-order
perturbation theory.  In technical language, this would be done by
keeping all such contributions to the wave vector and energy-dependent
self-energy.  Since we work to low order, a more naive approach
(which is entirely equivalent to calculating the self-energy) is
both convenient and easy to follow.  In this naive approach one
truncates all four operator terms by contracting out pairs of
operators in all possible ways.  This reproduces exactly the
results of the one-loop diagrams obtained by treating the four operator
vertices in first order perturbation theory.  In addition, we
would note that all nonHermitian terms at order $1/S$ do not
contribute to first order energies.  So, at order $1/S$ we simply
discard nonHermitian terms.  Since the three-operator terms are
of interest in producing small gaps, we will follow a calculational
method which is strictly correct only at zero wave vector.  The fact
that in our treatment the small perturbations have the wrong
dependence on wave vector is irrelevant because their effect is
only nonnegligible very near zero wave vector.  To avoid the
algebraic complexities due to the fact that the magnetic
structure has six sublattices, we simply construct, by the
methods mentioned above, the effective quadratic Hamiltonian
which includes all the self-energy corrections at order $1/S$.
As a check that our calculations are really as consistent as
we claim, we verify that the gaps have the expected dependence
on the perturbations.  In other words, when the perturbations
are known to not produce gaps, our calculations reproduce that
result.  This type of check indicates that, for instance, our
treatment of three-operator terms in second order perturbation
theory is consistent with our treatment of four-operator terms in
first order perturbation theory.

\subsection{Transformation to Bosons}

We make the following Dyson-Maleev transformation\cite{D-M} to bosons
($a, \ b , \ \dots f$):
\begin{eqnarray}
\label{BOSONEQ}
S_a^+ &=& \sqrt{2S} a \ , \ \ \ \ \ \
S_a^- = \sqrt{2S} a^\dagger \phi(a) \ , \ \ \ \ \ \
S_a^\xi = S - a^\dagger a \nonumber \\
S_b^+ &=& \sqrt{2S} b^\dagger \ , \ \ \ \ \ \
S_b^- = \sqrt{2S} \phi(b) b \ , \ \ \ \ \ \
S_b^\xi = -S + b^\dagger b \nonumber \\
S_c^+ &=& \sqrt{2S} c^\dagger \ , \ \ \ \ \ \
S_c^- = \sqrt{2S} \phi(c) c \ , \ \ \ \ \ \
S_c^\xi = -S + c^\dagger c \nonumber \\
S_d^+ &=& \sqrt{2S} d \ , \ \ \ \ \ \
S_d^- = \sqrt{2S} d^\dagger \phi(d) \ , \ \ \ \ \ \
S_d^\xi = S - d^\dagger d \nonumber \\
S_e^+ &=& \sqrt{2S} e^\dagger \ , \ \ \ \ \ \
S_e^- = \sqrt{2S} \phi(e) e \ , \ \ \ \ \ \
S_e^\xi = -S + e^\dagger e \nonumber \\
S_f^+ &=& \sqrt{2S} f \ , \ \ \ \ \ \
S_f^- = \sqrt{2S} f^\dagger \phi(f) \ , \ \ \ \ \ \
S_f^\xi = S - f^\dagger f \ ,
\end{eqnarray}
where $S^\pm = S^\eta \pm i S^z$, $\phi(x) = 1 -x^\dagger x/(2S)$, and
we have left the site labels implicit.  In
bosonic variables the isotropic interaction between spins assumes the form
\begin{eqnarray}
\label{VSEQ}
{\bf S}_{ai} \cdot {\bf S}_{bj} & = &
S \Biggl( a_i^\dagger a_i + b_j^\dagger b_j + a_i b_j + a_i^\dagger b_j^\dagger
\Biggr) - \case 1/2 \Biggl( b_j^\dagger b_j b_j a_i +
b_j^\dagger a_i^\dagger a_i^\dagger a_i
+ 2 a_i^\dagger a_i b_j^\dagger b_j \Biggr) \nonumber \\
{\bf S}_{ai} \cdot {\bf S}_{ej} & = &
S \Biggl( a_i^\dagger a_i + e_j^\dagger e_j + a_i e_j + a_i^\dagger e_j^\dagger
\Biggr) - \case 1/2 \Biggl( e_j^\dagger e_je_ja_i +
e_j^\dagger a_i^\dagger a_i^\dagger a_i
+ 2 a_i^\dagger a_i e_j^\dagger e_j \Biggr) \nonumber \\
{\bf S}_{ai} \cdot {\bf S}_{fj} & = &
S \Biggl(-a_i^\dagger a_i - f_j^\dagger f_j + a_i^\dagger f_j + a_i f_j^\dagger
\Biggr) - \case 1/2 \Biggl( a_i f_j^\dagger f_j^\dagger f_j
+ f_j a_i^\dagger a_i^\dagger a_i
- 2 a_i^\dagger a_i f_j^\dagger f_j \Biggr) \nonumber \\
{\bf S}_{bi} \cdot {\bf S}_{ej} & = &
S \Biggl( -b_i^\dagger b_i - e_j^\dagger e_j + b_i^\dagger e_j
+ b_i e_j^\dagger
\Biggr) - \case 1/2 \Biggl( b_i^\dagger e_j^\dagger e_j e_j
+ e_j^\dagger b_i^\dagger b_i b_i
- 2 b_i^\dagger b_i e_j^\dagger e_j \Biggr) \nonumber \\
{\bf S}_{bi} \cdot {\bf S}_{fj} & = &
S \Biggl( b_i^\dagger b_i + f_j^\dagger f_j + b_i^\dagger f_j^\dagger + b_i f_j
\Biggr) - \case 1/2 \Biggl( b_i^\dagger f_j^\dagger f_j^\dagger f_j
+ f_j b_i^\dagger b_i b_i
+ 2 b_i^\dagger b_i f_j^\dagger f_j \Biggr) \nonumber \\
{\bf S}_{ei} \cdot {\bf S}_{fj} & = &
S \Biggl( e_i^\dagger e_i + f_j^\dagger f_j + e_i^\dagger f_j^\dagger + e_i f_j
\Biggr) - \case 1/2 \Biggl( e_i^\dagger f_j^\dagger f_j^\dagger f_j
+ f_j e_i^\dagger e_i e_i + 2 e_i^\dagger e_i f_j^\dagger f_j \Biggr) \ .
\end{eqnarray}
The other interactions can be obtained by appropriate relabeling of
boson variables.

The effective bilinear spin-wave Hamiltonian is of the form (see below)
\begin{equation}
\label{BOSONH}
{\cal H} = \sum_{\bf q} \Biggl[ A({\bf q})_{\mu \nu} \xi_\mu({\bf q})^\dagger
\xi_\nu({\bf q}) + \case 1/2 B({\bf q})_{\mu \nu} \xi_\mu^\dagger ({\bf q})
\xi_\nu^\dagger (-{\bf q})
+ \case 1/2 B({\bf q})_{\mu \nu}^* \xi_\mu ({\bf q}) \xi_\nu (-{\bf q})
\Biggr] \ ,
\end{equation}
where $\xi_1({\bf q}) = a({\bf q})$ and so forth (in order $b$, $c$, $d$,
$e$, and $f$).  Here
\begin{equation}
\xi_\mu^\dagger (i) = {1 \over \sqrt N_{\rm uc} } \sum_{\bf q}
e^{i {\bf q} \cdot {\bf r}_i } \xi_\mu^\dagger ( {\bf q}) \ ,
\end{equation}
where $N_{\rm uc}$ is the number of unit cells.

\subsection{Spin-wave Spectrum. General Considerations}
The transformation to normal mode operators $\tau_k({\bf q})$ is
\begin{eqnarray}
\label{TRANSEQ}
\xi_i^\dagger ({\bf q}) & = & \sum_j P_{ij}({\bf q})^* \tau_j^\dagger ({\bf q})
+ \sum_j Q_{ij}({\bf q})^* \tau_j(-{\bf q}) \nonumber \\
\xi_i(-{\bf q}) & = & \sum_j Q_{ij}({\bf q})^* \tau_j^\dagger ({\bf q})
+ \sum_j P_{ij}({\bf q})^* \tau_j(-{\bf q})  \ .
\end{eqnarray}
To preserve the commutation relations we require that
\begin{equation}
\label{NORMEQ}
{\bf P}({\bf q}) {\bf P}^\dagger({\bf q})
- {\bf Q}({\bf q}) {\bf Q}^\dagger({\bf q}) = {\cal I} \ , \ \ \ \
{\bf P}({\bf q}) {\bf Q}^\dagger ({\bf q})
- {\bf Q}({\bf q}) {\bf P}^\dagger ({\bf q}) = 0 \ ,
\end{equation}
where ${\cal I}$ is the unit matrix.

The transformation inverse to Eq. (\ref{TRANSEQ}) is therefore
\begin{eqnarray}
\tau_j^\dagger ({\bf q}) & = & \sum_k P_{kj}({\bf q}) \xi_k^\dagger ({\bf q})
- \sum_k Q_{kj}({\bf q}) \xi_k(-{\bf q}) \nonumber \\
\tau_j({\bf q}) & = & - \sum_k Q_{kj}({\bf q})^* \xi_k^\dagger (-{\bf q})
+ \sum_k P_{kj}({\bf q})^* \xi_k ({\bf q}) \ .
\end{eqnarray}
The equation that determines the normal modes is
\begin{equation}
[ \tau_j({\bf q}) , {\cal H} ]_- = \omega_j({\bf q}) \tau_j({\bf q})
\end{equation}
which gives
\begin{equation} \left[ \begin{array} {cc}
{\bf A}({\bf q}) & {\bf B}({\bf q}) \\
-{\bf B}({\bf q}) & -{\bf A}({\bf q}) \\
\end{array} \right]
\left[ \begin{array} {c} {\bf P}_j({\bf q}) \\
{\bf Q}_j({\bf q}) \\ \end{array}
\right] = \omega_j({\bf q})  \left[ \begin{array} {c}
{\bf P}_j ({\bf q}) \\ {\bf Q}_j ({\bf q}) \\
\end{array} \right] \ .
\end{equation}
where ${\bf P}_j$ is the column vector with components
$P_{1j}, P_{2j}, \dots P_{nj}$ and

\noindent
${\bf P} = [ {\bf P}_1, {\bf P}_2, \dots {\bf P}_n]$
and similarly for the ${\bf Q}'s$.

From now on the arguments are always ${\bf q}$.  Then
\begin{eqnarray}
{[{\bf A} + {\bf B}] [{\bf P}_j + {\bf Q}_j ]} & = & \omega_j
[{\bf P}_j - {\bf Q}_j ] \nonumber \\
{[{\bf A} - {\bf B}] [{\bf P}_j - {\bf Q}_j ]} & = & \omega_j
[{\bf P}_j + {\bf Q}_j ] \ .
\end{eqnarray}
Therefore
\begin{equation}
\label{ABPQEQ}
[{\bf A} + {\bf B}] [{\bf A} - {\bf B}]
[{\bf P}_j - {\bf Q}_j ] = \omega_j^2 [{\bf P}_j - {\bf Q}_j ]\ .
\end{equation}
Hence,
the squares of the spin-wave energies are the eigenvalues of the matrix
\begin{equation}
\label{ABABEQ}
{\bf D} ( {\bf q} ) \equiv
[ {\bf A} ( {\bf q} ) + {\bf B} ( {\bf q} )]
\times [ {\bf A} ( {\bf q} ) - {\bf B} ( {\bf q} )] \ .
\end{equation}
Roughly speaking the matrices ${\bf A+B}$ and ${\bf A-B}$
reproduce the stiffnesses in the
two directions transverse to the sublattice magnetization.

As we shall see later, for the Hamiltonian of the form of
${\cal H}_1$ these dynamical matrices assume the form
\begin{mathletters}
\label{ABEQG}
\begin{eqnarray}
{\bf A} ( {\bf q}) = \left[ \begin{array} { c | c | c | c | c | c }
\hline
a_{11} & a_{12}c_+ & a_{12}c_- & 0 & a_{15}e_x & a_{16} e_x^*\ \ \\
a_{12}c_+ & a_{11} & 0 & a_{12}c_- &a_{16} e_y^* & a_{15} e_y \\
a_{12}c_- & 0 & a_{11} & a_{12}c_+ & a_{16}e_y & a_{15}e_y^* \\
0 & a_{12}c_- & a_{12}c_+ & a_{11} & a_{15}e_x^* & a_{16} e_x  \\
a_{15} e_x^* & a_{16} e_y & a_{16} e_y^* & a_{15} e_x & a_{55} &
a_{56}{c_x+c_y\over 2}  \\
a_{16} e_x & a_{15} e_y^* & a_{15} e_y & a_{16} e_x^*
& a_{56}{c_x+c_y \over 2} & a_{55} \\
\hline
\end{array} \right]
\end{eqnarray}
and
\begin{footnotesize}
\begin{eqnarray}
{\bf B} ( {\bf q}) = \left[ \begin{array} { c | c | c | c | c | c }
\hline
b_{11} & b_{12}c_+ +2J_3 S c_z & b_{12}c_- & 0 & b_{15}e_x & b_{16} e_x^*\ \ \\
b_{12}c_+ +2J_3 S c_z & b_{11} & 0 & b_{12}c_- & b_{16} e_y^* & b_{15} e_y \\
b_{12}c_- & 0 & b_{11} & b_{12}c_++2J_3Sc_z & b_{16}e_y & b_{15}e_y^* \\
0 & b_{12}c_- & b_{12}c_++2J_3 S c_z & b_{11} & b_{15}e_x^* & b_{16} e_x  \\
b_{15} e_x^* & b_{16} e_y & b_{16} e_y^* & b_{15} e_x & b_{55} &
b_{56}{c_x+c_y\over 2}  \\
b_{16} e_x & b_{15} e_y^* & b_{15} e_y & b_{16} e_x^*
& b_{56}{c_x+c_y\over 2} & b_{55} \\
\hline \end{array} \right] \ .
\end{eqnarray}
\end{footnotesize}
\end{mathletters}
where
\begin{eqnarray}
\label{PARAMEQ}
e_x & = & \exp( i q_x a/2) \ , \ \ \ \ \ \ e_y = \exp( i q_y a/2)\ , \ \ \ \ \
c_x = \cos (q_x a ) , \ \ \ \ \ \ c_y = \cos (q_y a )  \nonumber \\
c_+ &=& \cos [a (q_x+q_y)/2] \ , \ \ \ \ \
c_- = \cos [a (q_x-q_y)/2] \ , \ \ \ \ \
c_z = \cos (q_z c/2) \ .
\end{eqnarray}

From now on we will analyze the energies of the modes for wave vectors of the
form ${\bf G} + q_z \hat z$, where ${\bf G}$ is a reciprocal lattice
vector.  In that case the matrices ${\bf A}$ and ${\bf B}$ can be brought
into block diagonal form consisting of three $2 \times 2$ blocks.  The
unitary transformation such that ${\bf U}^\dagger {\bf A} {\bf U}$
and ${\bf U}^\dagger {\bf B} {\bf U}$ are block diagonal
depends on ${\bf G}$, although, of course, the mode energies do not.
For ${\bf G}=0$ we have
\begin{eqnarray}
{\bf U} &=& \left( \begin{array} {c c c c c c}
1/\sqrt 2 &  0 & 1/2 & 0 & 1/2 & 0 \\
0 & 1 /\sqrt 2 & 1/2 & 0 & - 1/2 & 0 \\
0 & - 1 /\sqrt 2 & 1/2 & 0 & - 1/2 & 0 \\
- 1/\sqrt 2 & 0 & 1/2 & 0 & 1/2 & 0 \\
0 & 0 & 0 & 1/\sqrt 2 & 0 & 1/\sqrt 2 \\
0 & 0 & 0 & 1/\sqrt 2 & 0 & -1/\sqrt 2 \\
\end{array} \right) \ .
\end{eqnarray}
${\bf U}({\bf G})$ for general ${\bf G}$ is given in Appendix \ref{INTAPP}.
For ${\bf G}=0$ the transformed block-diagonal matrices corresponding to 
columns 1 and 2, (labeled "12"), those for columns 3 and 4 
(labeled $\sigma=+1$), and those for columns 5 and 6
(labeled $\sigma=-1$) are
\begin{mathletters}
\begin{eqnarray}
{\bf A}_{12} & = & \left( \begin{array} {c c }
a_{11} & 0 \\
0 & a_{11} \\
\end{array} \right) \ , \\
{\bf A}_\sigma & = & \left( \begin{array} {c c }
a_{11} + 2\sigma a_{12} & \sqrt 2 (a_{15} + \sigma a_{16}) \\
\sqrt 2 (a_{15}+\sigma a_{16}) & a_{55} + \sigma a_{56} \\
\end{array} \right) \ ,
\end{eqnarray}
\end{mathletters}
\begin{mathletters}
\begin{eqnarray}
{\bf B}_{12} & = & \left( \begin{array} {c c }
b_{11} & 2J_3 S c_z \\
2J_3 S c_z & b_{11} \\
\end{array} \right) \ , \\
{\bf B}_\sigma & = & \left( \begin{array} {c c }
b_{11} + 2\sigma b_{12} + 2\sigma J_3 S c_z  & \sqrt 2 (b_{15}
+ \sigma b_{16}) \\
\sqrt 2 (b_{15}+\sigma b_{16}) & b_{55} + \sigma b_{56} \\
\end{array} \right) \ .
\end{eqnarray}
\end{mathletters}
These results remain valid when ${\cal H}_2$ is included,
providing it is evaluated at zero wave vector, which, as we have
said, is an excellent approximation.

\subsection{Isotropic Interactions}
For a qualitative understanding of the mode structure we start by
considering the results of linearized spin-wave theory when all
exchange interactions are isotropic.  Then one has 
\begin{eqnarray}
a_{11}=4JS+2J_3S \ , \ \  a_{16}=b_{15}=J_{12} S \ , \ \ 
a_{55}=b_{56}=4J_2S \ , \ \ b_{12}=2JS
\end{eqnarray}
and all the other matrix elements are zero.

In the "12" sector, we find two optical modes which are degenerate
for all $q_z$, with
\begin{equation}
\label{1234EQ}
(\omega/S)^2 = (4J+2J_3)^2 - (2J_3c_z)^2 \approx 16J^2 \ .
\end{equation}
Spin-wave interactions and anisotropic exchange interactions will
have only negligible effects on these optical modes and accordingly
we will generally not discuss these modes any further.\cite{AEP}
In the $\sigma=+1$ sector we find modes with energies 
\begin{eqnarray}
({\omega_+^>}/S)^2 & =& 2J_3 (1-c_z) \left[ 8J + 2 J_3 (1+c_z) \right]
\approx 16J J_3 (1-c_z) \ , \nonumber \\ {\omega_+^<}^2 & =& 0 \ .
\end{eqnarray}
Finally, the $\sigma=-1$ sector has modes whose energies are
\begin{eqnarray}
({\omega_-^>}/S)^2 & =& 2J_3 (1-c_z) \left[ 8J + 2 J_3 (1+c_z) \right]
\approx 16J J_3 (1-c_z) \ , \nonumber \\ {\omega_-^<}^2 & =& 0 \ .
\end{eqnarray}
Note that all modes are gapless  at zero wave vector
and that for both $\sigma=+1$ and
$\sigma=-1$ we have a dispersionless zero frequency mode
due to the frustration of the CuI - CuII interaction.  

Several aspects of the above results are noteworthy.
First of all, as we will see from our calculation of the dynamic
structure factor in Sec. VII, the $\sigma=+1$ ($\sigma=-1$)
sector corresponds to modes in which the spins move out of (within)
the basal plane and therefore we will refer to these modes as
out-of-plane (in-plane) modes.  (This identification can also
be deduced from the way the mode energies depend on the
out-of-plane and in-plane anisotropies.)  For both out-of-plane
and in-plane modes note the existence of a completely gapless mode:
when the CuI's rotate in phase, they produce zero coupling on the
CuII's, each plane of which can be rotated with zero cost in energy.  
The higher-energy out-of-plane and in-plane modes are degenerate
because we have not yet included any anisotropy and these modes
give rise to the usual twofold degenerate mode of the CuI subsystem.
Even when more general anisotropic interactions are included,
the higher-energy modes remain mostly on the CuI's and the
lower-energy modes remain mostly on the CuII's.

\subsection{Mode Energies for General Interactions}

Here we give the mode energies in terms of the matrix elements of
Eq. (\ref{ABEQG}) for general interactions for wave vectors
of the form ${\bf q}=(0,0,q_z)$.  (The eigenvalues, but not the matrices,
are invariant under addition of a reciprocal lattice vector
${\bf G}$ to ${\bf q}$.)  To evaluate Eq. (\ref{ABABEQ})
within the low-frequency sectors $\sigma=\pm 1$, we record the
form of the two by two blocks.  Since we need both ${\bf A}+{\bf B}$ and
${\bf A} - {\bf B}$, we write
\begin{small}
\begin{eqnarray}
\label{APLBEQ}
[{\bf A} + \eta {\bf B}]_\sigma &=& \begin{array} {| c | c |} \hline \ \
a_{11} + 2 \sigma a_{12} + \eta b_{11} + 2\sigma \eta J_3 S c_z
+ 2 \sigma \eta b_{12}
\ \ & \ \ \sqrt 2 [ a_{15} + \sigma a_{16} + \eta b_{15} + \sigma \eta b_{16}]
\ \  \\ \hline \sqrt 2 [ a_{15} + \sigma a_{16} + \eta b_{15} +
\sigma \eta b_{16}] &
a_{55} + \sigma a_{56} + \eta b_{55} + \sigma \eta b_{56} \\ \hline
\end{array}
\end{eqnarray}
\end{small}
In evaluating Eq. (\ref{ABABEQ}) it is useful to note that in the
$\sigma=+1$ sector the matrix element
$[{\bf A}_{11} + {\bf B}_{11}]_+ \sim 8JS$ is by far the largest
matrix element.  Similarly in the $\sigma=-1$ sector
$[{\bf A}_{11} - {\bf B}_{11}]_- \sim 8JS$ is by far the largest
matrix element.  In either case, then, Eq. (\ref{ABABEQ}) gives
the squares of the mode energies as the eigenvalues of a matrix
(or its transpose) of the form
\begin{eqnarray}
\begin{array} {|c|c|} \hline
\ \ U \ \ &\ \  V\ \  \\ \hline V & W \\ \hline \end{array}
\hspace{0.2 in} \begin{array} {|c|c|} \hline
\ \ u\  \ &\ \  v\ \  \\ \hline v & w \\ \hline \end{array} \ ,
\end{eqnarray}
where $\sqrt {Uu}$ dominates all other matrix elements.  In that case the
eigenvalues are
\begin{eqnarray}
\label{SOLN}
(\omega^>)^2 = Uu +2Vv + Wv^2/u \ ,
\ \ \ \ \ \ \ \ (\omega^<)^2 = (UW-V^2)(uw-v^2)/(\omega^>)^2 \ .
\end{eqnarray}
Explicitly, within the sectors $\sigma=\pm1$, we have
\begin{eqnarray}
\label{UVWEQ}
U_\sigma & = & a_{11} + 2 b_{12} +  \sigma (2a_{12} + b_{11}) + 2J_3S c_z \ ,
\nonumber \\
V_\sigma &=& \sqrt 2 [ a_{15} + b_{16} + \sigma (a_{16} + b_{15} ) ]  \ ,
\nonumber \\
W_\sigma & = & a_{55} + b_{56} + \sigma (a_{56} + b_{55}) \ , \nonumber \\
u_\sigma & = & a_{11} - 2b_{12} -2J_3S c_z +
\sigma (2a_{12} - b_{11})
\ , \nonumber \\
v_\sigma &=& \sqrt 2 [ a_{15} - b_{16} + \sigma (a_{16} - b_{15}) ] \ ,
\nonumber \\
w_\sigma & = & a_{55} - b_{56} + \sigma (a_{56} - b_{55}) \ .
\end{eqnarray}
Substituting these evaluations into Eq. (\ref{SOLN}) [or, if need be,
exactly implementing Eq. (\ref {ABABEQ})] gives the four
low energy modes for wave vectors along the $\underline c$ direction.
Obviously, since the mode energies are derived from a
two by two dynamical matrix, we can easily obtain exact expressions
for their energies.

\section{NONLINEAR SPIN WAVES}

\subsection {$1/S$ Corrections to $J$, $J_3$, and $J_2$}
When we include the effect of spin-wave interactions at order $1/S$
on the CuI-CuI interactions or on the CuII-CuII interactions, we
expect to get a simple renormalization.  For the exchange
interactions between neighbors in the same CuO plane, this
effect is well known.  As explained above, we decouple the fourth
order terms in ${\bf S}_{ai} \cdot {\bf S}_{bj}$ as
\begin{eqnarray}
- \case 1/2 [ b_j^\dagger b_j b_j a_i + b_j^\dagger a_i^\dagger a_i^\dagger a_i
+ 2 a_i^\dagger a_i b_j^\dagger b_j ] & \rightarrow &
- \langle a_i^\dagger a_i + a_i b_j \rangle [ a_i^\dagger a_i + b_j^\dagger b_j
+ a_i b_j + a_i^\dagger b_j^\dagger ] 
\end{eqnarray}
and those in ${\bf S}_{ei} \cdot {\bf S}_{fj}$ as
\begin{eqnarray}
- \case 1/2 [ e_i^\dagger f_j^\dagger f_j^\dagger f_j + f_j e_i^\dagger e_i e_i
+ 2 e_i^\dagger e_i f_j^\dagger f_j ] & \rightarrow &
- \langle e_i^\dagger e_i + e_i f_j \rangle [ e_i^\dagger e_i + f_j^\dagger f_j
+ e_i f_j + e_i^\dagger f_j^\dagger ] \ .
\end{eqnarray}
From this result we conclude that $J$ and $J_2$ should be replaced by
$Z_c J$ and $Z_2J_2$, respectively, with
$Z_c =1 - (1/S) \langle a_i^\dagger a_i + a_i b_j \rangle$,
where $i$ and $j$ are
nearest neighboring sites on the $\underline a$ and $\underline b$ 
sublattices, respectively, and
$Z_2 = 1 - (1/S) \langle e_i^\dagger e_i + e_i f_j \rangle$
in a similar notation, so that $Z_2\approx Z_c$ at zero temperature.
$Z_c$ has been calculated more accurately than this.
(In Ref. \onlinecite{1/S2} the value $Z_c \approx 1.17$ is given.)
For $J_3$ we note that $a_i$ and $b_j$ refer to sites in different
CuO planes, in which case $\langle a_i b_j \rangle \approx 0$.
So we should replace $J_3$ by $\tilde Z_3J_3$, where
\begin{eqnarray}
\label{Z3}
\tilde Z_3 = 1 - (1/S) \langle a_i^\dagger a_i \rangle \ ,
\end{eqnarray}
so that $\tilde Z_3/2$ is essentially the magnitude of the zero-point
staggered spin in the presence of quantum fluctuations.  (Thus
$\tilde Z_3\approx 0.6$ is very different from $Z_c$.)

\subsection{The Effect of Spin-Wave Interactions on $J_{12}$}

Now we discuss the effect of spin-wave interactions on $J_{12}$,
i. e. we consider the Shender interaction.\cite{EFS}
Correctly to order $1/S$
we construct the effective quadratic Hamiltonian by contracting two
operators in all possible ways.  I. e. we replace two operators by
the thermal expectation value (indicated by $\langle \dots \rangle$)
of their product.  Applying this
procedure to the relevant terms in Eq. (\ref{VSEQ}) we obtain the
effective interactions between a CuI spin $i$ on sublattice
$\underline a$ and nearest neighboring CuII spins as
\begin{eqnarray}
V_{ae}/J_{12} & = & a_i^\dagger a_i \Biggl( S - \langle a_i^\dagger e_j^+
\rangle - \langle e_j^\dagger  e_j \rangle \Biggr) + e_j^\dagger e_j \Biggl(
S - \langle e_j a_i \rangle - \langle a_i^\dagger a_i \rangle \Biggr)
\nonumber \\ &+ & a_i e_j \Biggl(S - \langle e_j^\dagger e_j \rangle -
\langle a_i^\dagger e_j^\dagger \rangle \Biggr) + a_i^\dagger e_j^\dagger
\Biggl( S - \langle a_i^\dagger a_i \rangle - \langle a_i e_j \rangle \Biggr)
\\ V_{af}/J_{12} & = & a_i^\dagger a_i \Biggl( -S - 
\langle a_i^\dagger  f_j \rangle + \langle f_j^\dagger f_j \rangle \Biggr)
+ f_j^\dagger f_j \Biggl( -S - \langle a_i f_j^\dagger \rangle 
+ \langle a_i^\dagger a_i \rangle \Biggr) \nonumber \\
&+& a_i^\dagger f_j \Biggl( S - \langle a_i^\dagger a_i \rangle
+ \langle a_i f_j^\dagger \rangle \Biggr) + f_j^\dagger a_i
\Biggl( S - \langle f_j^\dagger f_j \rangle + \langle a_i^\dagger f_j \rangle
\Biggr) \ .
\end{eqnarray}
Here to leading order in $1/S$ it suffices to evaluate the
various expectation values with respect to the original
quadratic Hamiltonian.  At quadratic order we have symmetry such that
$\langle a_i^\dagger e_j^\dagger \rangle = \langle a_i e_j \rangle$,
$\langle e_j^\dagger e_j \rangle = \langle f_j^\dagger f_j \rangle$, etc.
We define
\begin{eqnarray}
\label{VDMEQ}
J_{12}^{(1)} S/J_{12}  &=& S
+ \langle a_i f_j^\dagger \rangle - \langle a_i^\dagger a_i \rangle
\nonumber \\
J_{12}^{(2)} S/J_{12} &=& S
+ \langle a_i f_j^\dagger \rangle - \langle f_j^\dagger f_j \rangle
\nonumber \\
J_{12}^{(3)} S/J_{12} &=& S
- \langle a_i e_j \rangle - \langle a_i^\dagger a_i \rangle
\nonumber \\
J_{12}^{(4)} S/J_{12} &=& S
- \langle a_i e_j \rangle - \langle e_j^\dagger e_j \rangle \ .
\end{eqnarray}
Note that $J_{12}^{(3)} - J_{12}^{(4)} = J_{12}^{(1)}-J_{12}^{(2)}$. Then
\begin{eqnarray}
V_{ae} & = & J_{12}^{(4)} S a_i^\dagger a_i + J_{12}^{(3)} S e_j^\dagger e_j
+ J_{12}^{(4)} S a_i e_j + J_{12}^{(3)} S a_i^\dagger e_j^\dagger \nonumber \\
V_{af} & = & - J_{12}^{(2)} S a_i^\dagger a_i - J_{12}^{(1)} S f_j^\dagger f_j
+ J_{12}^{(1)} S a_i^\dagger f_j + J_{12}^{(2)} S f_j^\dagger a_i \ .
\end{eqnarray}

Since we only work to order $1/S$, we keep only the Hermitian part of
these perturbations:
\begin{eqnarray}
\label{VAEEQ}
V_{ae} & = & J_{12}^{(4)} S a_i^\dagger a_i + J_{12}^{(3)} S e_j^\dagger e_j
+ J_{12}^{(34)} S \Biggl( a_i e_j + a_i^\dagger e_j^\dagger \Biggr) \nonumber \\
V_{af} & = & - J_{12}^{(2)} S a_i^\dagger a_i - J_{12}^{(1)} S f_j^\dagger f_j
+ J_{12}^{(12)} S \Biggl( a_i^\dagger f_j + f_j^\dagger a_i \Biggr) \ ,
\end{eqnarray}
where
\begin{eqnarray}
\label{VDM2EQ}
J_{12}^{(12)} = \case 1/2 [J_{12}^{(1)}+J_{12}^{(2)} ] \ , \ \ \ \ \
J_{12}^{(34)} = \case 1/2 [J_{12}^{(3)}+J_{12}^{(4)} ] \ .
\end{eqnarray}
As it turns out, the energies of the modes we study depend only on
the single parameter
\begin{eqnarray}
\alpha  & = & (J_{12}^{(4)} - J_{12}^{(2)})S 
= (J_{12}^{(3)} - J_{12}^{(1)})S 
= - J_{12} (\langle a_i e_j \rangle + \langle a_i f_j^\dagger \rangle ) \ .
\end{eqnarray}
Note that the parameter $\delta$ in Ref. \onlinecite{KIM} is
$\delta =\alpha/S$.  We evaluate this parameter in Appendix \ref{APPSH}
and find
\begin{eqnarray}
\label{ALPHEQ}
\alpha= C_\alpha J_{12}^2/J \ ,
\end{eqnarray}
where $C_\alpha$ is a numerical factor which we found to be 0.1686.
The anharmonic effects of Eq. (\ref{VAEEQ}) give rise to contributions
to the dynamical matrix of
\begin{eqnarray}
\delta a_{11} &=& \alpha \ , \nonumber \\
\delta a_{16} &=&  J_{12}^{(12)}S - J_{12} S \ , \nonumber \\
\delta a_{55} &=&  2 \alpha \ , \nonumber \\
\delta b_{15} &=&  J_{12}^{(34)}S - J_{12} S \ .
\end{eqnarray}
It is known\cite{EFS,BCT,FCC} that in simpler problems these
anharmonic effects give rise {\it at zero momentum} to effective
biquadratic exchange interactions between sublattices which
otherwise are frustrated in harmonic theory.  To emphasize
this point we treat a biquadratic interaction between
nearest CuI - CuII neighbors (in the plane) which is of
the form
\begin{eqnarray}
H_{\rm BQ} &=& - { j_{\rm BQ} \over S^2}
\sum_{i \in II} \sum_{\delta_{2,1}}
({\bf S}_i \cdot {\bf S}_{i+\delta_{2,1}})^2 \ .
\end{eqnarray}
Then the contributions to the dynamical matrix are
\begin{eqnarray}
\delta a_{11} & = & 4 j_{\rm BQ} S \ , \nonumber \\
\delta a_{16} & = & -2 j_{\rm BQ} S \ , \nonumber \\
\delta a_{55} & = & 8 j_{\rm BQ} S \ , \nonumber \\
\delta b_{15} & = & 2 j_{\rm BQ} S \ .
\end{eqnarray}
Then using Eqs. (\ref{UVWEQ}) and (\ref{SOLN}) we find the mode
energies at zero transverse wave vector (for large $J$) are now
\begin{eqnarray}
(\omega_\sigma^>)^2 & = & 8JS [\alpha + 4 j_{\rm BQ} S
+ 2J_3S(1-c_z) ] \ \equiv 8JS [\alpha_{\rm eff} + 2J_3S(1-c_z) ]
, \nonumber \\ 
(\omega_\sigma^<)^2 & = &
{ 4 \alpha_{\rm eff}  J_3S (1 - c_z)(8J_2S + \alpha_{\rm eff} )
\over \alpha_{\rm eff} + 2J_3S(1-c_z)  } \ ,
\end{eqnarray}
where
\begin{eqnarray}
\label{AEFFEQ}
\alpha_{\rm eff} = \alpha + 4j_{\rm BQ} S \ .
\end{eqnarray}
These results demonstrate that the Shender interaction does
mimic a biquadratic exchange interaction at long wave length.
However, in view of the relation for spin 1/2 that
$({\bf S}_i \cdot {\bf S}_j)^2 = \case 3/{16} - \case 1/2
{\bf S}_i \cdot {\bf S}_j$, a biquadratic exchange interaction
between two spins 1/2 is equivalent to a Heisenberg exchange
interaction, and we may therefore assume that $j_{\rm BQ}$ vanishes. 

\begin{figure}
\centerline{\psfig{figure=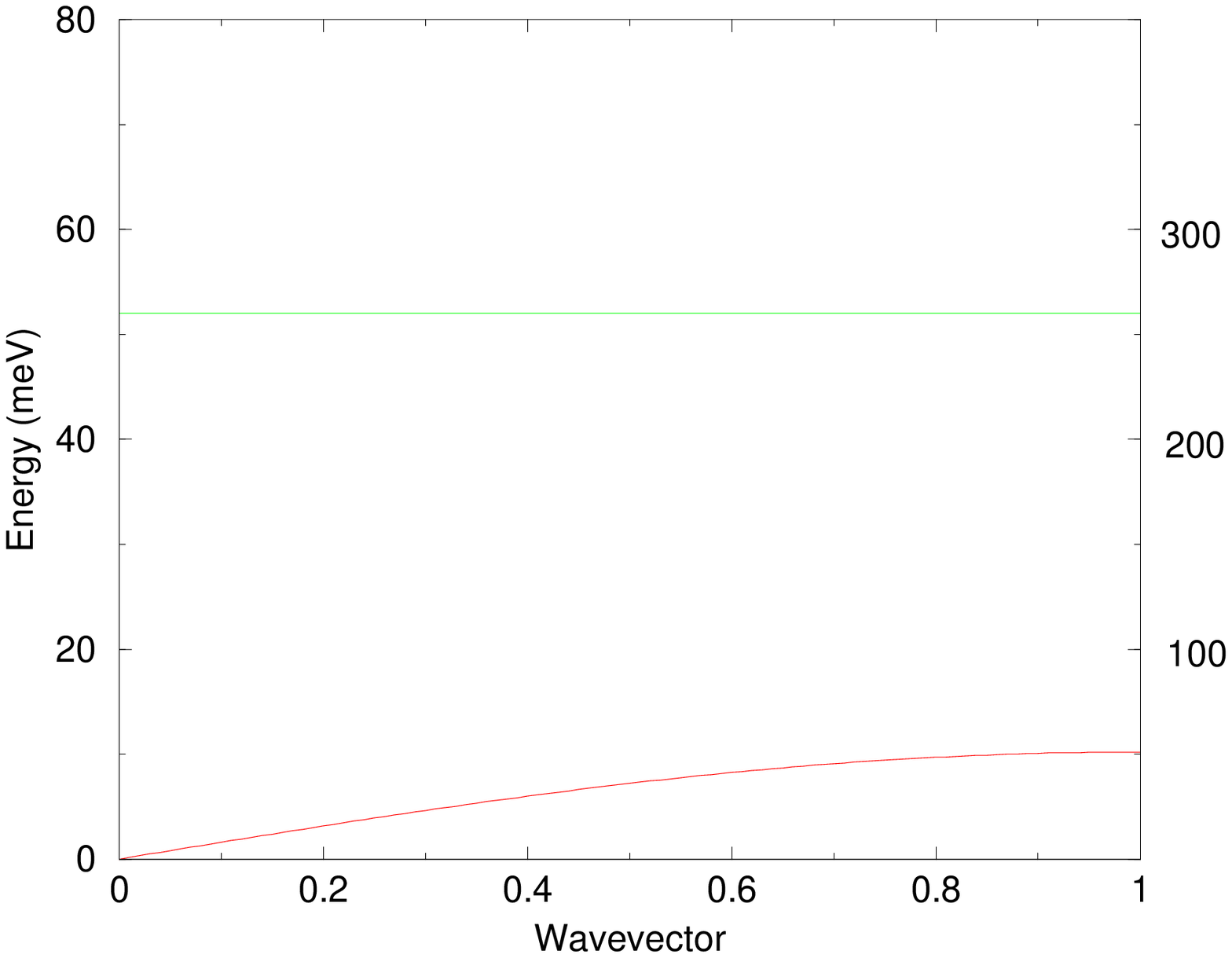,width=3in,height=3in}
\hspace{0.5 in}
\psfig{figure=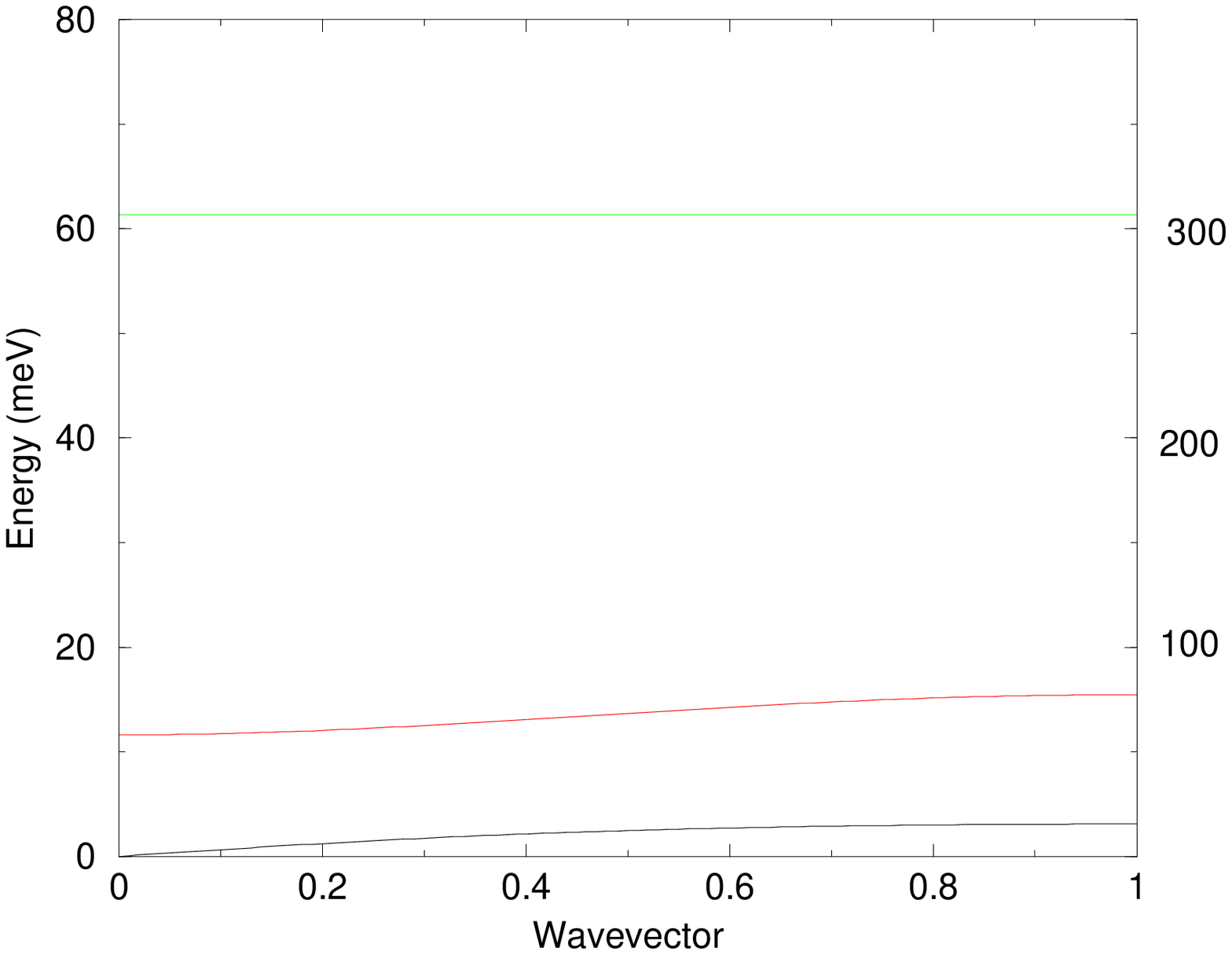,width=3in,height=3in}}

\caption{Spin-wave spectrum for wave vector $=q_zc/(2\pi )$ along the
$\underline c$ direction in the absence of anisotropy. Each mode is
two-fold degenerate.  The left-hand scale applies to the lower modes
and the right-hand scale applies to the optical mode.
Left: without spin-wave interactions. In this case one mode
has zero energy for arbitrary wave vector in the $\underline c$
direction.  Right: with spin-wave interactions. In the presence of
easy plane anisotropy, the two-fold degeneracy is removed and only
one mode (corresponding to rotation within the easy plane) is
gapless at zero wave vector.  When the four-fold in-plane
anisotropy is also included there are no gapless modes.}
\label{SWAVEFIG}
\end{figure}

As before, there is degeneracy between in-plane and out-of-plane 
energies because we have not yet included anisotropy.
However, by taking into account spin-wave interactions we
now have the mode structure one would expect for an isotropic
antiferromagnet: We have a doubly degenerate zero energy Goldstone
mode at zero wave vector, and doubly degenerate nonzero energy modes
for zero wave vector as shown in the right-hand panel of Fig.
\ref{SWAVEFIG}.  The quantum gap in the optical mode $\omega_\sigma^>$
at zero wave vector has been obtained for a number of other frustrated
systems in several theoretical studies\cite{CLH,BCT,FCC} beginning
with the work of Shender.\cite{EFS}  However, because we have two
subsystems which order at different temperatures,
the emergence of this gap has a very unique signature not present
in other experimental systems studied up to now.\cite{EFSGAP}

\section{INCLUSION OF ANISOTROPIES}

\subsection{Out -- of -- Plane Exchange Anisotropy}

To obtain the correct energy gaps at zero wave vector we must
add the anisotropy due to anisotropic exchange interactions.
(Since we are dealing with spin 1/2's, there can be no single
ion anisotropy.) In this subsection we include out--of--plane exchange
anisotropy.  This part of the anisotropic exchange energy 
between sublattices $\underline a$ and $\underline b$
of the CuI's is given as
\begin{equation}
V_{ab} \equiv - \Delta J_1  \sum_{i \in a, j \in b }
S_{ai}^z S_{bj}^z \Delta_{ij} \ ,
\end{equation}
where $\Delta_{ij}$ is defined so as to implement the nearest
neighbor restriction.  Thus, neglecting anharmonicity, we write
\begin{eqnarray}
V_{ab} & = & \case 1/4 \Delta J_1 \sum_{i \in a, j \in b}
[S_{ai}^+ - S_{ai}^-][S_{bj}^+ -S_{bj}^-] \Delta_{ij} \nonumber \\
&=& \case 1/2 \Delta J_1 S \sum_{i \in a, j \in b}
(a_i - a_i^\dagger ) (b_j^\dagger - b_j) \Delta_{ij} \nonumber \\
&= & \Delta J_1 S \sum_{\bf q} [a^\dagger({\bf q}) b({\bf q}) +
b^\dagger({\bf q}) a({\bf q}) - a^\dagger({\bf q}) b^\dagger(-{\bf q})
-a({\bf q}) b(-{\bf q})] c_+ \ .
\end{eqnarray}
This result allows us to identify the contribution to the
parameters of the dynamical matrix introduced in Eq. (\ref{ABEQG}) as
\begin{eqnarray}
\label{JIEQ}
\delta a_{12} = \Delta J_1S \ , \ \ \ \  \delta b_{12}=- \Delta J_1S \ ,
\end{eqnarray}
without having to explicitly consider the other CuI-CuI interactions.

Next we consider the out--of--plane anisotropy of the CuI -- CuII
interactions.  From the form of Eq. (\ref{ABEQG}) we see that we only need
construct the a-e and a-f interactions.  For the a-e interaction we have
\begin{eqnarray}
V_{ae} &=& - \Delta J_{12} \sum_{i \in a , j \in e }
S_{ai}^z S_{ej}^z \Delta_{ij} =
\case 1/4 \Delta J_{12} \sum_{i \in a , j \in e }
\Biggl( S_{ai}^+ - S_{ai}^- \Biggr)
\Biggl( S_{ej}^+ - S_{ej}^- \Biggr) \Delta_{ij} \nonumber \\ &=&
\case 1/2 \Delta J_{12} S \sum_{a \in i , j \in e }
\Biggl( a_i - a_i^\dagger \Biggr) \Biggl( e_j^\dagger - e_j \Biggr) \Delta_{ij}
\nonumber \\ &=&
\case 1/2 \Delta J_{12} S \sum_{\bf q} \Biggl[ a({\bf q}) e^\dagger ({\bf q})
e^{i{\bf q} \cdot ( {\bf r}_e - {\bf r}_a ) } -
a^\dagger(-{\bf q}) e^\dagger ({\bf q}) e^{i{\bf q} \cdot
( {\bf r}_e - {\bf r}_a ) } + {\rm h. c.} \Biggr] \nonumber \\ &=&
\case 1/2 \Delta J_{12} S \sum_{\bf q} \Biggl[ a({\bf q}) e^\dagger ({\bf q})
e_x^* - a^\dagger (-{\bf q}) e^\dagger ({\bf q}) e_x^*
+ a^\dagger ({\bf q}) e({\bf q}) e_x
- a (-{\bf q}) e({\bf q}) e_x \Biggr] \ ,
\end{eqnarray}
which gives a contribution to the dynamical matrix with
\begin{eqnarray}
\delta a_{15} = \case 1/2 \Delta J_{12} S \ , \ \ \ \ \ \
\delta b_{15} = - \case  1/2 \Delta J_{12} S \ .
\end{eqnarray}
Similarly
\begin{eqnarray}
V_{af} &=& \case 1/2 \Delta J_{12} S \sum_{\bf q} \Biggl[
a({\bf q}) f(-{\bf q}) e_x - a^\dagger ({\bf q}) f({\bf q}) e_x^*
+ a^\dagger ({\bf q}) f^\dagger(-{\bf q}) e_x^*
- a ({\bf q}) f^\dagger({\bf q}) e_x \Biggr] \ ,
\end{eqnarray}
from which we deduce that
\begin{eqnarray}
\delta a_{16} = - \case 1/2 \Delta J_{12} S \ , \ \ \ \ \ \
\delta b_{16} = \case 1/2 \Delta J_{12} S \ .
\end{eqnarray}

Finally we include the out--of--plane anisotropy of the CuII-CuII
interactions.  Thus
\begin{eqnarray}
V_{ef} & = & - \Delta J_2 \sum_{i \in e , j \in f} S_{ei}^z S_{fj}^z
\Delta_{ij} \nonumber \\ 
&=& \case 1/4 \Delta J_2 \sum_{i \in e , j \in f}
[ S_{ei}^+ - S_{ei}^-] [ S_{fj}^+ - S_{fj}^-] \Delta_{ij} \nonumber \\
&=& \case 1/2 \Delta J_2 S \sum_{i \in e , j \in f}
(e_i^\dagger - e_i)(f_j-f_j^\dagger) \Delta_{ij} \nonumber \\
&=& \Delta J_2 S \sum_{\bf q} \Biggl( e^\dagger({\bf q}) f({\bf q})
-e^\dagger({\bf q}) f^\dagger(-{\bf q}) - e({\bf q}) f(-{\bf q})
+ f^\dagger({\bf q}) e({\bf q}) \Biggr) [ c_x + c_y] \ ,
\end{eqnarray}
which leads to
\begin{eqnarray}
\delta a_{56} = 2 \Delta J_2 S \ , \ \ \ \ \
\delta b_{56} = -2 \Delta J_2 S \  .
\end{eqnarray}

The renormalization (at order $1/S$) of the out-of-plane anisotropy
is accomplished by replacing
$\sqrt{J\Delta J_1}$ by $Z_g \sqrt {J \Delta J_1}$.\cite{PETIT}

It is instructive to see the influence of this anisotropy on
the gaps at zero wave vector.  Referring to Eq. (\ref{SOLN}) we
see that the high energy mode gap due to the Shender fluctuation
term, causes $Uu$ to be nonzero.  To check for gaps in
the mode energies $\omega_\sigma^<$ at zero wave vector it
suffices to consider the quantity
\begin{eqnarray}
\Lambda \equiv uw-v^2 = [2\Delta J_1 S (1+ \sigma)+ \alpha ] 
[2 \alpha + 2 \Delta J_2 S (1 + \sigma)] - [ - \sqrt 2 \sigma \alpha ]^2 \ .
\end{eqnarray}
When we turn off both out-of-plane anisotropies, $\Delta J_1$ and
$\Delta J_2$, the two modes $\omega_\sigma^<$ are gapless.
When we allow the out-of-plane anisotropy
to be nonzero, we clearly introduce a gap ($\Lambda$ is nonzero) in
the out-of-plane ($\sigma=1$) sector but not in the in-plane
($\sigma =-1$) sector.  This result follows from the fact that
the spins can still undergo a global rotation within
the easy plane at no cost in energy.  Hence we still have
a single Goldstone mode with
zero energy at zero wave vector.  In order for this mode to have a gap,
we have to take account of effects which lead to a fourfold
in-plane anisotropy which we consider in the next subsection.

\subsection{In -- Plane Exchange Anisotropy}

\subsubsection{CuI-CuI Interactions}

In this subsection we discuss the effects of the in--plane anisotropy
of the CuI-CuI exchange interactions.  First of all, note that this
perturbation is extremely weak.  It gives rise to an effective
fourfold anisotropy.  This very small fourfold anisotropy only
has a nonnegligible effect within the low frequency sector and
even there only at zero wave vector.  The Hamiltonian describing
the in--plane anisotropy of the CuI-CuI interactions is
\begin{eqnarray}
V_{\rm in} &=& \delta J_1 \sum_{i \in a,d ; \delta} \sigma(\delta)
\Biggl( S_i^\xi S_{i + \delta}^\xi
- S_i^\eta S_{i+\delta}^\eta \Biggr) \ ,
\end{eqnarray}
where $j=i+\delta$, $\delta$ is summed over four values
(the two $\delta_+$'s and the two $\delta_-$'s), and
$\sigma(\delta_\pm ) =\pm 1$.  Then
\begin{eqnarray}
V_{\rm in} &=& \delta J_1 \sum_{i \in a,d ; \delta} \sigma(\delta)
\Biggl[  (S - \alpha_i^\dagger \alpha_i )(-S + \beta_j^\dagger \beta_j) -
\case 1/4 (2S) \Biggl( \alpha_i + \alpha_i^\dagger \phi(\alpha_i)
 \Biggr) \Biggl( \beta_j^\dagger + \phi(\beta_i)
 \beta_j \Biggr) \Biggr] \nonumber \\
&=& \delta J_1 \sum_{i \in a,d ; \delta} \sigma(\delta) \Biggl(
- \alpha_i^\dagger \alpha_i \beta_j^\dagger \beta_j - \case 1/2 S [\alpha_i + 
\alpha_i^\dagger ] [ \beta_j^\dagger + \beta_j ]
+ \case 1/4 \alpha_i^\dagger \alpha_i^\dagger
\alpha_i (\beta_j^\dagger + \beta_j) \nonumber \\ &&  \ \ 
+ \case 1/4 (\alpha_i^\dagger + \alpha_i) \beta_j^\dagger \beta_j \beta_j
- {1 \over 8S} \alpha_i^\dagger \alpha_i^\dagger \alpha_i \beta_j^\dagger
\beta_j \beta_j  \Biggr) \ ,
\end{eqnarray}
where $\alpha_i = a$ if site $i$ is an $a$ site and $\alpha_i=d$ if
$i$ is a $d$ site, and similarly for $\beta_j$.  We write
\begin{equation}
V_{\rm in} = V_{2, {\rm in}} + V_{4,{\rm in}} + V_{6,{\rm in}} \ ,
\end{equation}
where the subscript 2 (4 or 6) indicates terms quadratic (fourth or sixth)
order in boson operators.  Since we work systematically to first
order in $1/S$, we neglect $V_{6,{\rm in}}$.  Also
\begin{eqnarray}
V_{2, {\rm in}}  &=& - \case 1/2 \delta J_1 S \sum_{i \delta} \sigma ( \delta)
(\alpha_i + \alpha_i^\dagger ) (\beta_j + \beta_j^\dagger ) \nonumber \\ &=&
- \delta J_1 S \sum_{\delta , {\bf k} } \Biggl(
[a^\dagger({\bf k} ) + a(-{\bf k}) ] [b({\bf k} ) + b^\dagger(-{\bf k}) ]
+ [d^\dagger({\bf k} ) + d(-{\bf k}) ] [c({\bf k} ) + c^\dagger (-{\bf k})
]\Biggr) c_+ \nonumber \\ &&
+ \delta J_1 S \sum_{\delta , {\bf k} } \Biggl(
[a^\dagger ({\bf k} ) + a(-{\bf k}) ] [c({\bf k} ) + c^\dagger (-{\bf k}) ]
+ [d^\dagger ({\bf k} ) + d(-{\bf k}) ] [b({\bf k} )
+ b^\dagger (-{\bf k}) ]\Biggr) c_-
\end{eqnarray}
and
\begin{eqnarray}
V_{4,\rm in} &=& \delta J_1 \sum_{i \in a,d ; \delta} \sigma(\delta) \Biggl(
- \alpha_i^\dagger \alpha_i \beta_j^\dagger \beta_j
+ \case 1/4 \alpha_i^\dagger \alpha_i^\dagger
\alpha_i (\beta_j^\dagger + \beta_j) + \case 1/4 (\alpha_i^\dagger + \alpha_i)
\beta_j^\dagger \beta_j \beta_j \Biggr) \ .
\end{eqnarray}

We now consider the effect of $V_{2,{\rm in}}$ on the spectrum
for $k_x=k_y=0$, so that $c_+=c_-=1$.  In this case because
the perturbation is proportional to $b-c$ or to $b^\dagger-c^\dagger$,
one sees that $V_{2, {\rm in}}$ only couples to the optical mode
sector.  Accordingly, we do not consider $V_{2,{\rm in}}$ any further.

We expect that this in-plane anisotropy should give rise to a
macroscopic four-fold anisotropy.  In order to obtain this
anisotropy we must include anharmonic effects at relative order $1/S$.
Now we decouple the four operator terms into quadratic terms times
averages of the remaining quadratic factors.  This calculation is
done in Appendix \ref{APPI-I}.  In that calculation we naturally drop all
contributions to the optical mode sector and of the rest keep
only terms which have an effect on the mode energies at zero
wave vector.  The result is that contributions to the dynamical
matrices due to the in-plane CuI -- CuI interactions yield
\begin{mathletters}
\label{1INEQ}
\begin{eqnarray}
\delta a_{11} & = & 16 C_2 \tau  \\
\delta a_{12} & = & - 4 ( 6 C_2 - C_{2c} - 4 C_{2b}) \tau \\
\delta b_{11} & = & 8 C_{2c}  \tau  \\
\delta b_{12} & = & -16  C_{2b} \tau \ ,
\end{eqnarray}
\end{mathletters}
where $\tau \equiv (\delta J_1)^2/J$ and the $C$'s
are lattice sums defined in Eq. (\ref{C2EQ}) of
Appendix \ref{APPI-I}.  It turns out that because $\tau$ is so
small, the only evaluation we need is that $C_2=0.01$.
Note that the contributions in Eq. (\ref{1INEQ}) are of relative
order $1/S$ which is consistent with the fact that they 
represent the effect of quantum fluctuations.  The fact that
they represent a modification in the zero-point energy is
reflected by the appearance of the factor $C_2 \ll 1$.

\subsubsection{CuI-CuII Interactions}

Next we deal with the in-plane anisotropy of the CuI-CuII interactions.
The terms in Eq. (\ref{HPRIMEEQ2}) involving $\delta J_{12}$ are
\begin{eqnarray}
V_{\delta J_{12}} &=& - \delta J_{12}
\sum_{i \in II, \delta_x} \Biggl[ S_i^\xi S_{i+\delta_x}^\eta
+ S_i^\eta S_{i+\delta_x}^\xi \Biggr]
+ \delta J_{12} \sum_{i \in II, \delta_y} \Biggl[ S_i^\xi S_{i+\delta_y}^\eta
+ S_i^\eta S_{i+\delta_y}^\xi \Biggr] \ .
\end{eqnarray}
In terms of boson operators this is
\begin{eqnarray}
\label{V12EQ}
V_{\delta J_{12}} &=& \delta J_{12}\sqrt {S/2}
\sum_i \Biggl[ [S - e_i^\dagger e_i ] \Biggl( a_{i+x}
+ a_{i+x}^\dagger \phi(a_{i+x}) + d_{i-x} +
d_{i-x}^\dagger \phi(d_{i-x}) \Biggr) \nonumber \\ &&
+  \Biggl( e_i^\dagger + \phi(e_i) e_i \Biggr)
\Biggl( -2S + a_{i+x}^\dagger a_{i+x} + d_{i-x}^\dagger d_{i-x} \Biggr) \Biggr]
\nonumber \\ &&
- \delta J_{12}\sqrt {S/2} \sum_i \Biggl[ [S - e_i^\dagger e_i ]
\Biggl( b_{i-y}^\dagger + \phi( b_{i-y})b_{i-y} + c_{i+y}^\dagger +
\phi(c_{i+y})c_{i+y} \Biggr) \nonumber \\ &&
+  \Biggl( e_i^\dagger + \phi(e_i)e_i \Biggr)
\Biggl( 2S - b_{i-y}^\dagger b_{i-y} - c_{i+y}^\dagger c_{i+y} \Biggr) \Biggr]
\nonumber \\ &&
+ \delta J_{12}\sqrt {S/2} \sum_i \Biggl[ [-S + f_i^\dagger f_i ]
\Biggl( a_{i-x} + a_{i-x}^\dagger \phi(a_{i-x}) + d_{i+x} +
d_{i+x}^\dagger \phi(d_{i+x}) \Biggr) \nonumber \\ &&
+  \Biggl( f_i + f_i^\dagger \phi(f_i) \Biggr)
\Biggl( -2S + a_{i-x}^\dagger a_{i-x} + d_{i+x}^\dagger d_{i+x} \Biggr) \Biggr]
\nonumber \\ &&
- \delta J_{12}\sqrt {S/2} \sum_i \Biggl[ [-S + f_i^\dagger f_i ]
\Biggl( b_{i+y}^\dagger + \phi(b_{i+y})b_{i+y} + c_{i-y}^\dagger +
\phi(c_{i-y})c_{i-y} \Biggr) \nonumber \\ &&
+ \Biggl( f_i + f_i^\dagger \phi(f_i) \Biggr)
\Biggl( 2S - b_{i+y}^\dagger b_{i+y} - c_{i-y}^\dagger c_{i-y}
\Biggr) \Biggr] \ .
\end{eqnarray}

This perturbation contains terms linear and terms cubic in the
boson operators.  The linear terms and (at relative order $1/S$) the
cubic terms will shift the equilibrium so that the boson operators
are modified as
\begin{eqnarray}
e_i & \rightarrow & e_i + s \ , \ \ \ 
f_i \rightarrow f_i + s \ , \ \ \
a_i \rightarrow a_i + t \ , \ \ \ 
b_i \rightarrow b_i + t \ , \ \ \ 
c_i \rightarrow c_i + t \ , \ \ \ 
d_i \rightarrow d_i + t \ .
\end{eqnarray}
These shifts are evaluated in Appendix \ref{APPI-II}, where we find that
(to leading order in $1/S$) 
\begin{eqnarray}
\label{XYEQ}
s & = & {4\delta J_{12} \sqrt {S/2} \over 8J_2 } \ ,  \ \ \ \ \
t = - {2J_{12}s \over 8J + 4J_3}
= - {J_{12}\delta J_{12} \sqrt {S/2} \over J_2(8J+4J_3)} \ .
\end{eqnarray}
These are the expected results.  As one sees from Eq. (\ref{HPRIMEEQ2}),
the perpendicular field acting on an $e$ spin is $4\delta J_{12} S$ in the
positive $\eta$ direction, so that the perpendicular moment of the
$e$ spin is
$\Delta S_e = 4\delta J_{12} S \chi_{II} = 4\delta J_{12} S/(8J_2)$,
which agrees
with $\sqrt{2S} s$ when Eq. (\ref{XYEQ}) is used.  Further, due to
the isotropic exchange, the field  acting on an $a$ spin is
$2J_{12} \Delta S_e = J_{12} \delta J_{12} S/J_2$ in the negative $\eta$
direction.  Thus $\Delta S_a = - [J_{12} \delta J_{12}  S/J_2]\chi_I =
- [J_{12} \delta J_{12} S/J_2]/[8J+4J_3]$, which agrees with
$\sqrt {2S}t$ when Eq. (\ref{XYEQ}) is used.  Note that
$\Delta S_e$ and $\Delta S_a$ are both of order $S$, a result
which indicates that the effects here are completely classical.

To determine the effect of $V_{\delta J_{12}}$ on the spin-wave spectrum
we need to construct the effective quadratic Hamiltonian, which
results from introducing shifts into anharmonic terms.  This is done
in Appendix \ref{APPI-II}. When we insert these shifts into the cubic
terms of $V_{\delta J_{12}}$ we ignore $t$ in comparison to $s$ because
$J \gg J_{12}$.  Thereby we get contributions to the dynamical matrix of
\begin{eqnarray}
\label{DAIN1}
\delta a_{55} & = & \delta a_{11} = 2 \delta b_{55} =
\delta J_{12}^2 S/J_2 \equiv \zeta S \ , 
\nonumber \\
\delta a_{16} &=& - \delta a_{15} = \delta b_{16} = - 
\delta b_{15} = \case 1/4 \zeta S  \ .
\end{eqnarray}
We also insert these shifts into the four operator terms of the
isotropic Hamiltonian.  As before we only keep terms arising
from replacing two CuII operators by $\langle e \rangle$.  The
magnitude of other terms, {\it e.g.} CuI - CuI quartic terms
when CuI shifts $\langle a \rangle$ are  kept, are shown in
Appendix \ref{APPI-II} to be much smaller than those we have kept.
The result of the calculation in Appendix \ref{APPI-II} is that
we get the contributions to the dynamical matrix of
\begin{eqnarray}
\label{DAIN2}
\delta a_{55} & = & - \zeta S \ , \ \ \ \ \
\delta b_{55} = - \case 1/4 \zeta S \ , \nonumber \\
\delta a_{56} & = & - \case 3/4 \zeta S \ , \ \ \ \ \
\delta b_{56} = - \zeta S \ .
\end{eqnarray}
Note that these perturbative contributions from the CuI-CuII
in-plane anisotropy, are proportional to $S$, unlike the case for
the other in-plane anisotropies.   This indicates that the effect of
$\delta J_{12}$ (which we called $J_{pd}$ previously\cite{CHOU,KASTNER}),
is a classical effect which already appeared within mean field
theory.\cite{CHOU,KASTNER}  The
other in-plane anisotropies only have an effect when we consider
fluctuations.  However, since the effect of $\delta J_{12}$ is rather small,
we do not consider the effects of fluctuation corrections to it.

\subsubsection{CuII-CuII Intraplanar Interactions}

Here we consider the in-plane anisotropy of the interactions between
pairs of CuII spins in the same plane.  Their interaction is
\begin{eqnarray}
\label{INVEQ}
V & = & - \delta J_2 \sum_{i\in e} \Biggl[ \sum_{\delta_x: j=i+2\delta_x}
\Biggl( S_i^\xi S_j^\eta + S_i^\eta S_j^\xi \Biggr)
+ \sum_{\delta_y: j=i+2\delta_y}
\Biggl( S_i^\xi S_j^\eta + S_i^\eta S_j^\xi \Biggr) \Biggr] \nonumber \\
&=& - \delta J_2 \sqrt{{S \over 2}}
\sum_{i ,\delta_x} \Biggl\{ (S-e_i^\dagger e_i)
\Biggl[ f_j + f_j^\dagger \phi(f_j) \Biggr]
+ \Biggl[ e_i^\dagger + \phi(e_i)e_i
\Biggr] (-S + f_j^\dagger f_j ) \Biggr\} \nonumber \\
&& + \delta J_2 \sqrt{{S \over 2}}
\sum_{i ,\delta_y} \Biggl\{ (S-e_i^\dagger e_i)
\Biggl[ f_j + f_j^\dagger \phi(f_j) \Biggr]
+ \Biggl[ e_i^\dagger + \phi(e_i)e_i
\Biggr] (-S + f_j^\dagger f_j ) \Biggr\} \nonumber \\
&=& \delta J_2 \sqrt {S/2} \Biggl[ 
\sum_{i,\delta_x} \Biggl( e_i^\dagger e_i (f_j+f_j^\dagger) -
(e_i^\dagger + e_i) f_j^\dagger f_j \Biggr)
- \sum_{i,\delta_y} \Biggl( e_i^\dagger e_i (f_j+f_j^\dagger) -
(e_i^\dagger + e_i) f_j^\dagger f_j \Biggr) \Biggr] \nonumber \\
&=& \delta J_2 \sqrt{ {8S \over N}} \sum_{\bf q , k} \rho({\bf k})\Biggl(
[f({\bf k}) + f^\dagger (-{\bf k})] e^\dagger ({\bf q}) e({\bf q}-{\bf k})
- [e({\bf k}) + e^\dagger (-{\bf k})] f^\dagger ({\bf q}) f({\bf q}-{\bf k})
\Biggr),
\end{eqnarray}
where
\begin{equation}
\rho({\bf k}) = {1\over 2}[ \cos(ak_x) - \cos(ak_y) ].
\label{rho}
\end{equation}  

This Hamiltonian is treated in Appendix \ref{APPII-II}, where
the additional contributions to the spin-wave matrices (at $q_z=0)$
are found to be
\begin{eqnarray}
\delta a_{55} = - 16 [\delta J_2^2/J_2][ 2C_{2a} + C_{2b}] 
\equiv - 16 \xi [2C_{2a} + C_{2b}] \ , \ \ \ \ \
\delta a_{56} = 16 \xi [ 2C_{2a} - C_{2b}]
\end{eqnarray}
and
\begin{eqnarray}
\delta b_{55} = -16 \xi C_{2b} \ , \ \ \ \ \
\delta b_{56} = 48 \xi C_{2b} \ ,
\end{eqnarray}
where $C_{2a}$ and $C_{2b}$ are lattice sums defined in Appendix \ref{APPI-I}.

It is interesting to note that apart from a minus sign, these results
are exactly the same as in Yildirim et al.\cite{SOPR1}
This difference in sign is to be expected because the CuII's are
oriented in a hard direction with respect to only CuII-CuII
interactions.  Consequently, this term tends to decrease the gap. 

\subsubsection{ CuII-CuII Interplanar Interactions}

\begin{figure}
\centerline{\psfig{figure=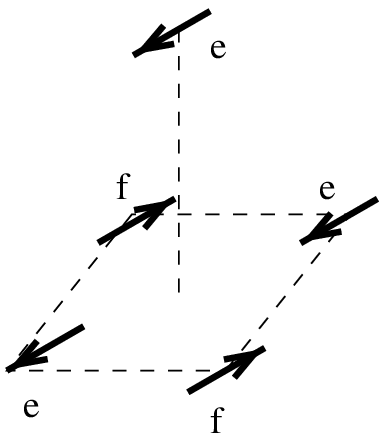} \psfig{figure=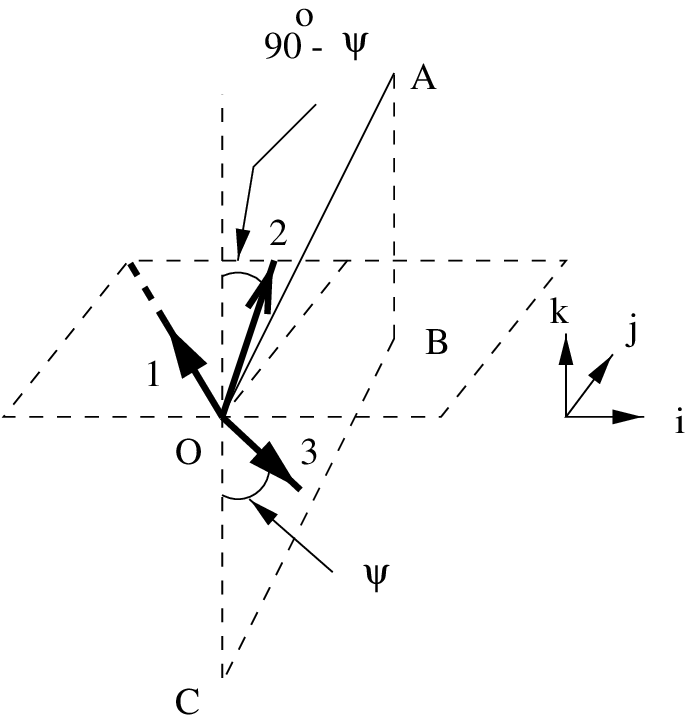}}
\caption{Interplanar CuII-CuII interactions.  Left: a plaquette
of CuII spins in one plane with a CuII neighbor in the adjacent
plane over the center of the plaquette such that the isotropic
CuII-CuII interaction is frustrated.  Right: The principal axes
for the exchange tensor of a spin in the $e$ sublattice at O
with a spin in the $e$ sublattice at A.  The directions of the
axes are given in Eq. (\ref{AXES}). The axes for the interactions
of the spin at A with other spins in the lower plane can be obtained
by a rotation of coordinates.}
\label{pseudo}
\end{figure}

Here we consider the effect of interactions between a pair of CuII
spins in adjacent planes.  The situation we consider is shown in
the left panel of Fig. \ref{pseudo}, where one sees that the
isotropic component of the CuII-CuII interplanar interaction is
frustrated.  To describe the anisotropy of this interaction we
introduce the principal axes (shown in the right panel of Fig.
\ref{pseudo}) as follows
\begin{eqnarray}
\label{AXES}
\hat n_1 = ( - \hat x + \hat y)/ \sqrt 2 \ . \ \ \
\hat n_2 = ( \hat x + \hat y) \cos \psi / \sqrt 2
+ \hat z \sin \psi  \ , \ \ \
\hat n_3 = ( \hat x + \hat y) \sin \psi / \sqrt 2
- \hat z \cos \psi  \ .
\end{eqnarray}
The angle $\psi$ is not fixed by symmetry.  We then write the
anisotropic CuII-CuII interaction ${\cal H}_{ij}^{II-II}$
between nearest-neighboring
spins $i$ and $j$ in adjacent planes as\cite{SPINPRL1}
\begin{eqnarray}
{\cal H}_{ij}^{II-II} &=& \sum_{k=1}^3 K_k
[{\bf S}_i \cdot \hat n_k^{(ij)}] [{\bf S}_j \cdot \hat n_k^{(ij)}] \ ,
\end{eqnarray}
where $\hat n_k^{(ij)}$ is the $k$th principal axes for the pair $ij$
which can by obtained from the right panel of Fig. \ref{pseudo}, by
a rotation of coordinates, if necessary, and $K_k$ is the associated
principal value of the exchange tensor.  The contributions
of this interaction to the dynamical matrix are evaluated for
$q_x=q_y=0$ in Appendix \ref{KII-II} as
\begin{mathletters}
\begin{eqnarray}
\label{A55EQ}
\delta a_{55} &=& \delta a_{66} = 4(K_1-K_2c^2-K_3s^2) S +
2(K_1+K_3c^2+K_2s^2)S c_z \ , \\
\label{A56EQ}
\delta a_{56} &=& \delta a_{65} = 2(K_2-K_3)(c^2-s^2)S c_z \ , \\
\label{B55EQ}
\delta b_{55} &=& \delta b_{66} = 2( K_1-K_2s^2-K_3c^2)Sc_z \ , \\
\label{B56EQ}
\delta b_{56} &=& \delta b_{65} = 2( K_2+K_3)Sc_z \ ,
\end{eqnarray}
\end{mathletters}
where $c \equiv \cos \psi$ and $s \equiv \sin \psi$.
As we will see later, this interaction can only contribute
significantly to the lowest energy in-plane mode, where
its effect is through the combination
\begin{eqnarray}
\label{DKEQ}
\delta (a_{55} - a_{56} + b_{55} - b_{56} ) =
4S( K_1 - K_2 c^2 - K_3 s^2 )(1+c_z) \equiv 4 \Delta K S (1 + c_z) \ .
\end{eqnarray}
Note that $\Delta K =0$ for isotropic exchange.

A closely related interaction is the long-range dipolar interaction,
whose contributions to the dynamical matrix are also evaluated
in Appendix \ref{KII-II}.  This interaction is dominant
in Sr$_2$CuO$_2$Cl$_2$\cite{RIKEN2}.  To include dipolar
interactions we obtain (in Appendix \ref{KII-II}) the result
\begin{eqnarray}
\delta (a_{55}-a_{56} +b_{55} - b_{56}) = 6 g^2 \mu_B^2 S (1+c_z)  X\ ,
\end{eqnarray}
where $X$ is the lattice sum
\begin{eqnarray}
\label{XEQ}
X & = & \sum_{j \in II: z_{ij}=c/2}
{x_{ij} y_{iy} \sigma_j \over r_{ij}^5 } \ ,
\end{eqnarray}
where $i$ labels a fixed CuII site,
$\sigma_j$ is +1 if spins $i$ and $j$ are parallel
and is $-1$ if they are antiparallel.  Numerical evaluation yields
\begin{eqnarray}
\label{XNUM}
X &=& 7 \times 10^{-4} \AA^{-3} \ .
\end{eqnarray}
Therefore we should replace $\Delta K$ by
\begin{eqnarray}
\label{KEFFEQ}
\Delta K_{\rm eff} = \Delta K + \case 3/2  g^2 \mu_B^2 X \ .
\end{eqnarray}

\subsubsection{ CuI-CuII Interplanar Interactions}

\begin{figure}
\centerline{\psfig{figure=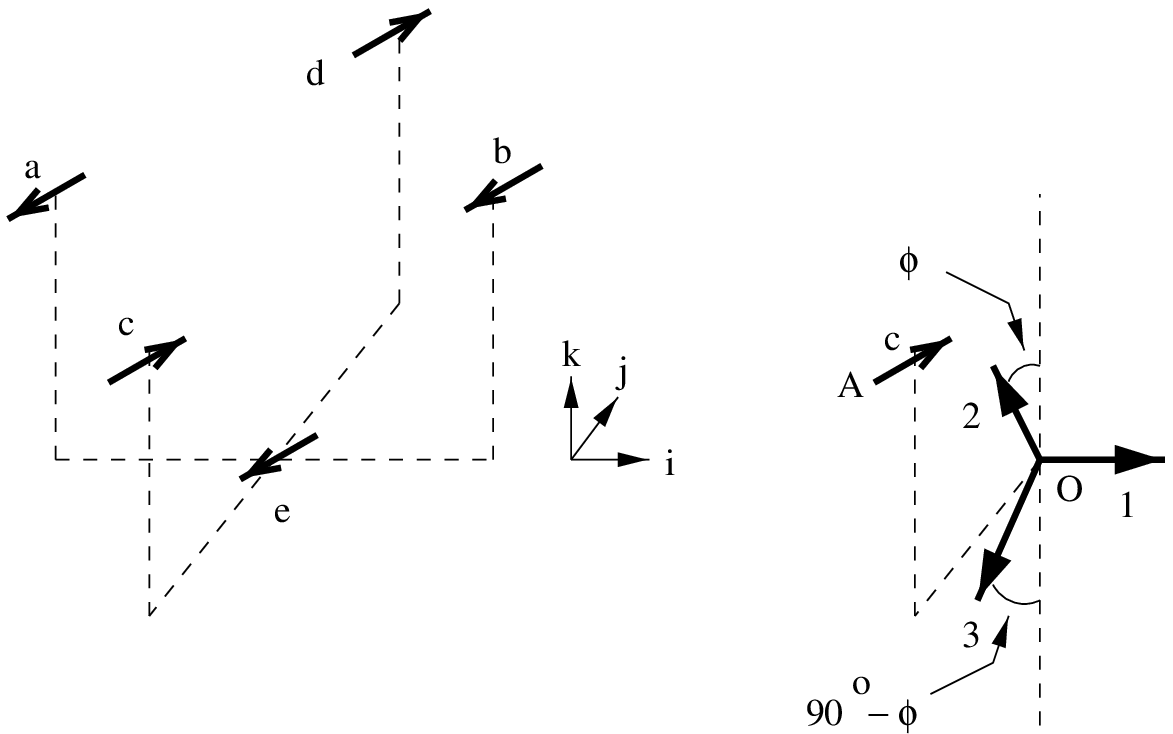}}

\vspace{0.2 in}
\caption{Interplanar CuI-CuII interactions.  Left: a plaquette
of CuI spins in one plane with a CuII neighbor in the adjacent
plane below the center of the plaquette such that the isotropic
CuI-CuII interaction is frustrated.  Right: The principal axes
for the exchange tensor of a spin in the $e$ sublattice at O
with a spin in the $c$ sublattice at A.  The directions of the
axes are given in Eq. (\ref{AXES}). The axes for the interactions
of other pairs of CuI-CuII nearest neighbors in adjacent planes
can be obtained by a rotation of coordinates.}
\label{pseudo3}
\end{figure}

Here we briefly summarize the results for a similar treatment of the
CuI-CuII anisotropic interactions.  The situation we consider is shown
in the left panel of Fig. \ref{pseudo3}, where one sees that the
isotropic component of the CuI-CuII interplanar interaction is
frustrated.  To describe the anisotropy of this interaction we
introduce the principal axes for the CuI-CuII pair 
$\underline a-\underline e$, shown in the
right panel of Fig. \ref{pseudo3}, as follows
\begin{eqnarray}
\hat m_1 = - \hat y \ . \ \ \
\hat m_2 = \hat z \cos \phi - \hat x \sin \phi \ , \ \ \ \
\hat m_3 = - \hat z \sin \phi  - \hat x \cos \phi \ .
\end{eqnarray}
The angle $\phi$ is not fixed by symmetry.  We then write the
anisotropic CuI-CuII interaction ${\cal H}_{ij}^{I-II}$ between
nearest-neighboring spins $i$ and $j$ in adjacent planes as
\begin{eqnarray}
{\cal H}_{ij}^{I-II} &=& \sum_{k=1}^3 K_k'
[{\bf S}_i \cdot \hat m_k(ij)] [{\bf S}_j \cdot \hat m_k(ij)] \ ,
\end{eqnarray}
where $\hat m_k(ij)$ is the $k$th principal axes for the pair $ij$
which can by obtained from the right panel of Fig. \ref{pseudo3}, by
a rotation of coordinates, if necessary, and $K_k'$ is associated
principal value of the exchange tensor.  In Appendix \ref{KI-II}
we obtained the following
contributions to the dynamical matrices for $q_x=q_y=0$
\begin{mathletters}
\label{AB15EQ}
\begin{eqnarray}
\delta a_{15} &=& \delta b_{16}= \case 1/2 \left[ K_1'
+ K_2'(1 - 3c^2) + K_3' (1-3s^2)\right]  \equiv G_{I-II} \\
\delta a_{16} &=& \delta b_{15} =
\case 1/2 \left[ K_1' + K_2'(1+c^2) + K_3'(1+ s^2) \right]
\equiv H_{I-II} \ ,
\end{eqnarray}
\end{mathletters}
where $c\equiv \cos \phi$ and $s \equiv \sin \phi$.
We will see later that these terms have a negligible effect on the
spin-wave spectrum.
 
\section{SPIN--WAVE SPECTRUM}

Explicitly, the dynamical matrices corresponding to the effective
quadratic Hamiltonian containing the above-mentioned anisotropies
are of the form of Eq. (\ref{ABEQG}) with 
\begin{eqnarray}
\label{ABVALUES}
a_{11} & = & 4JS + 2 J_3S + 16 C_2 \tau + \zeta S + \alpha
\ , \nonumber \\
a_{12} & = & \Delta J_1 S - 4 (6C_2-C_{2c} -4 C_{2b}) \tau \ , \nonumber \\
a_{15} &=& \case 1/2 \Delta J_{12}S - \case 1/4 \zeta S + G_{I-II}
\ , \nonumber \\
a_{16} &=& J_{12}^{(12)}S - \case 1/2 \Delta J_{12}S + \case 1/4 \zeta S
+ H_{I-II} \ , \nonumber \\
a_{55} &=& 4J_2S - 16 \xi (2C_2 - C_{2b}) + 2 \alpha \nonumber \\
&& \ \ + 4(K_1-K_2c^2-K_3s^2) S + 2(K_1+K_3c^2+K_2s^2)S c_z \ , \\
a_{56} &=& 2\Delta J_2S  + 16 \xi (2C_ 2- 3C_{2b}) - \case 3/4 \zeta S
+ 2(K_2-K_3)(c^2-s^2)S c_z \ , \nonumber \\
b_{11}&=& 8 C_{2c} \tau \ , \nonumber \\
b_{12} & = & 2JS -\Delta J_1 S - 16 C_{2b} \tau \ , \nonumber \\
b_{15} &=& J_{12}^{(34)}S - \case 1/2 \Delta J_{12}S - \case 1/4 \zeta S
+ H_{I-II} \ , \nonumber \\
b_{16} &=& \case 1/2 \Delta J_{12}S + \case 1/4 \zeta S + G_{I-II} \ ,
\nonumber \\ b_{55} &=& \case 1/4 \zeta S - 16 \xi C_{2b}
+ 2( K_1-K_2s^2-K_3c^2)Sc_z \ , \\
b_{56} &=& 4J_2S - 2\Delta J_2S - \zeta S + 48 \xi C_{2b}
+ 2( K_2+K_3)Sc_z \ .
\end{eqnarray}
(In the above tabulation we not have included dipolar interactions.
These are easiest to include when we give the mode energies
because these terms can then be combined via Eq. (\ref{KEFFEQ})
with the pseudodipolar terms which we treated explicitly.)

In Table \ref{PARAMS} we summarize the definitions of the various
parameters and in Table \ref{VALUES} we give estimates of their numerical
values.

\subsection{CuII's Ordered}

\subsubsection{Without $1/S$ Renormalizations}

Here we evaluate the energies of the four low-frequency modes in
the presence of CuII ordering without any $1/S$ renormalizations.
In what follows we will work to an accuracy of about 1\%.  That is,
the only corrections of relative order $1/J$ we will keep are those
of order $J_{12}/J$ or $J_2/J$.  Then, in the notation of Eqs.
(\ref{APLBEQ}) and (\ref{UVWEQ}) the components of the large matrix
$[{\bf A + \sigma B}]_\sigma$ are
\begin{eqnarray}
U_\sigma &=& 8JS \ ,  \hspace{0.5 in} V_\sigma= 2 \sqrt 2 \sigma J_{12} S\ ,
\hspace{0.5 in} W_\sigma = 8J_2 S \ .
\end{eqnarray}
We neglect terms which are small compared to $\alpha$ and obtain
\begin{eqnarray}
[{\bf A - B}]_{\sigma=+1} &=& \left[ \begin{array} { c| c}
4\Delta J_1 S + \alpha + x_3 & - \sqrt 2 \alpha \\ \hline
- \sqrt 2 \alpha & 4 \Delta J_2 S + 2\alpha
\end{array} \right]
\end{eqnarray}
for the out-of-plane sector, where $x_3 = 2J_3S (1 - c_z)$, and
\begin{eqnarray}
\label{INPLEQ}
[{\bf A + B}]_{\sigma=-1} &=& \left[ \begin{array} { c| c}
\zeta S + \alpha + 64 C_2 \tau + x_3 & \sqrt 2
(\alpha - \zeta S )\\
\hline \sqrt 2 (\alpha - \zeta S) & 2 ( \zeta S + \alpha) - 64 \xi C_2
+4 \Delta K_{\rm eff} S(1+ c_z) \end{array} \right]
\end{eqnarray}
for the in-plane sector, where $\Delta K_{\rm eff}$ was defined
by Eq. (\ref{KEFFEQ}).

From Eq. (\ref {SOLN}) we get the higher frequency modes as
\begin{mathletters}
\begin{eqnarray}
(\omega_+^>)^2 & = & 8JS (4 \Delta J_1S + \alpha  + x_3)
- 8J_{12}S \alpha + {16 J_2 S \alpha^2 \over \alpha
+ 4 \Delta J_1 S + x_3 } \\
\label{90b}
(\omega_-^>)^2 & = & 8JS (\alpha +x_3) + 8S\alpha (-J_{12} + 2 J_2)
- {16J_2 S \alpha x_3 \over \alpha + x_3 } \ ,
\end{eqnarray}
and the lower frequency modes as
\begin{eqnarray}
\label{90c}
(\omega_+^<)^2 & = & 64JJ_2S^2 \Biggl(
[4 \Delta J_1S + \alpha + x_3][
4 \Delta J_2 S + 2 \alpha ] - 2 \alpha^2 
\Biggr)/(\omega_+^>)^2 \\
(\omega_-^<)^2 & = & 64JJ_2S^2 \Biggl(
[\zeta S + \alpha + 64 C_2 \tau + x_3]
[2\zeta S + 2 \alpha -64 \xi C_2 + 4 \Delta K_{\rm eff}S(1+c_z)]
\nonumber \\ && \
- 2 (\alpha - \zeta S)^2 \Biggr)/ (\omega_-^>)^2 \nonumber
\\ & \approx & \left( {64JJ_2S^2 \alpha \over (\omega_-^>)^2 } \right)
\left[ 2 x_3 + 64 (2 \tau -\xi) C_2 + 8 \zeta S + 4 \Delta K_{\rm eff}S
(1+c_z) \right] \ .
\label{90d}
\end{eqnarray}
\end{mathletters}
In obtaining the above results we replaced $UW-V^2$ by $UW$ with an
error of order 1\%. To obtain the last line of Eq. (\ref{90d}) we
assumed that $\alpha$ dominates the other perturbations.

As we have already seen, quantum fluctuations of the frustrated
CuI - CuII interactions cause $\omega_\sigma^>$ to be nonzero
even if the exchange interactions are isotropic.
When we introduce easy plane anisotropy (by making
$\Delta J_1$ and/or $\Delta J_2$ nonzero) we introduce a gap into
$\omega_+^<$, but 
$\omega_-^<$ has no gap yet, because without in-plane anisotropy
a global rotation of spins within the easy plane costs no energy.
The lowest mode develops a gap when we introduce the in-plane 
anisotropy and take account of quantum fluctuations. 
One might imagine that the strongest such anisotropy,
namely that in $J$ (scaled by the parameter $\delta J_1$) would
dominate in $\omega_-^<$.  This effect is incorporated in the
term proportional to $\tau=\delta J_1^2/J$, and indeed when the
CuII's are {\it not} ordered  this term is the only one which
contributes at $q_z=0$.  However, when the CuII's are ordered,
the situation is different.  Notice that this factor has no
factor of $S$ and more importantly, it is accompanied
by the small numerical factor $C_2 \approx 0.01$.  These observations
remind us that this effect is another fluctuation effect.  Within
harmonic theory or mean-field theory the anisotropy of these CuI-CuI
in-plane interactions averages to zero.  In contrast, the weaker
in-plane interaction between CuI's and CuII's (scaled by
$\zeta \equiv (\delta J_{12})^2/J$) appears already in mean-field
theory.\cite{KASTNER} Thus, this term, which is proportional to $S$,
has no factor analogous to $C_2$ and it would dominate the term
proportional to $\tau$ except for the fact (see next section) that
its renormalization factor $Z_\zeta$ is quite small).  However, when
the CuII's are ordered, the interplanar CuII-CuII dipolar interactions
contained in $\Delta K_{\rm eff}$ are dominant, and lead to the
dramatic increase in the effective four-fold anisotropy observed
at low temperatures.  The isotropic interplanar nearest neighbor
CuI-CuII are frustrated.  The anisotropic 
CuI-CuII interlayer interactions (as embodied by the constants
$G$ and $H$) have only a negligible effect on the mode energies.

\subsubsection{$1/S$ Renormalizations}

In this subsection we summarize how we incorporate the
various renormalizations due to spin-wave interactions.  
We believe that the correct procedure is to
calculate the mode energies correctly at first order in
$1/S$ and then set $S=1/2$.  Following this prescription
we thereby obtain the following results
\begin{mathletters}
\begin{eqnarray}
\label{MODE4}
(\omega_+^>)^2 &=& 8JS \Biggl[ \alpha + 4\Delta J_1S Z_g^2
+ x_3 Z_3^2 \Biggr] - 8 J_{12} S \alpha
+ {16 J_2 S \alpha^2 \over \alpha + 4\Delta J_1S Z_g^2 + x_3 Z_3^2 } \\
\label{MODE3}
(\omega_-^>)^2 &=& 8JS \Biggl[ \alpha + x_3 Z_3^2 \Biggr] - 8J_{12}S \alpha
+ {16 J_2 S \alpha^2 \over \alpha + x_3 Z_3^2 } \\
\label{MODE2}
(\omega_+^<)^2 & = & 64JJ_2S^2 \Biggl( [4 \Delta J_1SZ_g^2 + \alpha
+ x_3 Z_3^2 ][ 4 \Delta J_2 S Z_g^2 + 2 \alpha ]
- 2 \alpha^2 \Biggr) /(\omega_+^>)^2 \\ 
\label{MODE1}
(\omega_-^<)^2 & = & 64JJ_2S^2 \Biggl( [\zeta S + \alpha
+ 64 C_2 \tau + Z_3^2 x_3] \nonumber \\ && \ \ \times [2\zeta S + 2 \alpha
-64 \xi C_2 + 4 \Delta K_{\rm eff} S (1+c_z)] -2 [\alpha - \zeta S]^2 \Biggr)/
(\omega_-^>)^2 \nonumber \\ & \approx &
\left( {64JJ_2S^2 \alpha \over (\omega_-^>)^2 } \right)
\left[ 2 Z_3^2 x_3 + 64 (2 \tau -\xi) C_2 + 8 \zeta S Z_\zeta
+ 4 \Delta K_{\rm eff}S (1+c_z) Z_3^2 \right] \ .
\end{eqnarray}
\end{mathletters}
Here we noted that spins not in the same plane are essentially uncorrelated
and hence we have
\begin{eqnarray}
\label{J3EQ}
J_3 \rightarrow \tilde Z_3 J_3 \ , \ \ \
\Delta K_{\rm eff} \rightarrow \tilde Z_3 \Delta K_{\rm eff} \ , \ \ \
\end{eqnarray}
where Eq. (\ref{Z3}) gives $\tilde Z_3 \approx 1- 0.2/S \rightarrow 0.6$.
But since $J_3$ and $\Delta K_{\rm eff}$ always enter the mode energies
in combination with an isotropic exchange constant, we associate with
them the renormalizations
\begin{eqnarray}
J_3 \rightarrow Z_3^2 J_3 \ , \ \ \
\Delta K_{\rm eff} \rightarrow Z_3^2 \Delta K_{\rm eff} \ ,
\end{eqnarray}
where $Z_3^2=\tilde Z_3 Z_c$.  Thus
$Z_3^2 = (1- 0.2/S)(1+0.085/S)=(1-0.115/S)\rightarrow 0.77$.
Also, we will determine $Z_\zeta$ by comparison, in
Eq. (\ref{kdeq}) below, with the phenomenological treatment\cite{KASTNER}
of the statics.  For convenience we summarize in Table \ref{RENORM}
the renormalizations of the various interactions which follow from our
treatment to order $1/S$.

\subsection{CuII's Disordered}

To get the energies of the spin-wave modes when the CuII's are
disordered one sets $J_{12}=J_2=0$ (i. e. modes $\omega_+^<$
and $\omega_-^<$ no longer exist as elementary excitations)
and $\alpha=0$, in which case we get
\begin{mathletters}
\begin{eqnarray}
\label{CUIENa}
(\omega_+)^2 &=& 8JS \Biggl[ 4\Delta J_1SZ_g^2  + 2J_3S Z_3^2 (1-c_z) \Biggr] \\
\label{CUIENb}
(\omega_-)^2 &=& 8JS \Biggl[ 64 \tau C_2 + 2J_3S Z_3^2 (1-c_z) \Biggr] \ .
\end{eqnarray}
\end{mathletters}
Note that in Eq. (\ref{90b}) we had dropped a term representing
the four-fold anisotropy which is proportional to $\tau$, because
such a term is negligible in comparison to $\alpha$.  Here, with
$\alpha$ not present, we restore this term in $\omega_-$.
Note also that the higher energy mode is the one which has fluctuations
out of the plane (as indicated by the dependence on $\Delta J_1$) and 
at zero wave vector is of
the expected form, $\omega^2 = 2H_E H_A$, with the exchange field
$H_E=4JS$ and the anisotropy field $H_A=4\Delta J_1S$.  The energy of this
out-of-plane gap is about 5 meV in many lamellar copper oxide
antiferromagnets.\cite{REV}
The lower energy mode involves motion of the spins within the plane and
would have no gap at zero wave vector except for the appearance of a
small effective four-fold anisotropy, which was obtained
previously\cite{SOPR1} from phenomenological considerations.  The same
result for the gap, namely
$\omega = 16 \delta J_{1} \sqrt {2C_2S}\approx 1.6 \delta J_{1}$,
is obtained from the microscopic calculation given in 
Appendix \ref{APPI-I} and also in Ref. \onlinecite{PETIT}.

\subsection{Comparison of Static and Dynamic Theories}

Here we briefly compare our results with those of a mean-field
treatment of the statics.\cite{KASTNER}  In that calculation the
four-fold anisotropy is included phenomenologically and the
anisotropic CuI - CuII interactions are included even when
the CuII sublattice is not antiferromagnetically ordered.
When the CuII sublattice is ordered, the static treatment
assumes that the Shender mechanism is strong enough that all
spins are essentially collinear.  So the dynamics
of the Goldstone mode should involve the static response
coefficients, although spin-wave hydrodynamics\cite{HH}
rigorously applies only in the limit of zero frequency.

Since the statics treat the four-fold anisotropy
phenomenologically, as did Yildirim {\it et al},\cite{SOPR1}
we identify their four-fold anisotropy constant $K$,
which scales the anisotropy energy per CuI spin, from
\begin{eqnarray}
E &=& - \case 1/2 K \cos (4 \theta ) \ ,
\end{eqnarray}
because there are two CuI's per unit cell.  Also
$\theta$ is the angle of the magnetic moment with respect
to the easy, (1,0,0), axis.  In Ref. \onlinecite{SOPR1}
the energy per CuI spin is (in the present notation)
\begin{eqnarray}
E  & = & 32 C_2 \tau S (S_x^2 S_y^2/S^4) \ . 
\end{eqnarray}
So we make the identification $K = 8 C_2 \tau S$, or, if
we include the effects of the CuII's,
\begin{eqnarray}
\label{KDEF}
K = 4 C_2 (2\tau - \xi ) S \ . 
\end{eqnarray}

We start by comparing the results of the two approaches 
when the CuII's are disordered.  There the spin-wave
calculation completely ignores the presence of the CuII's,
whereas in the statics the CuII's are characterized by their
susceptibility in the pseudo-dipolar field caused by the
small in-plane anisotropy of the CuI - CuII interactions.
In the statics for temperatures far below the ordering
temperature for the CuI sublattice (but still with
the CuII's disordered) one has the effective
fourth-order anisotropy constant $k_{\rm stat}$
from the statics as
\begin{eqnarray}
\label{KEQ}
k_{\rm stat} &=& 2K + 8M_0^2 J^2_{12}
\chi_I [ 1 - 8 \chi_{II}^2 J_{12}^2 ]^{-1} \ ,
\end{eqnarray}
where we introduce the Cu spin susceptibilities,
$\chi_I \approx 0.53 /(8J)$,
$\chi_{II} \approx 0.53 /(8J_2)$, and
(in the present notation)
\begin{eqnarray}
M_0 &=& 4 \delta J_{12} \langle S \rangle \chi_{II} \ ,
\end{eqnarray}
where $\langle \ \ \rangle$ denotes a thermal average.
If one takes $\delta J_{12} =  0.025$ meV, then $M_0 = 2 \times 10^{-4}$.
Then the second term on the right-hand side of Eq. (\ref{KEQ})
is about $6 \times 10^{-8}$ meV, compared with $2K$ which was
found\cite{KASTNER} to be $ 2 \times 10^{-6}$ meV.  So this
correction (due to paramagnetic CuII's) which is absent from
our spin-wave analysis is negligible.

When the CuII's are well ordered, Ref. \onlinecite{KASTNER} gives
approximately
\begin{eqnarray}
\label{kseq}
k_{\rm stat} & = &
2K + 8 (\delta J_{12} )^2\langle S \rangle^2 [0.53 /(8J_2)] \ .
\end{eqnarray}
Using Eq. (\ref{KDEF}) as the identification of $K$, we see
from Eq. (\ref{MODE1}) that the mode energy involves the combination
(for $\xi \ll \tau$ and $\zeta S \ll \alpha$) which we identify to be
the effective value of $k$ from the dynamics, $k_{\rm dyn}$, where
\begin{eqnarray}
\label{kdeq}
k_{\rm dyn} &=& 8 (2\tau - \xi ) C_2 S
+ \zeta S^2 Z_\zeta + \Delta K_{\rm eff} S^2 Z_3^2
= 2K + (\delta J_{12})^2 S^2 Z_\zeta / J_2 + \Delta K_{\rm eff} S^2 Z_3^2 \ . 
\end{eqnarray}
We see that the term $(0.53) \langle S \rangle^2$ in the statics
appears as $S^2 Z_\zeta$ in the spin-wave dynamics. With an
appropriate renormalization $Z_\zeta\approx 0.19$, these two terms are the same.
Thus, as far as the intralayer interactions are concerned the
comparison between statics and dynamics indicates that these
terms are correctly treated.  We also see that the treatment of the
statics did not include the interplanar anisotropic interaction,
$\Delta K_{\rm eff}$.  As we shall see, this term gives an
important contribution to the mode $\omega_-^<$, so it should
be included in a reanalysis of the statics.  In terms of the
constant $k_{\rm dyn}$ we may write Eq. (\ref{MODE1}) as
\begin{eqnarray}
\label{komeq}
(\omega_-^<)^2({\bf q}=0) &=& 64J_2k_{\rm dyn}
\Biggl( {J \over J - J_{12} + 2J_2 } \Biggr) \ .
\end{eqnarray}
Thus we conclude that except for the fact that the statics ignored
the interplanar anisotropic CuII-CuII interactions, the two theoretical
approaches are compatible with one another.  In the next section will
show that the {\it experimental} results from static and dynamic
measurements are also consistent with one another.

\subsection{Comparison to Experiments}

The comparison between the present theory and experiments has
been described briefly in several previous
publications.\cite{CHOU,KIM,RIKEN2}  Since a more detailed
comparison is given in I, we will
simply summarize the comparison of the theoretical and experimental
results.  First one has the estimate for $J$ which is nearly the
same for all cuprates.  This estimate has been refined by
Kim,\cite{YKTH} who gives $J=130$ meV.  The value
$J_2=10.5$ meV has been accurately determined\cite{KIM} by
comparing the experimental dispersion with respect to in-plane
wave vector of
CuII spin waves to various theoretical treatments which take
account of spin-wave interactions.\cite{ZBSW}

Now we discuss the analysis of the magnon gaps at zero wave vector
where values are listed in Table \ref{EXPT}.
We first fit the observed\cite{RIKEN2} in-plane gap when the CuII's are
disordered.  Equation (\ref{CUIENb}) yields
$\omega_-=16 \sqrt{2C_2 S} \delta J_1$ and
with\cite{RIKEN2} $\omega_-=0.066$ meV,
we get $|\delta J_1| = |J_\parallel-J_\perp|/2=0.042$ meV, a value which
is about twice the theoretical estimates.\cite{SOPR2}
Using Eqs. (\ref{KDEF}) and (\ref{KEQ}) this corresponds to
$k = 16 C_2 S\tau = 16(0.01)(0.5)(0.042)^2/130=1 \times 10^{-6}$meV,
compared to the value deduced from the statics,\cite{KASTNER}
$k=2 \times 10^{-6}$ meV, for 70K$\leq T \leq 120$K.
At low temperature ($T=1.4$K) , where the CuII's are well
ordered, the statics\cite{knew} gives $k=25 \times 10^{-6}$ meV.
From Eq. (\ref{komeq}) with $\omega_-^<=0.15$meV, we get
$k_{\rm dyn}=41 \times 10^{-6}$meV.
These results are listed in Table \ref{ktab}, where we see
only a qualitative consistency between the interpretation of the
static and dynamic experiments.  It is possible that the quantum
renormalizations (which affect the determination of $k$ from
the observed mode energy) are not quite correct.  Also, the
interpretation of the statics within which the CuII-CuII
interplanar anisotropy is subsumed into the four-fold anisotropy constant,
$k$, is not strictly correct.  If we fix $\delta J_1$ to fit the
value of $\omega_-^<$ at $T=100$K and assume that the interplanar
CuII-CuII interactions result from the actual dipole-dipole
interactions, then the temperature dependence of $k$ results
from the last term in Eq. (\ref{kdeq}).  With only dipolar
(i. e. no pseudo dipolar) interactions, Eqs. (\ref{KEFFEQ}) and 
(\ref{XNUM}) give (with $g=2.2$\cite{KASTNER}) 
$\Delta K_{\rm eff} = 273\times 10^{-6}$meV,
so that $\Delta K_{\rm eff} S^2 Z_3^2 = 53 \times 10^{-6}$ meV,
from which $k_{\rm dyn} = 56 \times 10^{-6}$ meV.
From Table \ref{ktab} it is clear that the experimentally
deduced temperature dependence of $k$ is qualitatively
accounted for by the intraplanar dipolar interactions,
especially if one increases $\Delta K$ by assuming it to have 
a small pseudodipolar component.

Now we consider the higher energy modes.  Fitting to the observed\cite{RJB}
energy $\omega_+=5.5$ meV of the out-of-plane gap when
the CuII's are disordered to Eq.  (\ref{CUIENa}) (with $Z_g=0.6$)
we obtain the value of $\Delta J_1=0.081$ meV.  As was the case
for $\delta J_1$, this result is also about twice the
theoretical estimates for a simple CuO plane.\cite{SOPR1,SOPR2}
Given the values of these parameters,
both higher-energy modes at low temperature involve only
the one additional parameter, $\alpha$.  If we determine
$\alpha$ from $\omega_+^>$ we get $\alpha=0.14$ meV, whereas
if we determine $\alpha$ from $\omega_-^>$ we get $\alpha=0.13$
meV.  These two values agree perfectly with one
another and their average coincides with the theoretical evaluation
of Appendix \ref{APPSH} that $\alpha=0.13$ meV.  Clearly
these agreements strongly support our interpretation
of the role of fluctuations embodied by the parameter $\alpha$.
Note that a biquadratic interaction between two spin 1/2's
can be subsumed into a ordinary Heisenberg exchange interaction.
Therefore biquadratic exchange can not contribute to $\Delta K_{\rm eff}$.

Finally we consider the lower energy out-of-plane mode in the
zero temperature limit.  The AFMR data\cite{RIKEN1,RIKEN2} gives
$\omega_+^< = 1.7473(4)$ meV, more
accurate than, but entirely consistent with, the data of
Ref. \onlinecite{RJB}.  Evaluating the expression in Eq.
(\ref{MODE2}) with $\Delta J_2 = 0$ gives $\omega_+^<=1.717$ meV.
If we fix $\Delta J_2$ to fit the experimental value of this gap,
we get $\Delta J_2= 0.004\pm 0.004$ meV.  We attribute a large
uncertainty to $\Delta J_2$ because its value changes significantly
if $\Delta J_1$ or $\alpha$ is slightly modified. To get the same
relative out-of-plane anisotropy, $\Delta J/J$, for the CuII-CuII
exchange as for the CuI-CuI exchange would require $\Delta J_1=0.008$ meV.

\section{DYNAMIC STRUCTURE FACTOR}

The cross section, $\sigma({\bf q},\omega)$, for inelastic neutron scattering
from magnetic ions is proportional to the dynamic structure factor
$S^{\alpha \beta} ({\bf q}, \omega)$ which in turn is related to the
spin-spin correlation function.  We have
\begin{eqnarray}
\sigma ({\bf q},\omega) \propto \sum_{\alpha \beta} ( \delta_{\alpha , \beta}
- q_\alpha q_\beta) S^{\alpha \beta}({\bf q}, \omega) \ .
\end{eqnarray}
According to the fluctuation-dissipation theorem, we may write
\begin{eqnarray}
S^{\alpha \beta}({\bf q}, \omega) = {1 \over \pi} n(\omega)
{\rm Im} \chi^{\alpha \beta} ({\bf q}, \omega - i0^+) \ ,
\end{eqnarray}
where $n(\omega) = [e^{\hbar \omega /(kT)} -1]^{-1}$ and, in the usual
notation,\cite{Z} the ${\bf A}-{\bf B}$ Green's function is
defined as
\begin{eqnarray}
\langle \langle {\bf A} ; {\bf B} \rangle \rangle_\omega &=&
\sum_{m,n} p_n \Biggl[ {\langle n | {\bf A} | m \rangle 
\langle m | {\bf B} | n \rangle  \over \omega -E_m + E_n } 
- { \langle n | {\bf B} | m \rangle \langle m | {\bf A} | n \rangle 
\over \omega +E_m -E_n } \Biggr] \ ,
\end{eqnarray}
where $|n\rangle$ and $|m\rangle$ are exact eigenstates with
respective energies $E_n$ and $E_m$ and $p_n$ is the Boltzmann
weight of the state $|n\rangle$.  Then $\chi$, the dynamic
susceptibility, is written as the Green's function
\begin{eqnarray}
\chi^{\alpha , \beta} ({\bf q}, \omega) =
\langle \langle S^\alpha ({\bf q}) ; S^\beta (-{\bf q})
\rangle  \rangle_{\omega}  \ .
\end{eqnarray}
We construct the dynamic susceptibility by writing the spin
operators in terms of boson operators at leading order in $1/S$:
\begin{eqnarray}
S^+({\bf q}) &=& \sqrt{2S} \left( a({\bf q}) + b^\dagger(-{\bf q})
+ c^\dagger(-{\bf q}) + d({\bf q}) + e^\dagger(-{\bf q}) + f({\bf q}) \right)
\nonumber \\
S^-(-{\bf q}) &=& \sqrt{2S} \left( a^\dagger({\bf q}) + b(-{\bf q})
+ c(-{\bf q}) + d^\dagger({\bf q}) + e(-{\bf q}) + f^\dagger({\bf q})
\right) \ .
\end{eqnarray}
Thus we have 
\begin{eqnarray}
S^\eta ({\bf q}) &=& [S^+{\bf q}) + S^-({\bf q})]/2 = \sqrt {S/2} \sum_m
\left[ V_m(\eta) \xi_m({\bf q}) + V_m(\eta)^* \xi_m^\dagger (-{\bf q})
\right] \ ,
\end{eqnarray}
and
\begin{eqnarray}
S^z ({\bf q}) &=& - i [S^+{\bf q}) - S^-({\bf q})]/2 = \sqrt{S/2} \sum_m
\left[ V_m(z) \xi_m({\bf q}) + V_m(z)^* \xi_m^\dagger  (-{\bf q}) \right] \ ,
\end{eqnarray}
where the operators are labelled as in Eq. (\ref{BOSONH}) and the
transpose of the column vectors ${\bf V}(\alpha)$ is
\begin{eqnarray}
\tilde V(\eta) = (1,1,1,1,1,1) , \hspace{1 in}
\tilde V(z) = i(1,-1,-1,1,-1,1) \ .
\end{eqnarray}
Thus we may write
\begin{eqnarray}
\chi^{\alpha \beta }({\bf q},\omega) &=& \case 1/2 S \sum_{mn} \langle \langle
\left[ V_m(\alpha) \xi_m({\bf q}) + V_m(\alpha)^* \xi_m^\dagger (-{\bf q})
\right] ; \left[ V_n(\beta)  \xi_n(-{\bf q}) +
V_n(\beta)^* \xi_n^\dagger ({\bf q}) \right] \rangle \rangle_{\omega} \ .
\end{eqnarray}

We may evaluate these response functions in terms of normal modes.
Suppose we have found the unnormalized right eigenvectors
of the dynamical matrix, Eq. (\ref{ABPQEQ}).
  That is we have the column vectors
$\Phi_j$ which satisfy
\begin{eqnarray}
[{\bf A} + {\bf B}] [{\bf A} - {\bf B}] \Phi_j & = & \omega_j^2 \Phi_j \ .
\end{eqnarray}
Then we make the identification that
\begin{equation}
{\bf P}_j - {\bf Q}_j = x_j \Phi_j \ .
\end{equation}
We can arbitrarily fix the phase of the normal mode operators so that
$x_j$ is real positive.  Then
\begin{eqnarray}
[{\bf A} - {\bf B}] x_j \Phi_j & = &
[{\bf A} - {\bf B}] [{\bf P}_j - {\bf Q}_j]  =
\omega_j [ {\bf P}_j + {\bf Q}_j ]  \ ,
\end{eqnarray}
or
\begin{eqnarray}
{\bf P}_j + {\bf Q}_j  & = &
(x_j / \omega_j) [{\bf A} - {\bf B}] \Phi_j \ ,
\end{eqnarray}
so that
\begin{eqnarray}
{\bf P}_j &=& {x_j \over 2} \Biggl[ {\cal I} + \omega_j^{-1}
[{\bf A} - {\bf B}] \Biggr] \Phi_j \nonumber \\
{\bf Q}_j &=& {x_j \over 2} \Biggl[ - {\cal I} + \omega_j^{-1}
[{\bf A} - {\bf B}] \Biggr] \Phi_j \ .
\end{eqnarray}
To use Eq. (\ref{NORMEQ}) we write
\begin{eqnarray}
{\bf P}_j^\dagger {\bf P}_j - {\bf Q}_j^\dagger {\bf Q}_j =
{x_j^2 \over \omega_j} \Biggl( \Phi_j^\dagger
[{\bf A} - {\bf B}] \Phi_j \Biggr) \ ,
\end{eqnarray}
so that
\begin{eqnarray}
x_j^2 &=& {\omega_j \over [\Phi_j^\dagger
[{\bf A} - {\bf B}] \Phi_j ] } \ .
\end{eqnarray}
Then we write the susceptibilities as
\begin{eqnarray}
&& \hspace{-0.2 in}
(2/S) \chi^{\alpha \beta} ({\bf q}, \omega) = \sum_{m,n,r}
\left[ V_m(\alpha) P_{mr} ({\bf q}) + V_m(\alpha)^* Q_{mr} ({\bf q}) \right]
\nonumber \\ && \hspace{0.5 in} \times
\left[ V_n(\beta)^* P_{nr} ({\bf q})^* + V_n(\beta) Q_{nr} ({\bf q})^* \right]
\langle \langle \tau_r({\bf q}) ; \tau_r^\dagger ({\bf q}) \rangle
\rangle_\omega \nonumber \\ && \hspace{0.5 in} + \sum_{mnr}
\left[ V_m(\alpha) Q_{mr} ({\bf q}) + V_m(\alpha)^* P_{mr} ({\bf q}) \right]
\nonumber \\ && \hspace{0.5 in} \times
\left[ V_n(\beta)^* Q_{nr} ({\bf q})^* + V_n(\beta) P_{nr} ({\bf q})^* \right]
\langle \langle \tau_r^\dagger({\bf q}) ; \tau_r({\bf q}) 
\rangle \rangle_\omega
\nonumber \\ && \hspace{0.5 in} =
\sum_r \Biggl\{ \Biggl[ \Biggl( \tilde {\bf V}(\alpha) {\bf P}_r \Biggr) +
\Biggl( \tilde {\bf V}(\alpha)^* {\bf Q}_r \Biggr) \Biggr]
\Biggl[ \Biggl( \tilde {\bf V}(\beta) {\bf P}_r \Biggr) +
\Biggl( \tilde {\bf V}(\beta)^* {\bf Q}_r \Biggr) \Biggr]^*
[\omega -\omega_r({\bf q})]^{-1} \nonumber \\ && \hspace{0.5 in}
+ \Biggl[ \Biggl( \tilde {\bf V}(\alpha) {\bf Q}_r \Biggr) +
\Biggl( \tilde {\bf V}(\alpha)^* {\bf P}_r \Biggr) \Biggr]
\Biggl[ \Biggl( \tilde {\bf V}(\beta) {\bf Q}_r \Biggr) +
\Biggl( \tilde {\bf V}(\beta)^* {\bf P}_r \Biggr) \Biggr]^*
[\omega + \omega_r({\bf q})]^{-1} \Biggr\} \nonumber \\
&& \hspace{0.5 in} \equiv \sum_r \Biggl[
{J^{\alpha \beta}_r ({\bf q}) \over \omega - \omega_r({\bf q})}
\ + \ {I^{\alpha \beta}_r ({\bf q}) \over \omega + \omega_r({\bf q})}
\Biggr]  \ ,
\end{eqnarray}
where we left the argument ${\bf q}$ implicit in several places.
We will refer to $I$ and $J$ as 'intensities', although to
get inelastic neutron scattering cross-sections one needs to include
several other factors.  At low temperature we only need
\begin{eqnarray}
I^{\alpha \beta}_r ({\bf q}) & = &  x_r^2 \Biggl[
\delta_{\alpha ,z} \Biggl( V(z)^\dagger \Phi_r ({\bf q}) \Biggr)
+ \omega_r({\bf q})^{-1} \delta_{\alpha, \eta} \Biggl(
V(\eta)^\dagger [{\bf A-B}] \Phi_r ({\bf q}) \Biggr) \Biggr] \nonumber \\
&& \times \Biggl[
\delta_{\beta ,z} \Biggl( V(z)^\dagger \Phi_r({\bf q}) \Biggr)
+ \omega_r({\bf q})^{-1} \delta_{\beta, \eta} \Biggl(
V(\eta)^\dagger [{\bf A-B}] \Phi_r ({\bf q}) \Biggr) \Biggr]^* \ .
\end{eqnarray}
In writing this result we used the fact that ${\bf V}(\eta)$
is real and ${\bf V}(z)$ is imaginary.  From now on,
we specialize to the case of wave vectors of the form
${\bf q} = {\bf G} + q_z \hat z$.  In that case
$I_r^{\eta z} + I_r^{z \eta}$ vanishes and
\begin{mathletters}
\label{INTOUT}
\begin{eqnarray}
I^{zz}_r &=&
{\left| \Biggl( {\bf V}(z)^\dagger \Phi_r({\bf q}) \Biggr) \right|^2
\omega_r ({\bf q})  \over \Biggl( \Phi_r({\bf q})^\dagger
[{\bf A-B}] \Phi_r({\bf q}) \Biggr) } \\
I^{\eta \eta}_r &=&
{\left| \Biggl( {\bf V}(\eta)^\dagger [{\bf A-B]} \Phi_r({\bf q})
\Biggr) \right|^2 \over \omega_r ({\bf q}) \Biggl( \Phi_r({\bf q})^\dagger
[{\bf A-B}] \Phi_r({\bf q}) \Biggr) } \ .
\end{eqnarray}
\end{mathletters}
The above results are useful for the out-of-plane ($\sigma=+1$)
modes in which case $[{\bf A-B}]$ is the small matrix.
Alternatively, for in-plane ($\sigma=-1$) modes when
${\bf A+B}$ is the small matrix the following forms are useful:
\begin{mathletters}
\label{INTIN}
\begin{eqnarray}
I^{z z }_r &=&
{\left| \Biggl( {\bf V}(z)^\dagger [{\bf A+B]} \Psi_r({\bf q})
\Biggr) \right|^2 \over \omega_r ({\bf q}) \Biggl( \Psi_r({\bf q})^\dagger
[{\bf A+B}] \Psi_r({\bf q}) \Biggr) } \\
I^{\eta \eta}_r &=&
{\left| \Biggl( {\bf V}(\eta)^\dagger \Psi_r({\bf q}) \Biggr) \right|^2
\omega_r ({\bf q})  \over \Biggl( \Psi_r({\bf q})^\dagger
[{\bf A+B}] \Psi_r({\bf q}) \Biggr) } \ .
\end{eqnarray}
\end{mathletters}
For high symmetry directions of the wave vector, the matrices
${\bf A}$ and ${\bf B}$ may be brought into block diagonal form by
a unitary transformation ${\bf U}$.  In that case we may apply the
above formulas in terms of the transformed quantities indicated
by primes:
\begin{eqnarray}
{\bf A}' & \equiv & {\bf U}^\dagger {\bf A} {\bf U} \ , \ \ \ \ 
{\bf B}' \equiv {\bf U}^\dagger {\bf B} {\bf U} \ , \nonumber \\
\Phi_r' & \equiv & {\bf U}^\dagger \Phi_r \ , \ \ \ \
\Psi_r' \equiv {\bf U}^\dagger \Psi_r \ , \ \ \ \
{\bf V}' (\alpha) \equiv {\bf U}^\dagger {\bf V} (\alpha ) \ .
\end{eqnarray}

For wave vectors which are equal modulo a reciprocal lattice
vector, the corresponding quantities,
${\bf A}'$, ${\bf B}'$. $\Phi'$, and $\Psi'$ are equal.  However,
the intensities at such equivalent points will differ because
${\bf U}$, and hence ${\bf V}'$, depend specifically on the
zone of the wave vector.   This can be seen explicitly in
Appendix \ref{INTAPP} where we obtain the results summarized in Tables
\ref{IHALF} and \ref{IINT}.  Note that the $\sigma =+1$ sector
does have intensity mainly in $I^{zz}$ in confirmation of our
identification of this as the out-of-plane sector.  Similarly,
the $\sigma=-1$ sector has its intensity mainly in $I^{\eta \eta}$
as expected for in-plane modes. These identifications are also
consistent with the fact that the $\sigma=+1$ modes depend on the
out-of-plane anisotropies scaled by the $\Delta J$'s, whereas the
$\sigma=-1$ modes do not involve these quantities.

\section{CONCLUSIONS}

Here we briefly summarize the significant conclusions from this work.

\noindent $\bullet$ 1.
The degeneracy, present within mean-field theory, in which the
CuII sublattice spins can be globally rotated with respect to the
CuI spins is removed by quantum fluctuations which cause the 
sublattice magnetizations to be collinear, as first indicated by
Shender.\cite{EFS}

\noindent $\bullet$ 2.
A degeneracy present within mean-field and linear spin-wave
theories, in which the magnetization can be globally rotated through
an arbitrary angle within the easy plane is similarly removed by
quantum fluctuations, as first proposed in Ref. \onlinecite{SOPR1}.

\noindent $\bullet$ 3.
These fluctuation effects, in addition to selecting the ground state
from among the classically degenerate configurations, also give rise
to nonzero energies of the corresponding spin-wave excitations.
The most dramatic evidences of this phenomenon are the
striking increases of the out-of-plane gap energy from 5 to 10 meV
and that of the in-plane gap from zero to 9 meV
when the CuII sublattice evolves from disorder to order.

\noindent $\bullet$ 4.
The experimental results of inelastic neutron scattering
for the lowest energy gaps are broadly consistent with the effective
four-fold anisotropy previously obtained from the statics
experiments.\cite{KASTNER}  More precise agreement may
depend on more accurate understanding of the various
renormalizations due to quantum and thermal fluctuations.

\noindent $\bullet$ 5.
Our improved theoretical treatment which now includes the
interlayer dipolar interactions resolves the mystery
surrounding the dramatic increase (first found in the
statics\cite{KASTNER}) in the effective
four-fold anisotropy as the temperature is reduced into
the regime where the CuII's order.  In fact the
dipolar interlayer interactions between the CuII's
dominates the effective four-fold anisotropy when the
CuII's develop long range order.

\noindent $\bullet$ 6.
Recent AFMR results\cite{RIKEN2} lead to an identification of the
small in-plane anisotropies and qualitatively confirm previous
theoretical estimates of the
exchange anisotropy induced by spin-orbit interactions.\cite{SOPR1,SOPR2}

\noindent 
{\bf ACKNOWLEDGEMENTS}
This work was supported by the US-Israel Binational Science Foundation.
RJB and YJK were also supported by NSF grant No.
DMR97-04532 and by the MRSEC Program of the NSF under award No.
DMR98-08941 (at MIT), under contract 
No. DE-AC02-98CH10886, Division of Material Science,
U. S. Department of Energy (at BNL), and by the NSF under agreement No.
DMR-9243101 (at NIST).  We thank M. A. Kastner for useful discussions.

\newpage
\appendix
\section{Intensity Calculations}
\label{INTAPP}

In this Appendix we evaluate the intensities for which formulas
are given in Sec. V.  We first give the unitary transformation
which brings the matrices ${\bf A}$ and ${\bf B}$ into block
diagonal form.  We do this for wave vectors
${\bf q} = {\bf G} + q_z\hat z$, where
\begin{eqnarray}
{\bf G} = 2 \pi \left[ { H\hat x \over a} + {K \hat y \over a}
+ { L \hat z \over c} \right] \
\end{eqnarray}
where $H$ and $K$ are either both half integral or both integral
and $H+K+L$ is an even integer.  Then
\begin{eqnarray}
{\bf U} &=& {1 \over 2} \left( \begin{array} {c c c c c c}
\sqrt 2 &  0 & 1 & 0 & 1 & 0 \\
0 & \sqrt 2 & (-1)^{H+K} & 0 & - (-1)^{H+K} & 0 \\
0 & - (-1)^{2H} \sqrt 2 & (-1)^{H-K} & 0 & - (-1)^{H-K} & 0 \\
- (-1)^{2H}\sqrt 2 & 0 & (-1)^{2H} & 0 & (-1)^{2H} & 0 \\
0 & 0 & 0 & (-i)^{2H}\sqrt 2 & 0 & (-i)^{2H}\sqrt 2 \\
0 & 0 & 0 & (i)^{2H}\sqrt 2 & 0 & - (i)^{2H}\sqrt 2 \\
\end{array} \right) \ .
\end{eqnarray}
The first two columns are the high frequency CuI optical modes.
Columns \#3 and 4 are the $\sigma=1$ out-of-plane modes and
columns \#5 and 6 are the $\sigma=-1$ in-plane modes.  The following
results hold for all wavevectors of the form
${\bf q} = {\bf G} + q_z\hat z$.

\subsection{Out-of-Plane Modes}

For the out-of-plane sector we have (for dominant $J$)
\begin{eqnarray}
{\bf A}' - {\bf B}' &=& \left( \begin{array} {c c }
x_3 + 4 \Delta J_1S + \case 1/2 \alpha \ \ & \alpha / \sqrt 2 \\
\alpha / \sqrt 2 & \ \ 4\Delta J_2S + \alpha \\
\end{array} \right) \ , \ \ \
{\bf A}'+ {\bf B}' = \left( \begin{array} {c c }
8JS & \sqrt 2 J_{12} S \\
\sqrt 2 J_{12} S & 8J_2 S \\
\end{array} \right) 
\end{eqnarray}
independent of ${\bf G}$, where $x_3=2J_3S[1-\cos(cq_z/2)]$.
Note that $q_z$ is measured relative to the reciprocal lattice
vector in question.  We now tabulate the right eigenvectors of
the block matrices $M_{+-}\equiv[{\bf A'+B'}][{\bf A'-B'}]$ associated with
the eigenvalues (the squares of the mode energies) $\omega_r^2$.
We have
\begin{eqnarray}
\tilde \Phi_+^> & = & [1,0] \ , \ \ \ (\omega_+^>)^2
= (8JS)(x_3+4\Delta J_1S + \case 1/2 \alpha ) \ ,  \nonumber \\
\tilde \Phi_+^< & = & [-\alpha /\sqrt 2 ,
x_3 + 4\Delta J_1S + \case 1/2 \alpha ] \ , \ \ (\omega_+^<)^2 =
(8J_2S) \left[ 4\Delta J_2S + \alpha - {\case 1/2 \alpha^2 \over
x_3+4\Delta J_1S + \case 1/2 \alpha } \right] \ .
\end{eqnarray}
Also we find that
\begin{mathletters}
\begin{eqnarray}
{\bf V}(z)' & = &
\left[ \begin{array} {c }  1 - (-1)^{H+K} \\ 0 \\ \end{array} \right]
\hspace{0.5 in} {\bf V} (\eta)' =
\left[ \begin{array} {c }  1 + (-1)^{H+K} \\ \sqrt 2 (-1)^H \\ \end{array}
\right] {\rm \ for \ integer \ } H  \\
{\bf V}(z)' & = &
\left[ \begin{array} {c }  0 \\ i^{2H} \sqrt 2 \\ \end{array} \right]
\hspace{0.5 in} {\bf V} (\eta)' =
\left[ \begin{array} {c }  0 \\ 0 \\ \end{array} \right] 
{\rm \ for \ half-integer \ } H
\end{eqnarray}
\end{mathletters}
Note that the vectors ${\bf V}(\alpha)'$ depend on ${\bf G}$.
Substituting these evaluations into Eq. (\ref{INTOUT}) we obtain
the results for the intensities in Tables \ref{IHALF} and \ref{IINT}
for the out-of-plane ($\sigma=+1$) modes.  

\subsection{In-Plane Modes}

For the in-plane sector we have (for dominant $J$)
\begin{eqnarray}
{\bf A}' + {\bf B}' &=& \left( \begin{array} {c c }
x_3 + \zeta S + \case 1/2 \alpha \ \ & \sqrt 2 (\zeta S - \case 1/2 \alpha ) \\
\sqrt 2 (\zeta S - \case 1/2 \alpha ) & \ \ 2 \zeta S + \alpha \\
\end{array} \right) \ , \ \ \
{\bf A}'- {\bf B}' = \left( \begin{array} {c c }
8JS & \sqrt 2 J_{12} S \\
\sqrt 2 J_{12} S & 8J_2 S \\
\end{array} \right) \ ,
\end{eqnarray}
and we now tabulate the right eigenvectors of the block matrices
$M_{-+} \equiv [{\bf A'-B'}][{\bf A'+B'}]$ associated with
the eigenvalues (the squares of the mode energies) $\omega_r^2$.
For dominant $J$ we have the approximate results
\begin{eqnarray}
{{\tilde \Psi}_-^>} & = & [1,0] \ , \ \ \ (\omega_-^> )^2
= (8JS)(x_3+\zeta S + \case 1/2 \alpha ) \ , \nonumber \\ 
{{\tilde \Psi}_-^<} &= & [-\sqrt 2 (\zeta S - \case 1/2 \alpha ) ,
x_3 + \zeta S + \case 1/2 \alpha ]
\ , \nonumber \\ (\omega_-^< )^2 & = & (8J_2S) \left[ 
2\zeta S + \alpha - 2 { (\zeta S - \case 1/2 \alpha )^2 \over
(x_3+\zeta S + \case 1/2 \alpha )} \right]
\end{eqnarray}
and
\begin{mathletters}
\begin{eqnarray}
{\bf V}(z)' & = &
\left[ \begin{array} {c }  1 + (-1)^{H+K} \\ - \sqrt 2 (-1)^H
\\ \end{array} \right] \hspace{0.3 in} {\bf V} (\eta)' =
\left[ \begin{array} {c }  1 - (-1)^{H+K} \\ 0 \\ \end{array}
\right] {\rm \ for \ integer \ } H  \\
{\bf V}(z)' & = &
\left[ \begin{array} {c }  0 \\ 0 \\ \end{array} \right]
\hspace{0.8 in} {\bf V} (\eta)' =
\left[ \begin{array} {c }  0 \\ - \sqrt 2 (i)^{2H} \\ \end{array} \right] 
{\rm \ for \ half-integer \ } H \ .
\end{eqnarray}
\end{mathletters}
As before, only the vectors ${\bf V}(\alpha)'$ depend on ${\bf G}$.
Substituting these evaluations into Eq. (\ref{INTIN}) we obtain
the results for the intensities in Tables \ref{IHALF} and \ref{IINT}
for the in-plane ($\sigma=-1$) modes.  

\section{Shender Parameters}
\label{APPSH}

In this Appendix we evaluate the averages
\begin{mathletters}
\begin{eqnarray}
A_1 & = & \langle a_m f_n^\dagger \rangle \\
A_2 & = & \langle a_m e_n \rangle \ ,
\end{eqnarray}
\end{mathletters}
where site $n$ is a nearest neighbor of site $m$.
The above quantities can be calculated perturbatively in
the frustrated coupling $J_{12}$ between CuI's and CuII's.
(See Fig. \ref{UCFIG}.)

\subsection{$A_1$}

Thus
\begin{eqnarray}
\label{PERT}
A_1 & = & - \left\langle 0 \left| V_{I-II} {1 \over {\cal E}} a_m
f_n^\dagger \right| 0 \right\rangle - \left\langle 0 \left|
a_m f_n^\dagger {1 \over {\cal E}} V_{I-II} \right| 0 \right\rangle \ ,
\end{eqnarray}
where ${\cal E}$ is the unperturbed energy of the virtual state
relative to the ground state.  Here we invoke perturbation theory
relative to decoupled CuI and CuII subsystems, and
\begin{eqnarray}
V_{I-II} &=& J_{12}S \Biggl[
  \sum_{i \in a, \delta} [ a_i^\dagger a_i + e_j^\dagger e_j + a_i e_j + a_i^\dagger e_j^\dagger]
+ \sum_{i \in a, \delta} [ a_i^\dagger a_i + f_j^\dagger f_j + a_i^\dagger f_j + f_j^\dagger a_i]
\nonumber \\ && \ \
+ \sum_{i \in b, \delta} [ b_i^\dagger b_i + e_j^\dagger e_j + b_i^\dagger e_j + e_j^\dagger b_i]
+ \sum_{i \in b, \delta} [ b_i^\dagger b_i + f_j^\dagger f_j + b_i f_j + b_i^\dagger f_j^\dagger]
\nonumber \\ && \ \
+ \sum_{i \in c, \delta} [ c_i^\dagger c_i + e_j^\dagger e_j + c_i^\dagger e_j + e_j^\dagger c_i]
+ \sum_{i \in c, \delta} [ c_i^\dagger c_i + f_j^\dagger f_j + c_i f_j + c_i^\dagger f_j^\dagger]
\nonumber \\ && \ \
+ \sum_{i \in d, \delta} [ d_i^\dagger d_i + e_j^\dagger e_j + d_i e_j + d_i^\dagger e_j^\dagger]
+ \sum_{i \in d, \delta} [ d_i^\dagger d_i + f_j^\dagger f_j + d_i^\dagger f_j + f_j^\dagger d_i]
\Biggr] \ .
\end{eqnarray}
Only terms in $V_{I-II}$ which have operators in both subsystems
contribute.  So, effectively
\begin{eqnarray}
\label{EFEQ}
V_{I-II} &=& J_{12}S \sum_{i,\delta} \Biggl[
a_i e_j + a_i^\dagger e_j^\dagger
+a_i^\dagger f_j + f_j^\dagger a_i
+b_i^\dagger e_j + e_j^\dagger b_i
+b_i f_j + b_i^\dagger f_j^\dagger \nonumber \\ && \ \
+c_i^\dagger e_j + e_j^\dagger c_i
+c_i f_j + c_i^\dagger f_j^\dagger
+d_i e_j + d_i^\dagger e_j^\dagger
+d_i^\dagger f_j + f_j^\dagger d_i \Biggr] \ ,
\end{eqnarray}
where site $j$ is the appropriate nearest neighbor of site $i$.
In fact, in Eq. (\ref{PERT}) we need to have only terms with $f$ or
$e^\dagger$ and $a^\dagger$, $d^\dagger$, $b$, or $c$.  So we set
\begin{eqnarray}
V_{I-II} &=& V_1 \equiv J_{12}S \sum_{i,\delta} \Biggl[
a_i^\dagger e_j^\dagger +a_i^\dagger f_j + e_j^\dagger b_i +b_i f_j
+ e_j^\dagger c_i +c_i f_j + d_i^\dagger e_j^\dagger +d_i^\dagger f_j \Biggr] \ .
\end{eqnarray}
Thus with $n=m+ \delta_{af}$ we have
\begin{eqnarray}
A_1 &=& - J_{12}S \sum_{i \in a } \left\langle 0 \left| 
a_i^\dagger {1 \over {\cal E}} a_m \right| 0 \right\rangle
\langle 0| [ e^\dagger_{i+\delta_{ae}}
+ f_{i+\delta_{af}}] f^\dagger_{m+\delta_{af}} | 0 \rangle 
\nonumber \\
&& - J_{12}S \sum_{i \in b } \left\langle 0 \left| 
b_i {1 \over {\cal E}} a_m \right| 0 \right\rangle
\langle 0| [ e^\dagger_{i+\delta_{be}}
+ f_{i+\delta_{bf}}] f^\dagger_{m+\delta_{af}} | 0 \rangle 
\nonumber \\
&& - J_{12}S \sum_{i \in c } \left\langle 0 \left| 
c_i {1 \over {\cal E}} a_m \right| 0 \right\rangle
\langle 0| [ e^\dagger_{i+\delta_{ce}}
+ f_{i+\delta_{cf}}] f^\dagger_{m+\delta_{af}} | 0 \rangle 
\nonumber \\
&& - J_{12}S \sum_{i \in d } \left\langle 0 \left| 
d_i^\dagger {1 \over {\cal E}} a_m \right| 0 \right\rangle
\langle 0| [ e^\dagger_{i+\delta_{de}}
+ f_{i+\delta_{df}}] f^\dagger_{m+\delta_{af}} | 0 \rangle 
\nonumber \\
&& - J_{12}S \sum_{i \in a } \left\langle 0 \left| 
a_m {1 \over {\cal E}} a_i^\dagger \right| 0 \right\rangle
\langle 0| f^\dagger_{m+\delta_{af}} [ e^\dagger_{i+\delta_{ae}}
+ f_{i+\delta_{af}}] | 0 \rangle 
\nonumber \\
&& - J_{12}S \sum_{i \in b } \left\langle 0 \left| 
a_m {1 \over {\cal E}} b_i \right| 0 \right\rangle
\langle 0| f^\dagger_{m+\delta_{af}} [ e^\dagger_{i+\delta_{be}}
+ f_{i+\delta_{bf}}] | 0 \rangle 
\nonumber \\
&& - J_{12}S \sum_{i \in c } \left\langle 0 \left| 
a_m {1 \over {\cal E}} c_i \right| 0 \right\rangle
\langle 0| f^\dagger_{m+\delta_{af}} [ e^\dagger_{i+\delta_{ce}}
+ f_{i+\delta_{cf}}] | 0 \rangle 
\nonumber \\
&& - J_{12}S \sum_{i \in d } \left\langle 0 \left| 
a_m {1 \over {\cal E}} d_i^\dagger \right| 0 \right\rangle
\langle 0| f^\dagger_{m+\delta_{af}} [ e^\dagger_{i+\delta_{de}}
+ f_{i+\delta_{df}}] | 0 \rangle \ . 
\end{eqnarray}
Here we neglected the energy of the CuII modes in comparison
to that of the CuI modes.  Also we used the unusual notation that
\begin{eqnarray}
\bbox{\delta}_{st} = {\bf r}_t - {\bf r}_s \ .
\end{eqnarray}
Then
\begin{eqnarray}
A_1 &=& - {J_{12}S \over N_{\rm uc}^2} \sum_{{\bf q},{\bf k}}
\sum_{i \in a } \left\langle 0 \left| 
a^\dagger({\bf q}) {1 \over {\cal E}} a({\bf q}) \right| 0 \right\rangle
e^{i({\bf q}+{\bf k}) \cdot {\bf r}_{im}}  \nonumber \\ && \ \ \times
\langle 0| [ e^\dagger({\bf k}) e^{i{\bf k} \cdot \bbox{\delta}_{ae}}
+ f(-{\bf k}) e^{i {\bf k} \cdot \bbox{\delta}_{af}} ]
f^\dagger(-{\bf k}) e^{-i{\bf k} \cdot \bbox{\delta}_{af}} | 0 \rangle 
\nonumber \\
&& - {J_{12}S  \over N_{\rm uc}^2} \sum_{{\bf q}, {\bf k}}
\sum_{i \in b } \left\langle 0 \left| 
b(-{\bf q}) {1 \over {\cal E}} a({\bf q}) \right| 0 \right\rangle
e^{i({\bf q}+{\bf k}) \cdot {\bf r}_{im}}  \nonumber \\ && \ \ \times
\langle 0| [ e^\dagger({\bf k}) e^{i {\bf k} \cdot \bbox{\delta}_{be}}
+ f(-{\bf k}) e^{i{\bf k} \cdot \bbox{\delta}_{bf}} ]
f^\dagger(-{\bf k})  e^{-i{\bf k} \cdot \bbox{\delta}_{af}} | 0 \rangle 
\nonumber \\
&& - {J_{12}S \over N_{\rm uc}^2} \sum_{{\bf q}, {\bf k}}
\sum_{i \in c } \left\langle 0 \left| 
c(-{\bf q}) {1 \over {\cal E}} a({\bf q}) \right| 0 \right\rangle
e^{i({\bf q}+{\bf k}) \cdot {\bf r}_{im}}  \nonumber \\ && \ \ \times
\langle 0| [ e^\dagger({\bf k}) e^{i {\bf k} \cdot \bbox{\delta}_{ce}}
+ f(-{\bf k}) e^{i {\bf k} \cdot \bbox{\delta}_{cf}} ]
f^\dagger(-{\bf k}) e^{-i{\bf k} \cdot \bbox{\delta}_{af}} ] | 0 \rangle 
\nonumber \\
&& - {J_{12}S \over N_{\rm uc}^2} \sum_{{\bf q}, {\bf k}}
\sum_{i \in d } \left\langle 0 \left| 
d^\dagger({\bf q}) {1 \over {\cal E}} a({\bf q}) \right| 0 \right\rangle
e^{i({\bf q}+{\bf k}) \cdot {\bf r}_{im}}  \nonumber \\ && \ \ \times
\langle 0| [ e^\dagger({\bf k}) e^{i {\bf k} \cdot \bbox{\delta}_{de}}
+ f(-{\bf k}) e^{i {\bf k} \cdot \bbox{\delta}_{df}} ]
f^\dagger(-{\bf k})  e^{-i{\bf k} \cdot \bbox{\delta}_{af}} | 0 \rangle 
\nonumber \\
&& - {J_{12}S \over N_{\rm uc}^2} \sum_{{\bf q}, {\bf k}}
\sum_{i \in a } \left\langle 0 \left| 
a({\bf q}) {1 \over {\cal E}} a^\dagger({\bf q}) \right| 0 \right\rangle
e^{i({\bf q}+{\bf k}) \cdot {\bf r}_{im}}  \nonumber \\ && \ \ \times
\langle 0| f^\dagger(-{\bf k}) e^{-i{\bf k}\cdot \bbox{\delta}_{af}}
[ e^\dagger({\bf k}) e^{i {\bf k} \cdot \bbox{\delta}_{ae}}
+ f(-{\bf k}) e^{i {\bf k} \cdot \bbox{\delta}_{af}} ] | 0 \rangle 
\nonumber \\
&& - {J_{12}S \over N_{\rm uc}^2} \sum_{{\bf q}, {\bf k}}
\sum_{i \in b } \left\langle 0 \left| 
a({\bf q}) {1 \over {\cal E}} b(-{\bf q}) \right| 0 \right\rangle
e^{i({\bf q}+{\bf k}) \cdot {\bf r}_{im}}  \nonumber \\ && \ \ \times
\langle 0| f^\dagger(-{\bf k}) e^{-i{\bf k}\cdot \bbox{\delta}_{af}}
[ e^\dagger({\bf k}) e^{i {\bf k} \cdot \bbox{\delta}_{be}}
+ f(-{\bf k}) e^{i {\bf k} \cdot \bbox{\delta}_{bf}} ] | 0 \rangle 
\nonumber \\
&& - {J_{12}S \over N_{\rm uc}^2} \sum_{{\bf q}, {\bf k}}
 \sum_{i \in c } \left\langle 0 \left| 
a({\bf q}) {1 \over {\cal E}} c(-{\bf q}) \right| 0 \right\rangle
e^{i({\bf q}+{\bf k}) \cdot {\bf r}_{im}}  \nonumber \\ && \ \ \times
\langle 0| f^\dagger(-{\bf k}) e^{-i{\bf k}\cdot \bbox{\delta}_{af}}
[ e^\dagger({\bf k}) e^{i {\bf k} \cdot \bbox{\delta}_{ce}}
+ f(-{\bf k}) e^{i {\bf k} \cdot \bbox{\delta}_{cf}} ] | 0 \rangle 
\nonumber \\
&& - {J_{12}S \over N_{\rm uc}^2} \sum_{{\bf q}, {\bf k}}
\sum_{i \in d } \left\langle 0 \left| 
a({\bf q}) {1 \over {\cal E}} d^\dagger({\bf q}) \right| 0 \right\rangle
e^{i({\bf q}+{\bf k}) \cdot {\bf r}_{im}}  \nonumber \\ && \ \ \times
\langle 0| f^\dagger(-{\bf k}) e^{-i{\bf k}\cdot \bbox{\delta}_{af}}
[ e^\dagger({\bf k}) e^{i {\bf k} \cdot \bbox{\delta}_{de}}
+ f(-{\bf k}) e^{i {\bf k} \cdot \bbox{\delta}_{df}} ] | 0 \rangle \ . 
\end{eqnarray}
Doing the sum over $i$ we get
\begin{eqnarray}
A_1 &=& - {J_{12}S \over N_{\rm uc}} \sum_{\bf q}
\left\langle 0 \left| 
a^\dagger({\bf q}) {1 \over {\cal E}} a({\bf q}) \right| 0 \right\rangle
\langle 0| [ e^\dagger(-{\bf q}) e^{-i{\bf q} \cdot \bbox{\delta}_{ae}}
+ f({\bf q}) e^{-i {\bf q} \cdot \bbox{\delta}_{af}} ]
f^\dagger({\bf q}) e^{i{\bf q} \cdot \bbox{\delta}_{af}} | 0 \rangle 
\nonumber \\
&& - {J_{12}S  \over N_{\rm uc}} \sum_{\bf q}
\left\langle 0 \left| 
b(-{\bf q}) {1 \over {\cal E}} a({\bf q}) \right| 0 \right\rangle
\langle 0| [ e^\dagger(-{\bf q}) e^{-i {\bf q} \cdot \bbox{\delta}_{be}}
+ f({\bf q}) e^{-i{\bf q} \cdot \bbox{\delta}_{bf}} ]
f^\dagger({\bf q})  e^{i{\bf q} \cdot \bbox{\delta}_{af}} | 0 \rangle 
\nonumber \\
&& - {J_{12}S \over N_{\rm uc}} \sum_{\bf q}
\left\langle 0 \left| 
c(-{\bf q}) {1 \over {\cal E}} a({\bf q}) \right| 0 \right\rangle
\langle 0| [ e^\dagger(-{\bf q}) e^{-i {\bf q} \cdot \bbox{\delta}_{ce}}
+ f({\bf q}) e^{-i {\bf q} \cdot \bbox{\delta}_{cf}} ]
f^\dagger({\bf q}) e^{i{\bf q} \cdot \bbox{\delta}_{af}} ] | 0 \rangle 
\nonumber \\
&& - {J_{12}S \over N_{\rm uc}} \sum_{\bf q}
\left\langle 0 \left| 
d^\dagger({\bf q}) {1 \over {\cal E}} a({\bf q}) \right| 0 \right\rangle
\langle 0| [ e^\dagger(-{\bf q}) e^{-i {\bf q} \cdot \bbox{\delta}_{de}}
+ f({\bf q}) e^{-i {\bf q} \cdot \bbox{\delta}_{df}} ]
f^\dagger({\bf q})  e^{i{\bf q} \cdot \bbox{\delta}_{af}} | 0 \rangle 
\nonumber \\
&& - {J_{12}S \over N_{\rm uc}} \sum_{\bf q}
\left\langle 0 \left| 
a({\bf q}) {1 \over {\cal E}} a^\dagger({\bf q}) \right| 0 \right\rangle
\langle 0| f^\dagger({\bf q}) e^{i{\bf q}\cdot \bbox{\delta}_{af}}
[ e^\dagger(-{\bf q}) e^{-i {\bf q} \cdot \bbox{\delta}_{ae}}
+ f({\bf q}) e^{-i {\bf q} \cdot \bbox{\delta}_{af}} ] | 0 \rangle 
\nonumber \\
&& - {J_{12}S \over N_{\rm uc}} \sum_{\bf q}
\left\langle 0 \left| 
a({\bf q}) {1 \over {\cal E}} b(-{\bf q}) \right| 0 \right\rangle
\langle 0| f^\dagger({\bf q}) e^{i{\bf q}\cdot \bbox{\delta}_{af}}
[ e^\dagger(-{\bf q}) e^{-i {\bf q} \cdot \bbox{\delta}_{be}}
+ f({\bf q}) e^{-i {\bf q} \cdot \bbox{\delta}_{bf}} ] | 0 \rangle 
\nonumber \\
&& - {J_{12}S \over N_{\rm uc}} \sum_{\bf q}
\left\langle 0 \left| 
a({\bf q}) {1 \over {\cal E}} c(-{\bf q}) \right| 0 \right\rangle
\langle 0| f^\dagger({\bf q}) e^{i{\bf q}\cdot \bbox{\delta}_{af}}
[ e^\dagger(-{\bf q}) e^{-i {\bf q} \cdot \bbox{\delta}_{ce}}
+ f({\bf q}) e^{-i {\bf q} \cdot \bbox{\delta}_{cf}} ] | 0 \rangle 
\nonumber \\
&& - {J_{12}S \over N_{\rm uc}} \sum_{\bf q}
\left\langle 0 \left| 
a({\bf q}) {1 \over {\cal E}} d^\dagger({\bf q}) \right| 0 \right\rangle
\langle 0| f^\dagger({\bf q}) e^{i{\bf q}\cdot \bbox{\delta}_{af}}
[ e^\dagger(-{\bf q}) e^{-i {\bf q} \cdot \bbox{\delta}_{de}}
+ f({\bf q}) e^{-i {\bf q} \cdot \bbox{\delta}_{df}} ] | 0 \rangle \ . 
\end{eqnarray}

For the CuII subsystem we have the usual relations
\begin{eqnarray}
e ({\bf q}) & = & l_{\bf q} \eta ({\bf q}) - m_{\bf q} \delta^\dagger (-{\bf q}) \ ,
\nonumber \\
f^\dagger (-{\bf q}) & = & -m_{\bf q} \eta ({\bf q}) + l_{\bf q} \delta^\dagger (-{\bf q})
\ ,
\end{eqnarray}
where $\eta({\bf q})$ and $\delta({\bf q})$ are the normal mode
operators for the CuII subsystem and $l_{\bf q}$ and
$m_{\bf q}$ are given in Eq. (\ref{LMEQ}).
In the ground state we evaluate the averages to get
\begin{eqnarray}
A_1 &=& - {J_{12}S \over N_{\rm uc}} \sum_{\bf q}
\left\langle 0 \left| 
a^\dagger({\bf q}) {1 \over {\cal E}} a({\bf q}) \right| 0 \right\rangle
[ - l_{\bf q} m_{\bf q} e^{-i{\bf q} \cdot \bbox{\delta}_{ae}}
+ l^2_{\bf q} e^{-i {\bf q} \cdot \bbox{\delta}_{af}} ]
e^{i{\bf q} \cdot \bbox{\delta}_{af}}
\nonumber \\
&& - {J_{12}S  \over N_{\rm uc}} \sum_{\bf q}
\left\langle 0 \left| 
b(-{\bf q}) {1 \over {\cal E}} a({\bf q}) \right| 0 \right\rangle
[ -l_{\bf q} m_{\bf q} e^{-i {\bf q} \cdot \bbox{\delta}_{be}}
+ l^2_{\bf q} e^{-i{\bf q} \cdot \bbox{\delta}_{bf}} ]
e^{i{\bf q} \cdot \bbox{\delta}_{af}}
\nonumber \\
&& - {J_{12}S \over N_{\rm uc}} \sum_{\bf q}
\left\langle 0 \left| 
c(-{\bf q}) {1 \over {\cal E}} a({\bf q}) \right| 0 \right\rangle
[ -l_{\bf q} m_{\bf q} e^{-i {\bf q} \cdot \bbox{\delta}_{ce}}
+ l^2_{\bf q} e^{-i {\bf q} \cdot \bbox{\delta}_{cf}} ]
e^{i{\bf q} \cdot \bbox{\delta}_{af}}
\nonumber \\
&& - {J_{12}S \over N_{\rm uc}} \sum_{\bf q}
\left\langle 0 \left| 
d^\dagger({\bf q}) {1 \over {\cal E}} a({\bf q}) \right| 0 \right\rangle
[ -l_{\bf q} m_{\bf q} e^{-i {\bf q} \cdot \bbox{\delta}_{de}}
+ l^2_{\bf q} e^{-i {\bf q} \cdot \bbox{\delta}_{df}} ]
e^{i{\bf q} \cdot \bbox{\delta}_{af}}
\nonumber \\
&& - {J_{12}S \over N_{\rm uc}} \sum_{\bf q}
\left\langle 0 \left| 
a({\bf q}) {1 \over {\cal E}} a^\dagger({\bf q}) \right| 0 \right\rangle
e^{i{\bf q}\cdot \bbox{\delta}_{af}}
[ -l_{\bf q} m_{\bf q} e^{-i {\bf q} \cdot \bbox{\delta}_{ae}}
+ m^2_{\bf q} e^{-i {\bf q} \cdot \bbox{\delta}_{af}} ]
\nonumber \\
&& - {J_{12}S \over N_{\rm uc}} \sum_{\bf q}
\left\langle 0 \left| 
a({\bf q}) {1 \over {\cal E}} b(-{\bf q}) \right| 0 \right\rangle
e^{i{\bf q}\cdot \bbox{\delta}_{af}}
[ -l_{\bf q} m_{\bf q} e^{-i {\bf q} \cdot \bbox{\delta}_{be}}
+ m^2_{\bf q} e^{-i {\bf q} \cdot \bbox{\delta}_{bf}} ]
\nonumber \\
&& - {J_{12}S \over N_{\rm uc}} \sum_{\bf q}
\left\langle 0 \left| 
a({\bf q}) {1 \over {\cal E}} c(-{\bf q}) \right| 0 \right\rangle
e^{i{\bf q}\cdot \bbox{\delta}_{af}}
[ -l_{\bf q} m_{\bf q} e^{-i {\bf q} \cdot \bbox{\delta}_{ce}}
+ m^2_{\bf q} e^{-i {\bf q} \cdot \bbox{\delta}_{cf}} ]
\nonumber \\
&& - {J_{12}S \over N_{\rm uc}} \sum_{\bf q}
\left\langle 0 \left| 
a({\bf q}) {1 \over {\cal E}} d^\dagger({\bf q}) \right| 0 \right\rangle
e^{i{\bf q}\cdot \bbox{\delta}_{af}}
[ -l_{\bf q} m_{\bf q} e^{-i {\bf q} \cdot \bbox{\delta}_{de}}
+ m^2_{\bf q} e^{-i {\bf q} \cdot \bbox{\delta}_{df}} ] \nonumber \\
&=& - {J_{12}S \over N_{\rm uc}} \sum_{\bf q}
\left\langle 0 \left| 
a^\dagger({\bf q}) {1 \over {\cal E}} a({\bf q}) \right| 0 \right\rangle
[ - l_{\bf q} m_{\bf q} e^{iaq_x} + l^2_{\bf q}]
\nonumber \\
&& - {J_{12}S  \over N_{\rm uc}} \sum_{\bf q}
\left\langle 0 \left| 
b(-{\bf q}) {1 \over {\cal E}} a({\bf q}) \right| 0 \right\rangle
[ -l_{\bf q} m_{\bf q} e^{-iq_ya/2} + l^2_{\bf q} e^{iq_ya/2} ] e^{iq_xa/2} 
\nonumber \\
&& - {J_{12}S \over N_{\rm uc}} \sum_{\bf q}
\left\langle 0 \left| 
c(-{\bf q}) {1 \over {\cal E}} a({\bf q}) \right| 0 \right\rangle
[ -l_{\bf q} m_{\bf q} e^{iq_ya/2} + l^2_{\bf q} e^{-iq_ya/2} ] e^{iq_xa/2}
\nonumber \\
&& - {J_{12}S \over N_{\rm uc}} \sum_{\bf q}
\left\langle 0 \left| 
d^\dagger({\bf q}) {1 \over {\cal E}} a({\bf q}) \right| 0 \right\rangle
[ -l_{\bf q} m_{\bf q} + l^2_{\bf q} e^{iq_xa} ]
\nonumber \\
&& - {J_{12}S \over N_{\rm uc}} \sum_{\bf q}
\left\langle 0 \left| 
a({\bf q}) {1 \over {\cal E}} a^\dagger({\bf q}) \right| 0 \right\rangle
[ -l_{\bf q} m_{\bf q} e^{iaq_x} + m^2_{\bf q} ]
\nonumber \\
&& - {J_{12}S \over N_{\rm uc}} \sum_{\bf q}
\left\langle 0 \left| 
a({\bf q}) {1 \over {\cal E}} b(-{\bf q}) \right| 0 \right\rangle
[ -l_{\bf q} m_{\bf q} e^{-iq_ya/2} + m^2_{\bf q} e^{iq_ya/2} ] e^{iq_xa/2}
\nonumber \\
&& - {J_{12}S \over N_{\rm uc}} \sum_{\bf q}
\left\langle 0 \left| 
a({\bf q}) {1 \over {\cal E}} c(-{\bf q}) \right| 0 \right\rangle
[ -l_{\bf q} m_{\bf q} e^{iq_ya/2} + m^2_{\bf q} e^{-iq_ya/2} ] e^{iq_xa/2}
\nonumber \\
&& - {J_{12}S \over N_{\rm uc}} \sum_{\bf q}
\left\langle 0 \left| 
a({\bf q}) {1 \over {\cal E}} d^\dagger({\bf q}) \right| 0 \right\rangle
s -l_{\bf q} m_{\bf q} + m^2_{\bf q} e^{iq_xa} ] \ .
\end{eqnarray}

For the CuI subsystem we have normal modes via the transformations,
\begin{eqnarray}
a({\bf q}) & = & (1/\sqrt 2) [ a_+({\bf q}) + a_-({\bf q})] \nonumber \\
d({\bf q}) & = & (1/\sqrt 2) [ a_+({\bf q}) - a_-({\bf q})] \nonumber \\
b({\bf q}) & = & (1/\sqrt 2) [ b_+({\bf q}) + b_-({\bf q})] \nonumber \\
c({\bf q}) & = & (1/\sqrt 2) [ b_+({\bf q}) - b_-({\bf q})] \ .
\end{eqnarray}
In terms of these operators (in the order $a_+$, $b_+$, $a_-$,
$b_-$) we have the matrices ${\bf A}$ and ${\bf B}$:
\begin{eqnarray}
{{\bf A} ( {\bf q}) \over S} = \left[
\begin{array} { c | c | c | c } \hline
4J + 2J_3 & 0 & 0 & 0 \\
0 & 4J + 2J_3 & 0 & 0 \\
0 & 0 & 4J + 2J_3 & 0 \\
0 & 0 & 0 & 4J + 2J_3 \\
\hline \end{array} \right]
\end{eqnarray}
and ${\bf B} ( {\bf q}) / S$ as
\begin{small}
\begin{eqnarray}
\left[ \begin{array} { c | c | c | c } \hline
0 & 2J (c_+ + c_-) + 2J_3 c_z & 0 & 0 \\
2J (c_+ + c_-) + 2J_3 c_z & 0 & 0 & 0 \\
0 & 0 & 0 & 2J (c_+ - c_-) + 2J_3 c_z \\
0 & 0 & 2J (c_+ - c_-) + 2J_3 c_z & 0 \\
\hline \end{array} \right] \ ,
\end{eqnarray}
\end{small}
where
\begin{eqnarray}
c_+ &=& \cos [a (q_x+q_y)/2] \ , \ \ \ \ \
c_- = \cos [a (q_x-q_y)/2] \ , \ \ \ \ \
c_z = \cos (q_z c/2) \ .
\end{eqnarray}
Now each sector has relations analogous to the CuII's:
\begin{eqnarray}
a_\sigma ({\bf q}) & = & l_{\sigma, {\bf q}} \alpha_\sigma({\bf q})
- m_{\sigma,{\bf q}} \beta_\sigma^\dagger (-{\bf q}) \nonumber \\
b_\sigma^\dagger (-{\bf q}) & = & -m_{\sigma,{\bf q}} \alpha_\sigma ({\bf q})
+ l_{\sigma,{\bf q}} \beta_\sigma^\dagger (-{\bf q}) \ ,
\end{eqnarray}
where $\alpha_\sigma({\bf q})$ and $\beta_\sigma({\bf q})$ are the
normal mode operators, and
\begin{eqnarray}
l_{\sigma,{\bf q}}^2 = {A + E_\sigma({\bf q})
\over 2 E_\sigma({\bf q}) } \ , \ \ \ \ \
m_{\sigma,{\bf q}}^2 = {A - E_\sigma({\bf q})
\over 2 E_\sigma({\bf q}) } \ , \ \ \ \ \
l_{\sigma,{\bf q}} m_{\sigma,{\bf q}} =
 {B_\sigma({\bf q}) \over 2 E_\sigma ({\bf q}) } \ .
\end{eqnarray}
Here
\begin{eqnarray}
E_\sigma({\bf q})^2 = A^2 - B_\sigma({\bf q})^2 \ ,
\end{eqnarray}
where
\begin{eqnarray}
A & = & 4J + 2J_3 \nonumber \\
B_\sigma ({\bf q}) &=& 2J \Biggl( \cos[(q_x+q_y)a/2]
+ \sigma \cos[(q_x-q_y)a/2]\Biggr) + 2J_3 \cos( q_z c) \ ,
\end{eqnarray}
so that
\begin{mathletters}
\begin{eqnarray}
B_+ ({\bf q}) &=&  2J_3 \cos (q_zc) +4J \cos (q_xa/2) \cos (q_ya/2) \\
B_- ({\bf q}) &=&  2J_3 \cos (q_zc) - 4J \sin (q_xa/2) \sin (q_ya/2) \ .
\end{eqnarray}
\end{mathletters}
Thus
\begin{eqnarray}
\langle 0 | a^\dagger({\bf q}) {1 \over {\cal E}} a({\bf q}) | 0 \rangle &=&
{1 \over 2S} \sum_\sigma \langle 0 | a_\sigma^\dagger({\bf q})
a_\sigma ({\bf q}) | 0 \rangle E_\sigma({\bf q})^{-1}
\nonumber \\ & = &
{1 \over 2S} \sum_\sigma m_{\sigma {\bf q}}^2 E_\sigma ({\bf q})^{-1} =
\sum_\sigma {A - E_\sigma({\bf q}) \over 4S E_\sigma({\bf q})^2 } \ .
\end{eqnarray}
Similarly
\begin{eqnarray}
\langle 0 | a({\bf q}) {1 \over {\cal E}} a^\dagger({\bf q}) | 0 \rangle &=&
{1 \over 2S} \sum_\sigma \langle 0 | a_\sigma({\bf q})
a_\sigma^\dagger ({\bf q}) | 0 \rangle E_\sigma({\bf q})^{-1}
\nonumber \\ & = &
{1 \over 2S} \sum_\sigma l_{\sigma {\bf q}}^2 E_\sigma ({\bf q})^{-1} =
\sum_\sigma {A + E_\sigma({\bf q}) \over 4S E_\sigma({\bf q})^2 }
\end{eqnarray}
\begin{eqnarray}
\langle 0 | d^\dagger({\bf q}) {1 \over {\cal E}} a({\bf q}) | 0 \rangle &=&
{1 \over 2S} \sum_\sigma \sigma \langle 0 | a_\sigma^\dagger({\bf q})
a_\sigma ({\bf q}) | 0 \rangle E_\sigma({\bf q})^{-1}
\nonumber \\ & = &
{1 \over 2S} \sum_\sigma \sigma m_{\sigma {\bf q}}^2 E_\sigma ({\bf q})^{-1}=
\sum_\sigma \sigma {A - E_\sigma({\bf q}) \over 4S E_\sigma({\bf q})^2 } 
\end{eqnarray}
\begin{eqnarray}
\langle 0 | a({\bf q}) {1 \over {\cal E}} d^\dagger({\bf q}) | 0 \rangle &=&
{1 \over 2S} \sum_\sigma \sigma \langle 0 | a_\sigma({\bf q})
a_\sigma^\dagger ({\bf q}) | 0 \rangle E_\sigma({\bf q})^{-1}
\nonumber \\ & = &
{1 \over 2S} \sum_\sigma \sigma l_{\sigma {\bf q}}^2 E_\sigma ({\bf q})^{-1}=
\sum_\sigma \sigma {A + E_\sigma({\bf q}) \over 4S E_\sigma({\bf q})^2 } 
\end{eqnarray}
\begin{eqnarray}
\langle 0 | b(-{\bf q}) {1 \over {\cal E}} a({\bf q}) | 0 \rangle &=&
{1 \over 2S} \sum_\sigma \langle 0 | b_\sigma(-{\bf q})
a_\sigma ({\bf q}) | 0 \rangle E_\sigma({\bf q})^{-1}
\nonumber \\ & = &
- {1 \over 2S} \sum_\sigma l_{\sigma,{\bf q}} m_{\sigma {\bf q}}
E_\sigma ({\bf q})^{-1}=
- \sum_\sigma {B_\sigma({\bf q}) \over 4S E_\sigma({\bf q})^2 } 
\end{eqnarray}
\begin{eqnarray}
\langle 0 | a({\bf q}) {1 \over {\cal E}} b(-{\bf q}) | 0 \rangle &=&
{1 \over 2S} \sum_\sigma \langle 0 | a_\sigma({\bf q})
b_\sigma (-{\bf q}) | 0 \rangle E_\sigma({\bf q})^{-1}
\nonumber \\ & = &
- {1 \over 2S} \sum_\sigma l_{\sigma,{\bf q}} m_{\sigma {\bf q}}
E_\sigma ({\bf q})^{-1}=
- \sum_\sigma {B_\sigma({\bf q}) \over 4S E_\sigma({\bf q})^2 } 
\end{eqnarray}
\begin{eqnarray}
\langle 0 | c(-{\bf q}) {1 \over {\cal E}} a({\bf q}) | 0 \rangle &=&
{1 \over 2S} \sum_\sigma \sigma \langle 0 | b_\sigma(-{\bf q})
a_\sigma ({\bf q}) | 0 \rangle E_\sigma({\bf q})^{-1}
\nonumber \\ &=&
- {1 \over 2S} \sum_\sigma \sigma l_{\sigma,{\bf q}} m_{\sigma {\bf q}}
E_\sigma({\bf q})^{-1} =
- \sum_\sigma \sigma {B_\sigma({\bf q}) \over 4S E_\sigma({\bf q})^2 } 
\end{eqnarray}
\begin{eqnarray}
\langle 0 | a({\bf q}) {1 \over {\cal E}} c(-{\bf q}) | 0 \rangle &=&
{1 \over 2S} \sum_\sigma \sigma \langle 0 | a_\sigma({\bf q})
b_\sigma (-{\bf q}) | 0 \rangle E_\sigma({\bf q})^{-1}
\nonumber \\ &=&
- {1 \over 2S} \sum_\sigma \sigma l_{\sigma,{\bf q}} m_{\sigma {\bf q}}
E_\sigma({\bf q})^{-1} =
- \sum_\sigma \sigma {B_\sigma({\bf q}) \over 4S E_\sigma({\bf q})^2 } \ .
\end{eqnarray}

Then
\begin{eqnarray}
\label{A1EQ}
A_1 &=& - {J_{12} \over 8N_{\rm uc}} \sum_{\bf q} \sum_\sigma
[E_\sigma({\bf q})^2 \epsilon({\bf q})]^{-1} \Biggl[ \nonumber \\ && \ \ 
[A - E_\sigma ({\bf q}) ] 
[ - \gamma({\bf q}) e^{iq_x a} + 1+\epsilon({\bf q}) ]
+ [A + E_\sigma ({\bf q}) ] 
[ - \gamma({\bf q}) e^{iq_x a} + 1-\epsilon({\bf q}) ]
\nonumber \\ && \ \ - B_\sigma ({\bf q})
[ -\gamma({\bf q}) e^{-iq_ya/2} + (1+\epsilon({\bf q})) e^{iq_ya/2}] e^{iq_xa/2}
\nonumber \\ && \ \ - B_\sigma ({\bf q})
[ -\gamma({\bf q}) e^{-iq_ya/2} + (1-\epsilon({\bf q})) e^{iq_ya/2}] e^{iq_xa/2}
\nonumber \\ && \ - \sigma B_\sigma ({\bf q})
[ - \gamma({\bf q}) e^{iq_ya/2} + (1+\epsilon({\bf q})) e^{-iq_ya/2}] e^{iq_xa/2}
\nonumber \\ && \ \ - \sigma B_\sigma ({\bf q})
[ - \gamma({\bf q}) e^{iq_ya/2} + (1-\epsilon({\bf q})) e^{-iq_ya/2}] e^{iq_xa/2}
\nonumber \\ && \ \ 
+ \sigma [A - E_\sigma ({\bf q}) ]
[ - \gamma({\bf q}) + (1 + \epsilon({\bf q})) e^{iq_xa} ]
\nonumber \\ && \ \ + \sigma [A + E_\sigma ({\bf q}) ]
[ - \gamma({\bf q}) + (1 - \epsilon({\bf q})) e^{iq_xa} ] \Biggr] \ ,
\end{eqnarray}
where
\begin{eqnarray}
\gamma({\bf q}) &=& \case 1/2 [ \cos (q_xa) + \cos (q_ya) ]
\end{eqnarray}
and
\begin{eqnarray}
\epsilon ({\bf q})^2 = 1 - \gamma({\bf q})^2 \ .
\end{eqnarray}

We use the fact that $J_3 \ll J$.  Only if a sum is divergent
will it make a difference if we retain nonzero $J_3$.  So we
tentatively assume no divergences and write
\begin{mathletters}
\begin{eqnarray}
B_+ ({\bf q}) &=&  4J \cos (q_xa/2) \cos (q_ya/2) \\
B_- ({\bf q}) &=&  -4J \sin (q_xa/2) \sin (q_ya/2) \ .
\end{eqnarray}
\end{mathletters}
We now simplify Eq. (\ref{A1EQ}).  We note that under the sum over
wavevectors we can replace $\exp(i q_xa)$ by $\gamma({\bf q})$.
Let us apply the same reasoning to $\exp[i (q_x \pm q_y )a/2]$:
\begin{eqnarray}
\exp[i(q_x \pm q_y )a/2] & = & \cos (q_xa/2) \cos (q_ya/2) \mp
\sin (q_xa/2) \sin(q_ya/2) \nonumber \\ && \
+ i \Biggl( \sin (q_xa/2) \cos(q_ya/2)a \pm \cos(q_xa/2) \sin)q_ya/2)
\Biggr) \ .
\end{eqnarray}
After summation over wavevectors the imaginary parts will drop out.  So
\begin{mathletters}
\begin{eqnarray}
\exp[i(q_x + q_y )a/2] & = & \cos (q_xa/2) \cos (q_ya/2) -
\sin (q_xa/2) \sin(q_ya/2) \nonumber \\ &=&
\left( {1 \over 4J} \right) \sum_\sigma B_\sigma ({\bf q}) \\
\exp[i(q_x - q_y )a/2] & = & \cos (q_xa/2) \cos (q_ya/2) +
\sin (q_xa/2) \sin(q_ya/2) \nonumber \\ &=&
\left( {1 \over 4J} \right) \sum_\sigma \sigma B_\sigma ({\bf q}) \ . 
\end{eqnarray}
\end{mathletters}
In this connection note that sums which are proportional to
$B_+({\bf q}) B_-({\bf q})$ vanish.  So
\begin{eqnarray}
A_1 &=& - {J_{12} \over 8N_{\rm uc}} \sum_{\bf q} \sum_\sigma
\left( { 1 \over E_\sigma({\bf q})^2 \epsilon({\bf q}) } \right) \Biggl(
\nonumber \\ && \ 
\left[ A - E_\sigma({\bf q}) \right] \left[ -\gamma({\bf q})^2
+ 1 + \epsilon({\bf q}) \right] +
\left[ A + E_\sigma({\bf q}) \right] \left[ -\gamma({\bf q})^2
+ 1 - \epsilon({\bf q}) \right] \nonumber \\ && \
+ \sigma \left[ A - E_\sigma({\bf q}) \right] \left[ \gamma({\bf q})
\epsilon({\bf q}) \right] + \sigma \left[ A + E_\sigma({\bf q}) \right]
\left[ -\gamma({\bf q}) \epsilon({\bf q}) \right] \nonumber \\ && \
- \left( {B_\sigma({\bf q}) \over 4J} \right)
\left[ - \gamma({\bf q}) \sigma B_\sigma({\bf q}) + (1 + \epsilon({\bf q}))
B_\sigma({\bf q}) \right] \nonumber \\ && \ \ 
- \left( {B_\sigma({\bf q}) \over 4J} \right)
\left[ - \gamma({\bf q}) \sigma B_\sigma({\bf q}) + (1 - \epsilon({\bf q}))
B_\sigma({\bf q}) \right] \nonumber \\ && \
- \left( {\sigma B_\sigma({\bf q}) \over 4J} \right)
\left[ - \gamma({\bf q}) B_\sigma({\bf q}) + (1 + \epsilon({\bf q}))\sigma
B_\sigma({\bf q}) \right] \nonumber \\ && \ \ 
- \left( {\sigma B_\sigma({\bf q}) \over 4J} \right)
\left[ - \gamma({\bf q}) B_\sigma({\bf q}) + (1 - \epsilon({\bf q})) \sigma
B_\sigma({\bf q}) \right] \Biggr) \nonumber \\ &=&
- {J_{12} \over 8N_{\rm uc}} \sum_{\bf q} \sum_\sigma
\left( { 1 \over E_\sigma({\bf q})^2 \epsilon({\bf q}) } \right) \Biggl(
2A \epsilon({\bf q})^2 - 2 E_\sigma({\bf q}) \epsilon ({\bf q}) 
- 2 E_\sigma({\bf q}) \epsilon ({\bf q}) \sigma \gamma({\bf q})
\nonumber \\ && \ 
- J^{-1} B_\sigma({\bf q})^2 \left[  1 - \sigma \gamma({\bf q})
\right] \Biggr) \ . 
\end{eqnarray}

Now we must understand how the wavevector sums are to be done.
The unit cell is
\begin{eqnarray}
{\bf a}_1 = a \hat x + a \hat y \ , \ \ \ \ \ 
{\bf a}_2 = - a \hat x + a \hat y \ , \ \ \ \ \ 
\end{eqnarray}
Thus the reciprocal lattice vectors are
\begin{eqnarray}
{\bf G}_1 &=& (\pi /a) ( \hat x + \hat y ) \ , \ \ \ \ \
{\bf G}_2 = (\pi /a) ( - \hat x + \hat y ) \ .
\end{eqnarray}
Thus the sums are carried over the first zone, shown below
in Fig. \ref{BZFIG}.

\subsection{$A_2$}

Thus
\begin{eqnarray}
A_2 & = & - \left\langle 0 \left| V_{I-II} {1 \over {\cal E}} a_m
e_n \right| 0 \right\rangle - \left\langle 0 \left|
a_m e_n {1 \over {\cal E}} V_{I-II} \right| 0 \right\rangle \ ,
\end{eqnarray}
where we invoke perturbation theory relative to decoupled
CuI and CuII subsystems.  As for $A_1$ effectively
we have Eq. (\ref{EFEQ}).  Thus
\begin{eqnarray}
A_2 &=& - J_{12}S \sum_{i \in a } \left\langle 0 \left| 
a_i^\dagger {1 \over {\cal E}} a_m \right| 0 \right\rangle
\langle 0| [ e^\dagger_{i+\delta_{ae}}
+ f_{i+\delta_{af}}] e_{m+\delta_{ae}} | 0 \rangle 
\nonumber \\
&& - J_{12}S \sum_{i \in b } \left\langle 0 \left| 
b_i {1 \over {\cal E}} a_m \right| 0 \right\rangle
\langle 0| [ e^\dagger_{i+\delta_{be}}
+ f_{i+\delta_{bf}}] e_{m+\delta_{ae}} | 0 \rangle 
\nonumber \\
&& - J_{12}S \sum_{i \in c } \left\langle 0 \left| 
c_i {1 \over {\cal E}} a_m \right| 0 \right\rangle
\langle 0| [ e^\dagger_{i+\delta_{ce}}
+ f_{i+\delta_{cf}}] e_{m+\delta_{ae}} | 0 \rangle 
\nonumber \\
&& - J_{12}S \sum_{i \in d } \left\langle 0 \left| 
d_i^\dagger {1 \over {\cal E}} a_m \right| 0 \right\rangle
\langle 0| [ e^\dagger_{i+\delta_{de}}
+ f_{i+\delta_{df}}] e_{m+\delta_{ae}} | 0 \rangle 
\nonumber \\
&& - J_{12}S \sum_{i \in a } \left\langle 0 \left| 
a_m {1 \over {\cal E}} a_i^\dagger \right| 0 \right\rangle
\langle 0| e_{m+\delta_{ae}} [ e^\dagger_{i+\delta_{ae}}
+ f_{i+\delta_{af}}] | 0 \rangle 
\nonumber \\
&& - J_{12}S \sum_{i \in b } \left\langle 0 \left| 
a_m {1 \over {\cal E}} b_i \right| 0 \right\rangle
\langle 0| e_{m+\delta_{ae}} [ e^\dagger_{i+\delta_{be}}
+ f_{i+\delta_{bf}}] | 0 \rangle 
\nonumber \\
&& - J_{12}S \sum_{i \in c } \left\langle 0 \left| 
a_m {1 \over {\cal E}} c_i \right| 0 \right\rangle
\langle 0| e_{m+\delta_{ae}} [ e^\dagger_{i+\delta_{ce}}
+ f_{i+\delta_{cf}}] | 0 \rangle 
\nonumber \\
&& - J_{12}S \sum_{i \in d } \left\langle 0 \left| 
a_m {1 \over {\cal E}} d_i^\dagger \right| 0 \right\rangle
\langle 0| e_{m+\delta_{af}} [ e^\dagger_{i+\delta_{de}}
+ f_{i+\delta_{df}}] | 0 \rangle \ . 
\end{eqnarray}
Then
\begin{eqnarray}
A_2 &=& - {J_{12}S \over N_{\rm uc}^2} \sum_{{\bf q},{\bf k}}
\sum_{i \in a } \left\langle 0 \left| 
a^\dagger({\bf q}) {1 \over {\cal E}} a({\bf q}) \right| 0 \right\rangle
e^{i({\bf q}+{\bf k}) \cdot {\bf r}_{im}}  \nonumber \\ && \ \ \times
\langle 0| [ e^\dagger({\bf k}) e^{i{\bf k} \cdot \bbox{\delta}_{ae}}
+ f(-{\bf k}) e^{i {\bf k} \cdot \bbox{\delta}_{af}} ]
e({\bf k}) e^{-i{\bf k} \cdot \bbox{\delta}_{ae}} | 0 \rangle 
\nonumber \\
&& - {J_{12}S  \over N_{\rm uc}^2} \sum_{{\bf q}, {\bf k}}
\sum_{i \in b } \left\langle 0 \left| 
b(-{\bf q}) {1 \over {\cal E}} a({\bf q}) \right| 0 \right\rangle
e^{i({\bf q}+{\bf k}) \cdot {\bf r}_{im}}  \nonumber \\ && \ \ \times
\langle 0| [ e^\dagger({\bf k}) e^{i {\bf k} \cdot \bbox{\delta}_{be}}
+ f(-{\bf k}) e^{i{\bf k} \cdot \bbox{\delta}_{bf}} ]
e({\bf k})  e^{-i{\bf k} \cdot \bbox{\delta}_{ae}} | 0 \rangle 
\nonumber \\
&& - {J_{12}S \over N_{\rm uc}^2} \sum_{{\bf q}, {\bf k}}
\sum_{i \in c } \left\langle 0 \left| 
c(-{\bf q}) {1 \over {\cal E}} a({\bf q}) \right| 0 \right\rangle
e^{i({\bf q}+{\bf k}) \cdot {\bf r}_{im}}  \nonumber \\ && \ \ \times
\langle 0| [ e^\dagger({\bf k}) e^{i {\bf k} \cdot \bbox{\delta}_{ce}}
+ f(-{\bf k}) e^{i {\bf k} \cdot \bbox{\delta}_{cf}} ]
e({\bf k}) e^{-i{\bf k} \cdot \bbox{\delta}_{ae}} ] | 0 \rangle 
\nonumber \\
&& - {J_{12}S \over N_{\rm uc}^2} \sum_{{\bf q}, {\bf k}}
\sum_{i \in d } \left\langle 0 \left| 
d^\dagger({\bf q}) {1 \over {\cal E}} a({\bf q}) \right| 0 \right\rangle
e^{i({\bf q}+{\bf k}) \cdot {\bf r}_{im}}  \nonumber \\ && \ \ \times
\langle 0| [ e^\dagger({\bf k}) e^{i {\bf k} \cdot \bbox{\delta}_{de}}
+ f(-{\bf k}) e^{i {\bf k} \cdot \bbox{\delta}_{df}} ]
e({\bf k})  e^{-i{\bf k} \cdot \bbox{\delta}_{ae}} | 0 \rangle 
\nonumber \\
&& - {J_{12}S \over N_{\rm uc}^2} \sum_{{\bf q}, {\bf k}}
\sum_{i \in a } \left\langle 0 \left| 
a({\bf q}) {1 \over {\cal E}} a^\dagger({\bf q}) \right| 0 \right\rangle
e^{i({\bf q}+{\bf k}) \cdot {\bf r}_{im}}  \nonumber \\ && \ \ \times
\langle 0| e({\bf k}) e^{-i{\bf k}\cdot \bbox{\delta}_{ae}}
[ e^\dagger({\bf k}) e^{i {\bf k} \cdot \bbox{\delta}_{ae}}
+ f(-{\bf k}) e^{i {\bf k} \cdot \bbox{\delta}_{af}} ] | 0 \rangle 
\nonumber \\
&& - {J_{12}S \over N_{\rm uc}^2} \sum_{{\bf q}, {\bf k}}
\sum_{i \in b } \left\langle 0 \left| 
a({\bf q}) {1 \over {\cal E}} b(-{\bf q}) \right| 0 \right\rangle
e^{i({\bf q}+{\bf k}) \cdot {\bf r}_{im}}  \nonumber \\ && \ \ \times
\langle 0| e({\bf k}) e^{-i{\bf k}\cdot \bbox{\delta}_{ae}}
[ e^\dagger({\bf k}) e^{i {\bf k} \cdot \bbox{\delta}_{be}}
+ f(-{\bf k}) e^{i {\bf k} \cdot \bbox{\delta}_{bf}} ] | 0 \rangle 
\nonumber \\
&& - {J_{12}S \over N_{\rm uc}^2} \sum_{{\bf q}, {\bf k}}
 \sum_{i \in c } \left\langle 0 \left| 
a({\bf q}) {1 \over {\cal E}} c(-{\bf q}) \right| 0 \right\rangle
e^{i({\bf q}+{\bf k}) \cdot {\bf r}_{im}}  \nonumber \\ && \ \ \times
\langle 0| e({\bf k}) e^{-i{\bf k}\cdot \bbox{\delta}_{ae}}
[ e^\dagger({\bf k}) e^{i {\bf k} \cdot \bbox{\delta}_{ce}}
+ f(-{\bf k}) e^{i {\bf k} \cdot \bbox{\delta}_{cf}} ] | 0 \rangle 
\nonumber \\
&& - {J_{12}S \over N_{\rm uc}^2} \sum_{{\bf q}, {\bf k}}
\sum_{i \in d } \left\langle 0 \left| 
a({\bf q}) {1 \over {\cal E}} d^\dagger({\bf q}) \right| 0 \right\rangle
e^{i({\bf q}+{\bf k}) \cdot {\bf r}_{im}}  \nonumber \\ && \ \ \times
\langle 0| e({\bf k}) e^{-i{\bf k}\cdot \bbox{\delta}_{ae}}
[ e^\dagger({\bf k}) e^{i {\bf k} \cdot \bbox{\delta}_{de}}
+ f(-{\bf k}) e^{i {\bf k} \cdot \bbox{\delta}_{df}} ] | 0 \rangle \ . 
\end{eqnarray}
Doing the sum over $i$ we get
\begin{eqnarray}
A_2 &=& - {J_{12}S \over N_{\rm uc}} \sum_{\bf q}
\left\langle 0 \left| 
a^\dagger({\bf q}) {1 \over {\cal E}} a({\bf q}) \right| 0 \right\rangle
\langle 0| [ e^\dagger(-{\bf q}) e^{-i{\bf q} \cdot \bbox{\delta}_{ae}}
+ f({\bf q}) e^{-i {\bf q} \cdot \bbox{\delta}_{af}} ]
e(-{\bf q}) e^{i{\bf q} \cdot \bbox{\delta}_{ae}} | 0 \rangle 
\nonumber \\
&& - {J_{12}S  \over N_{\rm uc}} \sum_{\bf q}
\left\langle 0 \left| 
b(-{\bf q}) {1 \over {\cal E}} a({\bf q}) \right| 0 \right\rangle
\langle 0| [ e^\dagger(-{\bf q}) e^{-i {\bf q} \cdot \bbox{\delta}_{be}}
+ f({\bf q}) e^{-i{\bf q} \cdot \bbox{\delta}_{bf}} ]
e(-{\bf q})  e^{i{\bf q} \cdot \bbox{\delta}_{ae}} | 0 \rangle 
\nonumber \\
&& - {J_{12}S \over N_{\rm uc}} \sum_{\bf q}
\left\langle 0 \left| 
c(-{\bf q}) {1 \over {\cal E}} a({\bf q}) \right| 0 \right\rangle
\langle 0| [ e^\dagger(-{\bf q}) e^{-i {\bf q} \cdot \bbox{\delta}_{ce}}
+ f({\bf q}) e^{-i {\bf q} \cdot \bbox{\delta}_{cf}} ]
e(-{\bf q}) e^{i{\bf q} \cdot \bbox{\delta}_{ae}} ] | 0 \rangle 
\nonumber \\
&& - {J_{12}S \over N_{\rm uc}} \sum_{\bf q}
\left\langle 0 \left| 
d^\dagger({\bf q}) {1 \over {\cal E}} a({\bf q}) \right| 0 \right\rangle
\langle 0| [ e^\dagger(-{\bf q}) e^{-i {\bf q} \cdot \bbox{\delta}_{de}}
+ f({\bf q}) e^{-i {\bf q} \cdot \bbox{\delta}_{df}} ]
e(-{\bf q})  e^{i{\bf q} \cdot \bbox{\delta}_{ae}} | 0 \rangle 
\nonumber \\
&& - {J_{12}S \over N_{\rm uc}} \sum_{\bf q}
\left\langle 0 \left| 
a({\bf q}) {1 \over {\cal E}} a^\dagger({\bf q}) \right| 0 \right\rangle
\langle 0| e(-{\bf q}) e^{i{\bf q}\cdot \bbox{\delta}_{ae}}
[ e^\dagger(-{\bf q}) e^{-i {\bf q} \cdot \bbox{\delta}_{ae}}
+ f({\bf q}) e^{-i {\bf q} \cdot \bbox{\delta}_{af}} ] | 0 \rangle 
\nonumber \\
&& - {J_{12}S \over N_{\rm uc}} \sum_{\bf q}
\left\langle 0 \left| 
a({\bf q}) {1 \over {\cal E}} b(-{\bf q}) \right| 0 \right\rangle
\langle 0| e(-{\bf q}) e^{i{\bf q}\cdot \bbox{\delta}_{ae}}
[ e^\dagger(-{\bf q}) e^{-i {\bf q} \cdot \bbox{\delta}_{be}}
+ f({\bf q}) e^{-i {\bf q} \cdot \bbox{\delta}_{bf}} ] | 0 \rangle 
\nonumber \\
&& - {J_{12}S \over N_{\rm uc}} \sum_{\bf q}
\left\langle 0 \left| 
a({\bf q}) {1 \over {\cal E}} c(-{\bf q}) \right| 0 \right\rangle
\langle 0| e(-{\bf q}) e^{i{\bf q}\cdot \bbox{\delta}_{ae}}
[ e^\dagger(-{\bf q}) e^{-i {\bf q} \cdot \bbox{\delta}_{ce}}
+ f({\bf q}) e^{-i {\bf q} \cdot \bbox{\delta}_{cf}} ] | 0 \rangle 
\nonumber \\
&& - {J_{12}S \over N_{\rm uc}} \sum_{\bf q}
\left\langle 0 \left| 
a({\bf q}) {1 \over {\cal E}} d^\dagger({\bf q}) \right| 0 \right\rangle
\langle 0| e(-{\bf q}) e^{i{\bf q}\cdot \bbox{\delta}_{ae}}
[ e^\dagger(-{\bf q}) e^{-i {\bf q} \cdot \bbox{\delta}_{de}}
+ f({\bf q}) e^{-i {\bf q} \cdot \bbox{\delta}_{df}} ] | 0 \rangle \ . 
\end{eqnarray}
This is
\begin{eqnarray}
&& A_2 = - {J_{12}S \over N_{\rm uc}} \sum_{\bf q}
\left\langle 0 \left| 
a^\dagger({\bf q}) {1 \over {\cal E}} a({\bf q}) \right| 0 \right\rangle
\left[ \langle 0|  e^\dagger(-{\bf q}) e(-{\bf q}) | 0 \rangle 
+ \langle 0 | f({\bf q}) e(-{\bf q}) | 0 \rangle 
e^{i{\bf q} \cdot \bbox{\delta}_{fae}} \right]
\nonumber \\
&& - {J_{12}S  \over N_{\rm uc}} \sum_{\bf q}
\left\langle 0 \left| 
b(-{\bf q}) {1 \over {\cal E}} a({\bf q}) \right| 0 \right\rangle
\left[ \langle 0| [ e^\dagger(-{\bf q}) e(-{\bf q})  | 0 \rangle 
e^{i {\bf q} \cdot \bbox{\delta}_{aeb}}
+ \langle 0| f({\bf q}) e(-{\bf q})  | 0 \rangle
e^{i{\bf q} \cdot ( \bbox{\delta}_{ae} -\bbox{\delta}_{bf} ) } \right]
\nonumber \\
&& - {J_{12}S \over N_{\rm uc}} \sum_{\bf q}
\left\langle 0 \left| 
c(-{\bf q}) {1 \over {\cal E}} a({\bf q}) \right| 0 \right\rangle
\left[ \langle 0| [ e^\dagger(-{\bf q}) e(-{\bf q}) | 0 \rangle
e^{i{\bf q} \cdot \bbox{\delta}_{aec}} 
+ \langle 0| f({\bf q}) e(-{\bf q}) |0 \rangle
e^{i{\bf q} \cdot \bbox{\delta}_{ae}
-i {\bf q} \cdot \bbox{\delta}_{cf}} \right]
\nonumber \\
&& - {J_{12}S \over N_{\rm uc}} \sum_{\bf q}
\left\langle 0 \left| 
d^\dagger({\bf q}) {1 \over {\cal E}} a({\bf q}) \right| 0 \right\rangle
\left[ \langle 0| [ e^\dagger(-{\bf q}) e(-{\bf q}) |0 \rangle
e^{i{\bf q} \cdot \bbox{\delta}_{aed}}
+ \langle 0| f({\bf q}) e(-{\bf q}) |0 \rangle
e^{i{\bf q} \cdot (\bbox{\delta}_{ae} - \bbox{\delta}_{df})} \right]
\nonumber \\
&& - {J_{12}S \over N_{\rm uc}} \sum_{\bf q}
\left\langle 0 \left| 
a({\bf q}) {1 \over {\cal E}} a^\dagger({\bf q}) \right| 0 \right\rangle
\left[ \langle 0| e(-{\bf q}) e^\dagger (-{\bf q}) |0 \rangle +
\langle 0| e(-{\bf q})  f({\bf q}) |0 \rangle
e^{i {\bf q} \cdot \bbox{\delta}_{fae}} \right]
\nonumber \\
&& - {J_{12}S \over N_{\rm uc}} \sum_{\bf q}
\left\langle 0 \left| 
a({\bf q}) {1 \over {\cal E}} b(-{\bf q}) \right| 0 \right\rangle
\left[ \langle 0| e(-{\bf q}) e^\dagger(-{\bf q}) | 0 \rangle
e^{i {\bf q} \cdot \bbox{\delta}_{aeb}}
+ \langle 0 | e(-{\bf q}) f({\bf q}) |0 \rangle 
e^{i {\bf q} \cdot (\bbox{\delta}_{ae} - \bbox{\delta}_{bf})} \right]
\nonumber \\
&& - {J_{12}S \over N_{\rm uc}} \sum_{\bf q}
\left\langle 0 \left| 
a({\bf q}) {1 \over {\cal E}} c(-{\bf q}) \right| 0 \right\rangle
\left[ \langle 0| e(-{\bf q}) e^\dagger(-{\bf q}) |0\rangle
e^{i {\bf q} \cdot \bbox{\delta}_{aec}} +
\langle 0| e(-{\bf q}) f({\bf q}) |0 \rangle
e^{i {\bf q} \cdot (\bbox{\delta}_{ae} - \bbox{\delta}_{cf})} \right]
\nonumber \\
&& - {J_{12}S \over N_{\rm uc}} \sum_{\bf q}
\left\langle 0 \left| 
a({\bf q}) {1 \over {\cal E}} d^\dagger({\bf q}) \right| 0 \right\rangle
\left[ \langle 0| e(-{\bf q}) e^\dagger(-{\bf q}) |0 \rangle
e^{i{\bf q}\cdot \bbox{\delta}_{aed}}
\langle 0| e(-{\bf q}) f({\bf q}) |0 \rangle
e^{i {\bf q} \cdot (\bbox{\delta}_{ae} - \bbox{\delta}_{df})} \right] \ .
\end{eqnarray}
Here the symbol $\delta_{fae}$ denotes the vector which goes from
an f site to an e site via an a site, such that fae is a sequence
of nearest neighboring sites.  So
\begin{eqnarray}
&& A_2 = - {J_{12} \over 4N_{\rm uc}} \sum_{\bf q} \sum_\sigma \Biggl(
E_\sigma ({\bf q})^{-2} \nonumber \\ && \ \
\left[ A - E_\sigma ({\bf q}) \right] \left[ m_{\bf q}^2
- l_{\bf q} m_{\bf q} e^{-i q_x a} \right]
+ \left[ A + E_\sigma ({\bf q}) \right] \left[ l_{\bf q}^2
- l_{\bf q} m_{\bf q} e^{-iq_xa} \right] \nonumber \\ && \ \
- B_\sigma ({\bf q}) \left[ e^{-i(q_x + q_y)a/2} m_{\bf q}^2 -
l_{\bf q} m_{\bf q} e^{i(q_y-q_x)a/2} \right]
- B_\sigma ({\bf q}) \left[ e^{-i(q_x + q_y)a/2} l_{\bf q}^2 -
l_{\bf q} m_{\bf q} e^{i(q_y-q_x)a/2} \right]
\nonumber \\ && \ \
-\sigma B_\sigma({\bf q}) \left[ e^{i(q_y-q_x)a/2} m_{\bf q}^2
-e^{-i(q_x+q_y)a/2} l_{\bf q}m_{\bf q} \right]
-\sigma B_\sigma({\bf q}) \left[ e^{i(q_y-q_x)a/2} l_{\bf q}^2
-e^{-i(q_x+q_y)a/2} l_{\bf q}m_{\bf q} \right]
\nonumber \\ && \ \
+\sigma \left[A-E_\sigma \right] \left[ e^{-iq_xa} m_{\bf q}^2
- l_{\bf q} m_{\bf q} \right]
+\sigma \left[A+E_\sigma \right] \left[ e^{-iq_xa} m_{\bf q}^2
- l_{\bf q} m_{\bf q} \right] \Biggr)\ .
\end{eqnarray}
Making the same replacements as in $A_1$ we get
\begin{eqnarray}
A_2 &=& - {J_{12} \over 4N_{\rm uc}} \sum_{\bf q} \sum_\sigma 
E_\sigma({\bf q})^{-2} \Biggl(
\left[ A - E_\sigma ({\bf q}) \right] \left[ m_{\bf q}^2
- l_{\bf q} m_{\bf q} \gamma({\bf q}) \right]
+ \left[ A + E_\sigma ({\bf q}) \right] \left[ l_{\bf q}^2
- l_{\bf q} m_{\bf q} \gamma({\bf q}) \right] \nonumber \\ && \ \
- B_\sigma ({\bf q})B_\sigma ({\bf q})(4J)^{-1} \left[ m_{\bf q}^2 -
\sigma l_{\bf q} m_{\bf q} \right]
- B_\sigma ({\bf q})B_\sigma ({\bf q})(4J)^{-1} \left[ l_{\bf q}^2 -
\sigma l_{\bf q} m_{\bf q} \right]
\nonumber \\ && \ \
-\sigma B_\sigma({\bf q}) B_\sigma ({\bf q})(4J)^{-1}
\left[ \sigma m_{\bf q}^2 - l_{\bf q}m_{\bf q} \right]
-\sigma B_\sigma({\bf q}) B_\sigma ({\bf q})(4J)^{-1}
\left[ \sigma l_{\bf q}^2 - l_{\bf q}m_{\bf q} \right]
\nonumber \\ && \ \
+\sigma \left[A-E_\sigma \right] \left[ \gamma({\bf q}) m_{\bf q}^2
- l_{\bf q} m_{\bf q} \right]
+\sigma \left[A+E_\sigma \right] \left[ \gamma({\bf q}) l_{\bf q}^2
-  l_{\bf q} m_{\bf q} \right] \Biggr) \ .
\end{eqnarray}
This is
\begin{eqnarray}
&& A_2 = - {J_{12} \over 8N_{\rm uc}} \sum_{\bf q} \sum_\sigma 
E_\sigma({\bf q})^{-2} \epsilon({\bf q})^{-1} \Biggl( \nonumber \\ && \ \
\left[ A - E_\sigma ({\bf q}) \right] \left[ 1 - \epsilon({\bf q})
- \gamma({\bf q})^2 \right]
+ \left[ A + E_\sigma ({\bf q}) \right] \left[ 1 + \epsilon({\bf q})
- \gamma({\bf q})^2 \right] \nonumber \\ && \ \
- \left( {B_\sigma ({\bf q})^2 \over 4J} \right)
\left[ 2 - 2 \sigma \gamma ({\bf q}) \right]
- \left( {\sigma B_\sigma ({\bf q})^2 \over 4J} \right)
\left[ 2 \sigma  - 2 \gamma ({\bf q}) \right]
\nonumber \\ &&
+\sigma \left[A-E_\sigma ({\bf q}) \right]
\Biggl[ \gamma({\bf q}) [1 - \epsilon({\bf q})]
- \gamma({\bf q}) \Biggr]
+\sigma \left[A+E_\sigma ({\bf q}) \right]
\Biggl[ \gamma({\bf q}) [1 + \epsilon({\bf q})]
+  \gamma({\bf q}) \Biggr] \Biggr) \ .
\end{eqnarray}
So
\begin{eqnarray}
A_2 &=& - {J_{12} \over 8N_{\rm uc}} \sum_{\bf q} \sum_\sigma 
\left( {1 \over E_\sigma({\bf q})^2 \epsilon({\bf q})} \right) \Biggl( 
2A \epsilon({\bf q})^2 + 2 E_\sigma({\bf q}) \epsilon({\bf q})
+ 2 E_\sigma({\bf q}) \sigma \gamma({\bf q}) \epsilon({\bf q}) \nonumber \\
&& \ - B_\sigma ({\bf q})^2 J^{-1} \left[ 1 - \sigma \gamma ({\bf q})
\right] \Biggr) \ .
\end{eqnarray}

\subsection{Summary}

So
\begin{mathletters}
\begin{eqnarray}
A_1 = - \left( {J_{12} \over 2J} \right) \left( C_\alpha - C_\beta) \right) \\
A_2 = - \left( {J_{12} \over 2J} \right) \left( C_\alpha + C_\beta) \right) \ ,
\end{eqnarray}
\end{mathletters}
where
\begin{mathletters}
\begin{eqnarray}
C_\alpha &=& {J \over 4N_{\rm uc}} \sum_{\bf q} \sum_\sigma 
\left( {1 \over E_\sigma({\bf q})^2 \epsilon({\bf q})} \right) \Biggl( 
2A \epsilon({\bf q})^2 - B_\sigma({\bf q})^2 J^{-1}
[1 - \sigma \gamma({\bf q}) ] \Biggr) \\
C_\beta &=& {J \over 4N_{\rm uc}} \sum_{\bf q} \sum_\sigma 
\left( {1 \over E_\sigma({\bf q})} \right) \Biggl( 
2 [1 + \sigma \gamma({\bf q}) ] \Biggr) \ .
\end{eqnarray}
\end{mathletters}
If we extend the sum over $- \pi /a < q_x, q_y < \pi /a$, then we may write
these as
\begin{mathletters}
\begin{eqnarray}
C_\alpha &=& {J \over 4N_{\rm uc}} \sum_{\bf q}
\left( {1 \over E_+({\bf q})^2 \epsilon({\bf q})} \right) \Biggl( 
2A \epsilon({\bf q})^2 - B_+({\bf q})^2 J^{-1}
[1 - \gamma({\bf q}) ] \Biggr) \\
C_\beta &=& {J \over 4N_{\rm uc}} \sum_{\bf q}
\left( {1 \over E_+({\bf q})} \right) \Biggl( 
2 [1 + \gamma({\bf q}) ] \Biggr) \ .
\end{eqnarray}
\end{mathletters}
Of course, note that now $\sum_{\bf q} = 2 N_{uc}$.  So it is convenient
to introduce the notation $\langle \ \ \rangle_{\bf q}$ to denote
$(2N_{uc})^{-1} \sum_{\bf q}$.  Then
\begin{mathletters}
\begin{eqnarray}
C_\alpha  &=& {J \over 2} \left \langle
\left( {1 \over E_+({\bf q})^2 \epsilon({\bf q})} \right) \Biggl( 
2A \epsilon({\bf q})^2 - B_+({\bf q})^2 J^{-1}
[1 - \gamma({\bf q}) ] \Biggr) \right \rangle_{\bf q} \\
C_\beta &=& {J \over 2}
\left \langle \left( {1 \over E_+({\bf q})} \right) \Biggl( 
2 [1 + \gamma({\bf q}) ] \Biggr) \right \rangle_{\bf q} \ .
\end{eqnarray}
\end{mathletters}
Or
\begin{mathletters}
\begin{eqnarray}
C_\alpha & = & {1 \over 4} \left \langle {1 - \gamma ({\bf q})^2 - 2
\cos^2 (aq_x/2) \cos^2 (aq_y/2) [ 1 - \gamma ({\bf q}) ] \over
\left[ 1 - \cos^2 (aq_x/2) \cos^2 (aq_y/2) \right]
\sqrt{1 - \gamma ({\bf q})^2} } \right\rangle_{\bf q} \\
C_\beta & = & {1 \over 4} \left \langle {1 + \gamma ({\bf q}) \over
\sqrt { 1 - \cos^2 (aq_x/2) \cos^2 (aq_y/2) } } \right \rangle_{\bf q} \ .
\end{eqnarray}
\end{mathletters}
In the approximation that $\gamma({\bf q})=0$, etc.
$C_\alpha = C_\beta = \case 1/4$.
Numerical evaluation yields
\begin{eqnarray}
C_\alpha  = 0.1686 \ , \ \ \ \ \ \ \ C_\beta = 0.4210 \ .
\end{eqnarray}

\section{IN-PLANE CuI -- CuI INTERACTION}
\label{APPI-I}
\begin{figure}
\centerline{\psfig{figure=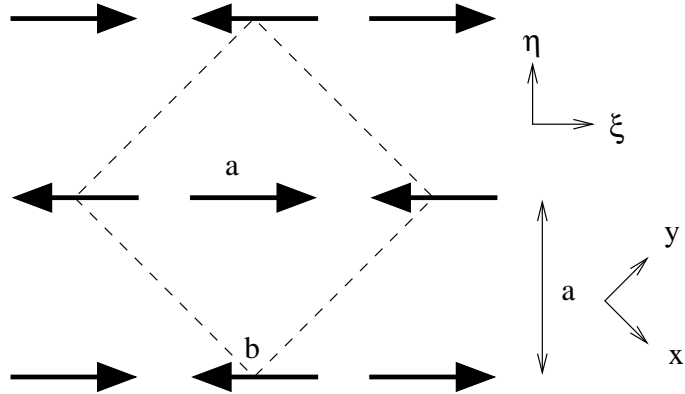}}

\vspace{0.2 in}
\caption{Unit cell of the square lattice.}
\label{APPF1FIG}
\end{figure}

Here we reproduce by perturbation theory the gap
found phenomenologically by Yildirim et al.\cite{SOPR1}
We treat an antiferromagnet on a square lattice (of lattice constant $a$),
in which there are
two sublattices, $a$ and $b$.  The lattice is shown in the
Fig. \ref{APPF1FIG} with the magnetic unit cell within dashed lines.
The magnetic unit cell has basis vectors
\begin{eqnarray}
{\bf a}_1 & = & a \hat \xi + a \hat \eta \ , \nonumber \\
{\bf a}_2 & = & - a \hat \xi + a \hat \eta \ .
\end{eqnarray}
We transform to bosons using Eq. (\ref{BOSONEQ}).

First we consider terms ${\cal H}$ in the Hamiltonian which are quadratic
in boson operators.  We write
\begin{eqnarray}
{\cal H} &=& {\cal H}_J + {\cal H}_\delta \ .
\end{eqnarray}
Here
\begin{eqnarray}
\label{AHZEROEQ}
{\cal H}_J = 4JS \sum_{\bf q} \Biggl( a^\dagger ({\bf q}) a({\bf q}) +
b^\dagger ({\bf q}) b({\bf q}) + \gamma({\bf q}) [
 a^\dagger ({\bf q}) b^\dagger (-{\bf q}) + a({\bf q}) b(-{\bf q}) ] \Biggr) \ ,
\end{eqnarray}
with 
\begin{equation}
\gamma({\bf q}) = \case 1/2 [ \cos q_x a + \cos q_y a] \ .
\end{equation}
and the sum over wave vectors is over the
Brillouin zone associated with the magnetic unit cell.   Also
\begin{eqnarray}
\label{AV2EQ}
{\cal H}_\delta &=& \delta J_1 S 
\sum_{\bf k} \Biggl( c_x({\bf k}) - c_y ({\bf k}) \Biggr)
\Biggl[ a({\bf k}) + a^\dagger (-{\bf k}) \Biggr]
\Biggl[ b^\dagger({\bf k}) + b (-{\bf k}) \Biggr] \ ,
\end{eqnarray}
where $c_x({\bf k}) = \cos k_x a$ and $c_y({\bf k}) = \cos k_y a$.

Since the effect we wish to treat involves energies of relative order
$(1/S)$, we now consider the fourth-order terms, $V_4$, in the boson
Hamiltonian, which we write as
\begin{eqnarray}
V_4 = V_J + V_\delta \ ,
\end{eqnarray}
where
\begin{eqnarray}
V_J & = & - \case 1/2 J \sum_{\langle ij \rangle}
b_j^\dagger (a_i^\dagger + b_j)^2 a_i  \ ,
\end{eqnarray}
where $\langle ij \rangle$ indicates that
$i$ is summed over a sites and $j$ over nearest neighboring b sites
and
\begin{eqnarray}
V_\delta  &=& \delta J_1 \sum_{\langle ij \rangle} \sigma_\delta
\Biggl[ - \case 1/4 a_i^\dagger a_i^\dagger a_i (b_j^\dagger + b_j)
- \case 1/4 (a_i^\dagger + a_i) b_j^\dagger b_j b_j + a_i^\dagger a_i b_j^\dagger b_j \Biggr] \ ,
\end{eqnarray}
where  $\sigma_\delta$ is $+1$ for $x$ bonds and $-1$ for $y$ bonds.

We construct the effective quadratic Hamiltonian by taking all
possible averages of pairs of operators out of the fourth order terms.
Thus we have the effective quadratic terms
\begin{eqnarray}
\Delta H_J &=& - \case 1/2 J \sum_{\langle ij \rangle} \Biggl[ 
a_i b_j^\dagger \langle (b_j + a_i^\dagger)^2 \rangle
+ 2 a_i (b_j + a_i^\dagger)  \langle b_j^\dagger (b_j + a_i^\dagger) \rangle \nonumber \\ && \ \
+ 2 b_j^\dagger (b_j + a_i^\dagger)  \langle (b_j + a_i^\dagger)a_i \rangle
+ (b_j+a_i^\dagger)^2 \langle a_i b_j^\dagger \rangle \Biggr]
\end{eqnarray}
and
\begin{eqnarray}
\Delta H_\delta &=&  \delta J_1 \sum_{\langle ij \rangle} \sigma_\delta
\Biggl[ -\case 1/2 a_i^\dagger (b_j^\dagger + b_j) \langle a_i^\dagger a_i \rangle
-\case 1/2 a_i^\dagger a_i \langle a_i^\dagger (b_j^\dagger + b_j) \rangle
-\case 1/4 a_i^\dagger a_i^\dagger \langle a_i (b_j^\dagger + b_j) \rangle \nonumber \\ && \ \
-\case 1/4 a_i (b_j^\dagger + b_j) \langle a_i^\dagger a_i^\dagger \rangle
-\case 1/2 (a_i^\dagger + a_i)b_j \langle b_j^\dagger b_j \rangle
-\case 1/2 b_j^\dagger b_j \langle b_j (a_i^\dagger + a_i) \rangle \nonumber \\ && \ \
-\case 1/4 (a_i^\dagger + a_i)b_j^\dagger \langle b_j b_j \rangle
-\case 1/4 b_j b_j \langle (a_i^\dagger + a_i)b_j^\dagger \rangle 
+ a_i^\dagger a_i \langle b_j^\dagger b_j \rangle 
+ b_j^\dagger b_j \langle a_i^\dagger a_i \rangle \nonumber \\ && \ \ 
+ a_i^\dagger b_j \langle b_j^\dagger a_i \rangle 
+ a_i^\dagger b_j^\dagger \langle a_i b_j \rangle
+ b_j^\dagger a_i \langle a_i^\dagger b_j \rangle 
+ b_j a_i \langle b_j^\dagger a_i^\dagger \rangle \Biggr]  \ ,
\end{eqnarray}
where $\langle X \rangle$ denotes an average with respect to the
quadratic Hamiltonian.

Since the quadratic Hamiltonian is real and Hermitian we can equate
averages like $\langle a_i^\dagger b_j^\dagger \rangle$ and
$\langle a_i b_j \rangle $.  Also at this order of $(1/S)$ we only
need keep Hermitian contributions to the effective Hamiltonian.
Therefore we write
\begin{eqnarray}
\Delta H_J &=& - \case 1/4 J \sum_{\langle ij \rangle} \Biggl[ 
(a_i b_j^\dagger + a_i^\dagger b_j) \langle (b_j + a_i^\dagger)^2 \rangle
+ 4 (a_i + b_j^\dagger ) (b_j + a_i^\dagger)  \langle b_j^\dagger (b_j + a_i^\dagger) \rangle
\nonumber \\ && \ \
+ (b_j^\dagger+a_i)^2 \langle a_i b_j^\dagger \rangle
+ (b_j+a_i^\dagger)^2 \langle a_i b_j^\dagger \rangle \Biggr] \ .
\end{eqnarray}

Next we consider $\Delta H_\delta$.  Here we can eliminate any terms
which involve local averages (e. g. $\langle a_i^\dagger a_i \rangle$) 
because they multiply a function whose Fourier coefficient
vanishes at zero wave vector.  Thereby we have
\begin{eqnarray}
\Delta H_\delta &=&  \delta J_1 \sum_{\langle ij \rangle}
\sigma_\delta \Biggl[
-\case 1/2 a_i^\dagger a_i \langle a_i^\dagger (b_j^\dagger + b_j) \rangle
-\case 1/4 a_i^\dagger a_i^\dagger \langle a_i (b_j^\dagger + b_j) \rangle 
-\case 1/2 b_j^\dagger b_j \langle b_j (a_i^\dagger + a_i) \rangle \nonumber \\ && \ \
-\case 1/4 b_j b_j \langle (a_i^\dagger + a_i)b_j^\dagger \rangle 
+ a_i^\dagger b_j \langle b_j^\dagger a_i \rangle 
+ a_i^\dagger b_j^\dagger \langle a_i b_j \rangle
+ b_j^\dagger a_i \langle a_i^\dagger b_j \rangle 
+ b_j a_i \langle b_j^\dagger a_i^\dagger \rangle \Biggr]  \ .
\end{eqnarray}
Taking the Hermitian part of this we get
\begin{eqnarray}
\Delta H_\delta &=& \delta J_1 \sum_{\langle ij \rangle} \sigma_\delta
\Biggl[ -\case 1/2 a_i^\dagger a_i \langle a_i^\dagger (b_j^\dagger + b_j) \rangle
-\case 1/8 (a_i^\dagger a_i^\dagger + a_i a_i) \langle a_i (b_j^\dagger + b_j) \rangle 
-\case 1/2 b_j^\dagger b_j \langle b_j (a_i^\dagger + a_i) \rangle \nonumber \\ && \ \
-\case 1/8 (b_j^\dagger b_j^\dagger +  b_j b_j ) \langle (a_i^\dagger + a_i)b_j^\dagger \rangle 
+ (a_i^\dagger b_j + a_i b_j^\dagger ) \langle b_j^\dagger a_i \rangle 
+ ( a_i^\dagger b_j^\dagger + a_i b_j) \langle a_i b_j \rangle \Biggr]  \ .
\end{eqnarray}

Thus we need the averages
\begin{mathletters}
\begin{eqnarray}
X_1 & \equiv & \langle b_j^2 \rangle =
\langle (b_j^\dagger)^2 \rangle =
\langle a_i^2 \rangle =
\langle (a_i^\dagger)^2 \rangle \\
X_2 & \equiv & \langle b_j^\dagger b_j \rangle =
\langle a_i^\dagger a_i \rangle \\
Y_{ij} &\equiv & \langle b_j a_i^\dagger \rangle =
\langle b_j^\dagger a_i \rangle 
\equiv Y_0 + \sigma_\delta Y\\
Z_{ij} &\equiv & \langle b_j^\dagger a_i^\dagger \rangle =
\langle b_j a_i \rangle 
\equiv Z_0 + \sigma_\delta Z \ ,
\end{eqnarray}
\end{mathletters}
where
\begin{mathletters}
\begin{eqnarray}
Y_0 & = & \case 1/4 \sum_j Y_{ij} =
\case 1/4 \sum_j \langle b_j^\dagger a_i \rangle \\
Y & = & \case 1/4 \sum_j \sigma_\delta Y_{ij} =
\case 1/4 \sum_j \sigma_\delta \langle b_j^\dagger a_i \rangle \\
Z_0 & = & \case 1/4 \sum_j Z_{ij} =
\case 1/4 \sum_j \langle b_j a_i \rangle \\ 
Z & = & \case 1/4 \sum_j \sigma_\delta Z_{ij} =
\case 1/4 \sum_j \sigma_\delta \langle b_j a_i \rangle \ , 
\end{eqnarray}
\end{mathletters}
where the sums over $j$ are restricted to sites that are
nearest neighbors of site $i$.
Now drop terms which sum to zero because of $\sigma_\delta$ and also
those (like $\sum_{ij} \sigma_\delta a_i^\dagger b_j$) which do not
contribute at zero wavevector.  Then we get
\begin{eqnarray}
\label{HEFFEQA}
\Delta H_J + \Delta H_\delta &=& - \case 1/4 J
\sum_{ \langle ij \rangle} \Biggl[ (2X_1+4Y_0) (a_ib_j^\dagger + a_i^\dagger b_j)
+ Y_0 [(b_j^\dagger)^2 + b_j^2 + a_i^2 + (a_i^\dagger)^2 ] \nonumber \\
&& \ \ + 4(X_2+Z_0)(a_i^\dagger + b_j)(a_i + b_j^\dagger ) \Biggr] \nonumber \\
&& \ \ + \case 1/8 \delta J_1 \sum_{ij} \Biggl[
(Y + Z)[ -4 a_i^\dagger a_i -4 b_j^\dagger b_j - a_i^2 - (a_i^\dagger)^2
-b_j^2 -(b_j^\dagger)^2 ] \nonumber \\ && \ \
+ 8 Y (a_i^\dagger b_j + a_ib_j^\dagger) + 8 Z (a_i^\dagger b_j^\dagger +a_i b_j) \Biggr] \ .
\end{eqnarray}
The coefficients can be evaluated straightforwardly.  For instance,
if one considers ${\cal H}_J$ as the unperturbed Hamiltonian and
treats ${\cal H}_\delta$ as a perturbation, then one has
\begin{eqnarray}
Y &=& \case 1/4 \sum_j \sigma_\delta \langle b_j^\dagger a_i \rangle \nonumber
\\ &=& \sum_j \sigma_\delta \Biggl[
\langle 0 | b_j^\dagger a_i {1 \over {\cal E}} {\cal H}_\delta | 0 \rangle
+ \langle 0 | {\cal H}_\delta {1 \over {\cal E}} b_j^\dagger a_i | 0 \rangle
\Biggr] \ ,
\end{eqnarray}
where $|0>$ is the spin--wave vacuum and ${\cal E}$ is the unperturbed
energy of the virtual state.  We give the evaluations
\begin{mathletters}
\begin{eqnarray}
X_1 & = & 4C_{2c} (\delta J_1/J)^2  \\
Y_0 & = & 4C_{2d} (\delta J_1/J)^2  \\
Y   & = & -8 C_{2a} (\delta J_1/J)  \\
Z   & = & -8 C_{2b} (\delta J_1/J) \ ,
\end{eqnarray}
\end{mathletters}
where 
\begin{mathletters}
\label{C2EQ}
\begin{eqnarray}
C_{2a} & = & {1 \over 128 N} \sum_{\bf q}
{ [ c_x({\bf q}) - c_y ({\bf q})]^2 \over \epsilon ({\bf q})^3} \\
C_{2b} & = & {1 \over 128 N} \sum_{\bf q}
{ [ c_x({\bf q}) - c_y ({\bf q})]^2 \over \epsilon ({\bf q})^3}
\gamma ({\bf q})^2 \\
C_{2c} & = & {1 \over 128 N} \sum_{\bf q}
{ [ c_x({\bf q}) - c_y ({\bf q})]^2 \over \epsilon ({\bf q})^5}
[ 1  + 2\gamma ({\bf q})^2 ] \\
C_{2d} &=& C_2 - C_{2c} \ ,
\end{eqnarray}
\end{mathletters}
where $\epsilon({\bf q})^2 = 1 - \gamma({\bf q})^2$,
$C_2 = C_{2a} + C_{2b}$.

To summarize: the effect of quantum fluctuations of the in-plane
exchange anisotropy are contained in the effective Hamiltonian
of Eq. (\ref{HEFFEQA}).  Since the result is given in real space, we
can apply it now to the 2342 structure where it gives rise to
contributions to the dynamical matrices written in Eq. (\ref{1INEQ}).
The terms proportional to $X_2+Z_0$ are taken into account by
the spin-wave renormalization incorporated in $Z_c$.

\section{In -- Plane Anisotropic I -- II Interaction}
\label{APPI-II}

We start from Eq. (\ref{V12EQ}), which can be written as
\begin{eqnarray}
V_{12} &=& - \delta J_{12}
\sqrt{S/2} \sum_i e_i^\dagger e_i \Biggl( a_{i+x} + a_{i+x}^\dagger
+ d_{i-x} + d_{i-x}^\dagger - b_{i-y}^\dagger - b_{i-y} - c_{i+y} - c_{i+y}^\dagger \Biggr)
\nonumber \\ &&
+\delta J_{12} \sqrt{S/2} \sum_i f_i^\dagger f_i \Biggl( a_{i-x} + a_{i-x}^\dagger
+ d_{i+x} + d_{i+x}^\dagger - b_{i+y}^\dagger
- b_{i+y} - c_{i-y} - c_{i-y}^\dagger \Biggr) \nonumber \\ &&
- 4\delta J_{12} S \sqrt {S/2} \sum_i 
[ e_i^\dagger + e_i - {e_i^\dagger e_i e_i \over 2S} ]
- 4\delta J_{12} S \sqrt {S/2}
\sum_i [ f_i^\dagger + f_i - {f_i^\dagger f_i^\dagger f_i \over 2S} ]
\nonumber \\ &&
+ \delta J_{12} \sqrt {S/2} \sum_i (e_i^\dagger + e_i ) [ a_{i+x}^\dagger a_{i+x}
+ d_{i-x}^\dagger d_{i-x} + b_{i-y}^\dagger b_{i-y} + c_{i+y}^\dagger c_{i+y} ] 
\nonumber \\ &&
+ \delta J_{12} \sqrt {S/2} \sum_i (f_i^\dagger + f_i ) [ a_{i-x}^\dagger a_{i-x}
+ d_{i+x}^\dagger d_{i+x} + b_{i+y}^\dagger b_{i+y} + c_{i-y}^\dagger c_{i-y} ] \ . 
\end{eqnarray}
Eliminate terms linear in the boson operators by the shifts
\begin{eqnarray}
e_i & \rightarrow & e_i + s \ , \ \ \ \ \ \
f_i \rightarrow f_i + s \ , \nonumber \\
a_i & \rightarrow & a_i + t \ , \ \ \ \ \ \
b_i \rightarrow b_i + t \ , \ \ \ \ \ \
c_i \rightarrow c_i + t \ , \ \ \ \ \ \
d_i \rightarrow d_i + t \ .
\end{eqnarray}
The corresponding Fourier transforms are shifted by a factor $\sqrt {N_{uc}}$.
\begin{eqnarray}
\label{SHIFTEQ}
\sum_i e_i = e(0) \rightarrow e(0) + \sqrt {N_{uc}} s \ , \ \ \ \ \
\sum_i a_i = a(0) \rightarrow a(0) + \sqrt {N_{uc}} t \ .
\end{eqnarray}
In what follows $e$ will denote $e({\bf q}=0)$ and similarly for other
operators.  Then the linear terms in the Hamiltonian are
\begin{eqnarray}
V_1 = - 4 \delta J_{12}
S \sqrt{ N_{uc}S/2} \Biggl( e^\dagger + e + f^\dagger + f \Biggr) \ .
\end{eqnarray}
The quadratic terms in the isotropic part of the Hamiltonian are
\begin{eqnarray}
V_2 & = & (4J+2J_3) S (a^\dagger a + b^\dagger b + c^\dagger c + d^\dagger d )
\nonumber \\ && + 4J_2S(e^\dagger e + f^\dagger f)
+ J_{12}S \Biggl( [a^\dagger + d^\dagger ]f + [b^\dagger + c^\dagger ] e
+ [a + d ]f^\dagger + [b + c ] e^\dagger  \Biggr) \nonumber \\
&+&(2J+2J_3)S (a^\dagger b^\dagger + c^\dagger d^\dagger + ab + cd) + 2JS (a^\dagger c^\dagger + b^\dagger d^\dagger +ac
+bd) + 4J_2S (e^\dagger f^\dagger + ef) \nonumber \\ &+& J_{12}S
\Biggl( [a^\dagger + d^\dagger ]e^\dagger + [b^\dagger + c^\dagger ]f^\dagger + [a+d]e + [b+c]f \Biggr) \ .
\end{eqnarray}
Now
\begin{eqnarray}
{\partial (V_1 + V_2) \over \partial e } &=& 0 =
\sqrt {N_{uc}} S \Biggl( - 4\delta J_{12} \sqrt{S/2}
+ 2J_{12}t + 4J_2s + 2J_{12}t + 4J_2s \Biggr)
\nonumber \\
{\partial (V_1 + V_2) \over \partial a } &=& 0 =
\sqrt {N_{uc}} S \Biggl( (4J+2J_3)t + J_{12}s + 
(2J+2J_3)t + 2Jt + J_{12}s \Biggr) \ .
\end{eqnarray}
For $J_{12}^2 \ll 4J J_2$ we have
\begin{eqnarray}
\label{XXEQ}
s & = & {4\delta J_{12} \sqrt {S/2} \over 8J_2 } \ , \\
\label{YYEQ}
t & = & - {2J_{12}s \over 8J + 4J_3} =
- {J_{12} \delta J_{12} \sqrt {S/2} \over J_2 (8J+4J_3)} \ .
\end{eqnarray}
As discussed in the text, these are the expected results.

Now we record the terms in the Hamiltonian which are cubic in boson operators:
\begin{eqnarray}
&& {{\cal H}^{(3)}} = \delta J_{12} \sqrt {S/2} \Biggl\{
- \sum_{i \in e} e_i^\dagger e_i
\Biggl( a_{i+x}^\dagger + a_{i+x} + d_{i-x}^\dagger
+d_{i-x} - b_{i-y}^\dagger - b_{i-y} - c_{i+y}^\dagger - c_{i+y} \Biggr)
\nonumber \\
&& + \sum_{i \in f} f_i^\dagger f _i
\Biggl( a_{i-x}^\dagger + a_{i-x} + d_{i+x}^\dagger
+d_{i+x} - b_{i+y}^\dagger - b_{i+y} - c_{i-y}^\dagger - c_{i-y} \Biggr)
\nonumber \\ && + \sum_{i \in e} e_i^\dagger e_i e_i
+ \sum_{i \in f} f_i^\dagger f_i^\dagger f_i \nonumber \\ &&
+ \sum_{i \in e} (e_i^\dagger + e_i) \Biggl( a_{i+x}^\dagger a_{i+x} + d_{i-x}^\dagger d_{i-x}
+b_{i-y}^\dagger b_{i-y} + c_{i+y}^\dagger c_{i+y} \Biggr) \nonumber \\ && +
\sum_{i \in f} (f_i^\dagger + f_i) \Biggl( a_{i-x}^\dagger a_{i-x} + d_{i+x}^\dagger d_{i+x}
+b_{i+y}^\dagger b_{i+y} + c_{i-y}^\dagger c_{i-y} \Biggr) \Biggr\} \ .
\end{eqnarray}
Now make the replacements of Eq. (\ref{SHIFTEQ}) to get
\begin{eqnarray}
&& {{\cal H}^{(3)}} = \langle e \rangle \delta J_{12} \sqrt{S/2}  \Biggl\{ -
\sum_{i \in e} (e_i^\dagger + e_i) \Biggl( a_{i+x}^\dagger + a_{i+x} + d_{i-x}^\dagger
+d_{i-x} - b_{i-y}^\dagger - b_{i-y} - c_{i+y}^\dagger - c_{i+y} \Biggr)
\nonumber \\ && + \sum_{i \in f}
(f_i^\dagger + f_i) \Biggl( a_{i-x}^\dagger + a_{i-x} + d_{i+x}^\dagger
+d_{i+x} - b_{i+y}^\dagger - b_{i+y} - c_{i-y}^\dagger - c_{i-y} \Biggr)
\nonumber \\ && + \sum_{i \in e} [ e_i^2 + 2e_i^\dagger e_i ] + \delta J_{12}
\sqrt{2S} \sum_i [(f_i^\dagger)^2 + 2 f_i^\dagger f_i] \nonumber \\ &&
+ \sum_{i \in e} \Biggl( a_{i+x}^\dagger a_{i+x} + d_{i-x}^\dagger d_{i-x}
+b_{i-y}^\dagger b_{i-y} + c_{i+y}^\dagger c_{i+y} \Biggr) \nonumber \\ &&
+ \sum_{i \in f} \Biggl( a_{i-x}^\dagger a_{i-x} + d_{i+x}^\dagger d_{i+x}
+b_{i+y}^\dagger b_{i+y} + c_{i-y}^\dagger c_{i-y} \Biggr)  \Biggr\} \ .
\end{eqnarray}
Here we dropped the terms proportional to $\langle a \rangle$.
They are smaller than those
in $\langle e \rangle$ by $J_{12}/(4J) \approx 1/50$.  Also, as before, to
this order in $1/S$ we may replace the perturbation by its Hermitian
part.  Then the effective quadratic terms above are
\begin{eqnarray}
&& {{\cal H}^{(3)}} = {(\delta J_{12})^2 S \over 4J_2 } \Biggl\{
- \sum_{i \in e} (e_i^\dagger + e_i) \Biggl( a_{i+x}^\dagger + a_{i+x} + d_{i-x}^\dagger
+d_{i-x} - b_{i-y}^\dagger - b_{i-y} - c_{i+y}^\dagger - c_{i+y} \Biggr)
\nonumber \\ &&
+ \sum_{i \in f} (f_i^\dagger + f_i) \Biggl( a_{i-x}^\dagger + a_{i-x} + d_{i+x}^\dagger
+d_{i+x} - b_{i+y}^\dagger - b_{i+y} - c_{i-y}^\dagger - c_{i-y} \Biggr) \nonumber \\
&& + \sum_{i \in e} [ e_i^2 + (e_i^\dagger)^2 + 4e_i^\dagger e_i ]
+ \sum_{i \in f} [(f_i^\dagger)^2 + f_i^2 + 4 f_i^\dagger f_i] \nonumber \\ &&
+ 2 \sum_{i \in e} \Biggl( a_{i+x}^\dagger a_{i+x} + d_{i-x}^\dagger d_{i-x}
+b_{i-y}^\dagger b_{i-y} + c_{i+y}^\dagger c_{i+y} \Biggr) \nonumber \\ &&
+ 2 \sum_{i \in f} \Biggl( a_{i-x}^\dagger a_{i-x} + d_{i+x}^\dagger d_{i+x}
+b_{i+y}^\dagger b_{i+y} + c_{i-y}^\dagger c_{i-y} \Biggr)  \Biggr\} \nonumber \\ && =
{(\delta J_{12})^2 S \over 4J_2 } \sum_{\bf q} \Biggl\{
4 \Biggl[ a^\dagger({\bf q}) a({\bf q}) + b^\dagger({\bf q}) b({\bf q})
+ c^\dagger({\bf q}) c({\bf q}) + d^\dagger({\bf q}) d({\bf q})
+ e^\dagger({\bf q}) e({\bf q}) + f^\dagger({\bf q}) f({\bf q}) \Biggr]
\nonumber \\ &&
+ \Biggl[ - a^\dagger ({\bf q}) + b^\dagger ({\bf q}) + c^\dagger({\bf q})
- d^\dagger({\bf q}) \Biggr]
\Biggl[ e({\bf q}) - f({\bf q}) + e^\dagger(-{\bf q}) - f^\dagger(-{\bf q})
\Biggr] + e^\dagger({\bf q}) e^\dagger(-{\bf q})
\nonumber \\ &&
+ \Biggl[ - a ({\bf q}) + b ({\bf q}) + c({\bf q}) - d({\bf q}) \Biggr]
\Biggl[ e^\dagger({\bf q}) - f^\dagger({\bf q}) + e(-{\bf q})
- f(-{\bf q}) \Biggr] + f^\dagger({\bf q}) f^\dagger(-{\bf q}) \Biggr\}  \ ,
\end{eqnarray}
which leads to Eq. (\ref{DAIN1}).

Now we look at the fourth order terms in the CuII -- CuII
isotropic exchange interaction.  These are
\begin{eqnarray}
V_{DM} = - \case 1/2 J_2 \sum_{i \in e , \delta} \Biggl(
e_i^\dagger f_{i+\delta}^\dagger f_{i+\delta}^\dagger f_{i+\delta}
+ f_{i+\delta} e_i^\dagger e_i e_i
+ 2 e_i^\dagger e_i f_{i+\delta}^\dagger f_{i+\delta} \Biggr) \ .
\end{eqnarray}
Substituting in two shifts of $\langle e \rangle$, this is
\begin{eqnarray}
V_{DM} & = & - \case 1/2 \langle e \rangle^2 J_2 \sum_{i, \delta} \Biggl[
e_i^\dagger f_{i+\delta} + 2 f_{i+\delta}^\dagger f_{i+\delta} + 2 e_i^\dagger f_{i+\delta}^\dagger
+ (f_{i+\delta}^\dagger )^2 + 2 f_{i+\delta} e_i \nonumber \\ && +
f_{i+\delta} e_i^\dagger + 2 e_i^\dagger e_i + e_i^2 + 2 e_i^\dagger e_i + 
2f_{i+\delta}^\dagger f_{i+\delta} + 2(e_i^\dagger + e_i)(f_{i+\delta}^\dagger
+ f_{i+\delta} )
\Biggr] \ .
\end{eqnarray}
Taking the Hermitian part of this, we get
\begin{eqnarray}
V_{DM}&=& - {(\delta J_{12})^2S \over 16J_2 }
\sum_{i , \delta} \Biggl[ e_i^\dagger f_{i+\delta}
+ e_i f_{i+\delta}^\dagger + 2 e_i^\dagger f_{i+\delta}^\dagger + 2 e_i f_{i+\delta}
+ 4 f_{i+\delta}^\dagger f_{i+\delta} \nonumber \\ && + 4 e_i^\dagger e_i
+ \case 1/2 e_i^2 + \case 1/2 (e_i^\dagger)^2 + \case 1/2 f_{i+\delta}^2 +
\case 1/2 (f_{i+\delta}^\dagger )^2 + 2 (e_i^\dagger + e_i) (f_{i+\delta}^\dagger + f_{i+\delta}
) \Biggr] \nonumber \\ &=& - {(\delta J_{12})^2 S \over 16J_2}
\sum_{\bf q} \Biggl[
\Biggl( 6 e({\bf q}) f^\dagger({\bf q}) + 6 e^\dagger({\bf q}) f({\bf q})
+ 8 e({\bf q}) f(-{\bf q}) + 8 e^\dagger({\bf q}) f^\dagger(-{\bf q}) \Biggr) (c_x+c_y)
\nonumber \\ && + 16 f^\dagger ({\bf q}) f({\bf q}) + 16 e^\dagger ({\bf q}) e({\bf q}) +
2e^\dagger({\bf q}) e^\dagger(-{\bf q}) + 2e({\bf q}) e(-{\bf q})
\nonumber \\ && +
2f^\dagger({\bf q}) f^\dagger(-{\bf q}) + 2f({\bf q}) f(-{\bf q}) \Biggr] \ ,
\end{eqnarray}
which leads to Eq. (\ref{DAIN2}).

Contributions from quartic terms in the CuI -- CuII interaction are
smaller, i. e. of order, $(\delta J_{12})^2 J_{12}^2 /(JJ_2^2)$,
if we take out one
factor of $\langle e \rangle$ and one factor of $\langle a \rangle$.
Taking out two $\langle a \rangle$ factors gives an even smaller result.
Taking out two $\langle a \rangle$ shifts from the CuI -- CuI
anharmonic term gives a contribution of order
$J_{12}^2 \delta J_{12}^2/(JJ_2^2)$.  All these terms are neglected.

\section{In -- Plane Anisotropic II -- II Interaction}
\label{APPII-II}

\subsection{Self-Energy Due to Cubic Perturbations}

We start by discussing how one constructs the self--energy due
to cubic perturbations.  The point is that we wish to avoid the
complexities involving Matsubara sums etc.  Let us suppose that
we have an unperturbed Hamiltonian in terms of normal mode
operators, $E({\bf q})$ and $F({\bf q})$:
\begin{equation}
{\cal H} = \sum_{\bf q}  \omega({\bf q}) \Biggl( E^\dagger ({\bf q}) E({\bf q})
+ F^\dagger({\bf q}) F({\bf q}) \Biggr) \ .
\end{equation}
Now we want to identify the perturbative contributions to the matrices
${\bf A}({\bf q})$ and ${\bf B}({\bf q})$.  Suppose we wish to calculate
perturbative contributions leading to an effective quadratic Hamiltonian
of the form
\begin{equation}
\case 1/2 B({\bf q}) E^\dagger({\bf q}) E^\dagger(-{\bf q}) \ .
\end{equation}
For this purpose we make the identification
\begin{eqnarray}
\delta B({\bf q}) = \langle 0| E({\bf q}) E(-{\bf q}) V
{1 \over {\cal E}} V | 0 \rangle \ .
\end{eqnarray}
Thus for $\omega ({\bf q}) \rightarrow 0$ and considering only the ground
state, we may write
\begin{eqnarray}
\delta B({\bf q}) & = & \langle 0| \partial V / \partial E^\dagger({\bf q})
{1 \over {\cal E}} \partial V / \partial E^\dagger (-{\bf q}) | 0 \rangle \nonumber
\\ && + \langle 0| \partial V / \partial E^\dagger(-{\bf q})
{1 \over {\cal E}} \partial V / \partial E^\dagger ({\bf q}) | 0 \rangle \ .
\end{eqnarray}
Similarly for the term in the Hamiltonian
\begin{equation}
A({\bf q}) E^\dagger({\bf q}) F({\bf q})
\end{equation}
we make the identification
\begin{eqnarray}
\delta A({\bf q}) = \langle 0| E({\bf q}) V
{1 \over {\cal E}} V F^\dagger({\bf q}) | 0 \rangle \ .
\end{eqnarray}
Thus for $\omega ({\bf q}) \rightarrow 0$ and considering only the ground
state, we may write
\begin{eqnarray}
\delta A({\bf q}) & = & \langle 0| \partial V / \partial E^\dagger({\bf q})
{1 \over {\cal E}} \partial V / \partial F ({\bf q}) | 0 \rangle \nonumber
\\ && + \langle 0| \partial V / \partial F({\bf q})
{1 \over {\cal E}} \partial V / \partial E^\dagger ({\bf q}) | 0 \rangle \ .
\end{eqnarray}
This type of relation holds generally under the two assumptions:
a) we consider the perturbation to modes whose energy can be neglected
in the energy denominators, and b) we consider only the ground and
low lying excited states, so that boson occupation numbers are zero.

We have made the identification in terms of the normal mode operators,
but equally we may transform to any set of modes.

\subsection{Application to CuII - CuII In-Plane Interactions}

We start from Eq. (\ref{INVEQ}) and
implement the results of the preceding subsection.  For small
${\bf k}$ we write
\begin{eqnarray}
T_1 & \equiv & \partial V/\partial e({\bf k}) = \delta J_ 2\sqrt{2S/N_{\rm uc}}
\sum_{\bf q} [f({\bf q}) + f^\dagger(-{\bf q})] e^\dagger ({\bf q}) (c_x-c_y) \nonumber \\
T_2 & \equiv & \partial V/\partial e^\dagger({\bf k}) = \delta J_2
\sqrt{2S/N_{\rm uc}}
\sum_{\bf q} [f({\bf q}) + f^\dagger(-{\bf q})] e (-{\bf q}) (c_x-c_y) \nonumber \\
T_3 & \equiv & \partial V/\partial f({\bf k}) = - \delta J_2
\sqrt{2S/N_{\rm uc}}
\sum_{\bf q} [e({\bf q}) + e^\dagger(-{\bf q})] f^\dagger ({\bf q}) (c_x-c_y) \nonumber \\
T_4 & \equiv & \partial V/\partial f^\dagger({\bf k}) = - \delta J_2
\sqrt{2S/N_{\rm uc}}
\sum_{\bf q} [e({\bf q}) + e^\dagger(-{\bf q})] f (-{\bf q}) (c_x-c_y) \ ,
\end{eqnarray}
where $c_x= \cos (aq_x)$ and $c_y=\cos (aq_y)$. Thus if
$\bar {\bf p}$ denotes $-{\bf p}$, then
\begin{eqnarray}
\langle & T_1 &{1 \over {\cal E}} T_1 \rangle =
{2S (\delta J_2)^2 \over N_{\rm uc} } \sum_{{\bf q},{\bf p}}
\left\langle [f({\bf q}) + f^\dagger(\bar {\bf q})] e^\dagger ({\bf q}) (c_x-c_y)
{1 \over {\cal E}} [f({\bf p}) + f^\dagger(\bar {\bf p})]
e^\dagger ({\bf p}) (c_x-c_y) \right\rangle \nonumber
\\ &=& {2S (\delta J_2)^2 \over N_{\rm uc} } \sum_{{\bf q},{\bf p}}
\left\langle [l_{\bf q} F_{\bf q} - m_{\bf q} E_{\bf q}]
(-m_{\bf q} F_{\bar {\bf q}}) (c_x-c_y) {1 \over {\cal E}} 
[-m_{\bf p} E^\dagger_{\bar {\bf p}}+l_{\bf p} F^\dagger_{\bar {\bf p}}]
l_{\bf p} E^\dagger_{\bf p} (c_x-c_y) \right\rangle \nonumber \\ &=&
- {2S (\delta J_2)^2 \over N_{\rm uc}} \sum_{\bf q} {(c_x-c_y)^2 l^2_{\bf q}
m^2_{\bf q} \over 8J_2S \epsilon ({\bf q}) } = 
- {(\delta J_2)^2 \over 4J_2 N_{\rm uc}} \sum_{\bf q} \Biggl(
{(c_x-c_y)^2 \over \epsilon ({\bf q})^3} \Biggr)
{\gamma ({\bf q})^2 \over 4} \ ,
\end{eqnarray}
where we introduced the normal mode operators
\begin{eqnarray}
E_{\bf q} &=& l_{\bf q} e({\bf q}) - m_{\bf q} f^\dagger ({\bf q}) \ ,
\nonumber \\
F_{\bf q} &=& l_{\bf q} f({\bf q}) - m_{\bf q} e^\dagger ({\bf q}) \ ,
\end{eqnarray}
where
\begin{eqnarray}
\label{LMEQ}
l_{\bf q}^2 = {1 + \epsilon({\bf q}) \over 2 \epsilon({\bf q}) } \ , \ \ \ \ \
m_{\bf q}^2 = {1 - \epsilon({\bf q}) \over 2 \epsilon({\bf q}) } \ , \ \ \ \ \
l_{\bf q} m_{\bf q} = - {\gamma({\bf q}) \over 2 \epsilon({\bf q}) } \ ,
\end{eqnarray}
where $\gamma ({\bf q}) = \case 1/2[\cos(aq_x) + \cos (aq_y)]$.  Similarly
\begin{eqnarray}
\langle & T_1 &{1 \over {\cal E}} T_2 \rangle =
\langle T_2 {1 \over {\cal E}} T_1 \rangle =
- {(\delta J_2)^2 \over 4J_2 N_{\rm uc}} \sum_{\bf q} \Biggl(
{(c_x-c_y)^2 \over \epsilon ({\bf q})^3} \Biggr) \Biggr(
{1 - \epsilon ({\bf q}) \over 2} + {\gamma ({\bf q})^2 \over 4} \Biggr)
\end{eqnarray}
\begin{eqnarray}
\langle & T_1 &{1 \over {\cal E}} T_3 \rangle =
\langle  T_3 {1 \over {\cal E}} T_1 \rangle =
{(\delta J_2)^2 \over 4J_2 N_{\rm uc}} \sum_{\bf q} \Biggl(
{(c_x-c_y)^2 \over \epsilon ({\bf q})^3} \Biggr)
{3\gamma ({\bf q})^2 \over 4} 
\end{eqnarray}
\begin{eqnarray}
\langle & T_1 &{1 \over {\cal E}} T_4 \rangle =
\langle T_4 {1 \over {\cal E}} T_1 \rangle =
{ (\delta J_2)^2 \over 4J_2 N_{\rm uc}}
\sum_{\bf q} \Biggl( {(c_x-c_y)^2 \over \epsilon ({\bf q})^3} \Biggr)
{[1-\epsilon ({\bf q})]^2 \over 4} \ .
\end{eqnarray}

Now we have the contribution to the coefficient of $e^\dagger e$,
which we denote $\delta a_{55}$, as
\begin{eqnarray}
\delta a_{55} & = & \langle T_1 {1 \over {\cal E}} T_2 \rangle
+ \langle T_2 {1 \over {\cal E}} T_1 \rangle =
\left[\delta J_2^2 /J_2 \right]  \left[ - 32 C_{2a} - 16 C_{2b} \right] \ ,
\end{eqnarray}
where $C_{2a}$ and $C_{2b}$ were defined in Eq. (\ref{C2EQ}).

Likewise the contribution to the
coefficient of $e^\dagger f$ which we denote $\delta a_{56}$ is
\begin{eqnarray}
\delta a_{56} & = & \langle T_1 {1 \over {\cal E}} T_4 \rangle
+ \langle T_4 {1 \over {\cal E}} T_1 \rangle =
\left[(\delta J_2 )^2 /J_2 \right]  \left[ 32 C_{2a} - 16 C_{2b} \right] \ .
\end{eqnarray}
Similarly, $\delta b_{5}$ is the contribution to the coefficient of
$\case 1/2 e^\dagger e^\dagger$, so that
\begin{eqnarray}
\delta b_{55} & = & 2 \langle T_1 {1 \over {\cal E}} T_1 \rangle =
\left[(\delta J_2)^2 /J_2 \right]  \left[ - 16 C_{2b} \right]
\end{eqnarray}
and likewise
\begin{eqnarray}
\delta b_{56} & = & \langle T_1 {1 \over {\cal E}} T_3 \rangle
+ \langle T_3 {1 \over {\cal E}} T_1 \rangle =
\left[(\delta J_2)^2 /J_2 \right]  \left[ 48 C_{2b} \right] \ .
\end{eqnarray}

\section{Interplanar Anisotropic CuII - CuII Interaction}
\label{KII-II}

\subsection{Pseudodipolar Interactions}

In order to facilitate the evaluation of the lattice sums
we parametrize the anisotropic exchange interactions
bewteen the $i$th CuII spin in one plane and the nearest
neighboring $j$th CuII spin in an adjacent layer.  We
introduce the indicator variable $\sigma_i$ which is unity
if $i$ is on the $e$-sublattice and is $-1$ if $i$ is on the
$f$-sublattice.  We also introduce a variable $\mu_i$ to distinguish
between the two nearest neighboring sites with the same value
of $\sigma_i$.  Then for the interaction between nearest
neighboring CuII spins $i$ and $j$ in adjacent CuO layers we
use Fig. 4 to write the principal axes as
\begin{mathletters}
\begin{eqnarray}
\hat n_1^{(ij)} &=& \left[ \case 1/2 (1 + \sigma_i \sigma_j) \hat \eta
- \case 1/2 (1 - \sigma_i \sigma_j) \hat \xi \right] \mu_i \mu_j \\
\hat n_2^{(ij)} &=& \left[ \case 1/2 (1 + \sigma_i \sigma_j)
\hat \xi \cos \psi
+ \case 1/2 (1 - \sigma_i \sigma_j) \hat \eta \cos \psi \right] \mu_i \mu_j
+ \hat z \sin \psi \\
\hat n_3^{(ij)} &=& \left[ \case 1/2 (1 + \sigma_i \sigma_j)
\hat \xi \sin \psi
+ \case 1/2 (1 - \sigma_i \sigma_j) \hat \eta \sin \psi \right] \mu_i \mu_j
- \hat z \cos \psi \ .
\end{eqnarray}
\end{mathletters}
We also write
\begin{eqnarray}
\label{SEQ}
{\bf S}_i = - \sigma_i (S - a_i^\dagger a_i) \hat \xi
+ \sqrt{S/2} (a_i^\dagger + a_i) \hat \eta + i \sqrt{S/2}
(a_i - a_i^\dagger) \hat z \sigma_i \ ,
\end{eqnarray}
where, in this appendix, $a_i$ is the boson operator for spin $i$.
Then we have
\begin{mathletters}
\begin{eqnarray}
{\bf S}_i \cdot \hat n_1^{(ij)} &=&
\case 1/2 (\sigma_i - \sigma_j) \mu_i \mu_j (S- a_i^\dagger a_i)
+ \case 1/2 (1 + \sigma_i \sigma_j) \mu_i \mu_j
\sqrt{S/2} (a_i^\dagger + a_i) \\
{\bf S}_i \cdot \hat n_2^{(ij)} &=& - \case 1/2 (S-a_i^\dagger a_i) c
(\sigma_i + \sigma_j)\mu_i \mu_j
\nonumber \\ && \ \  + \case 1/2 (1- \sigma_i \sigma_j)
\mu_i \mu_j c \sqrt{S/2} (a_i^\dagger + a_i)
+ i s \sigma_i \sqrt{S/2} (a_i - a_i^\dagger) \\
{\bf S}_i \cdot \hat n_3^{(ij)} &=& - \case 1/2 (S-a_i^\dagger a_i)
s (\sigma_i + \sigma_j)\mu_i \mu_j 
\nonumber \\ && \ \ + \case 1/2 (1- \sigma_i \sigma_j)
\mu_i \mu_j s \sqrt{S/2} (a_i^\dagger + a_i)
- i c \sigma_i \sqrt{S/2} (a_i - a_i^\dagger) \ ,
\end{eqnarray}
\end{mathletters}
where $c \equiv \cos \psi$ and $s \equiv \sin \psi$.
Then we have
\begin{eqnarray}
\label{F4EQ}
{\cal H}_{ij} & = & S \sum_{m=1}^3 K_m [ {\bf S}_i \cdot \hat n_m]
[{\bf S}_j \cdot \hat n_m] \equiv S \sum_{m=1}^3 K_m T_m \ ,
\end{eqnarray}
where, at quadratic order,
\begin{eqnarray}
T_1 &=& \case 1/2 (1- \sigma_i \sigma_j)
( a_i^\dagger a_i + a_j^\dagger a_j ) + \case 1/4
(1 + \sigma_i \sigma_j) (a_i^\dagger + a_i) (a_j^\dagger + a_j) \nonumber \\
T_2 &=& - \case 1/2 (1+ \sigma_i \sigma_j)c^2 (a_i^\dagger a_i
+ a_j^\dagger a_j) +
\case 1/2 \Biggl(\case 1/2  (1- \sigma_i \sigma_j) (a_i^\dagger + a_i)
c \mu_i \mu_j + i \sigma_i s (a_i - a_i^\dagger) \Biggr) \nonumber \\
&& \ \ \times \Biggl(\case 1/2  (1- \sigma_i \sigma_j) (a_j^\dagger + a_j)
c \mu_i \mu_j  + i \sigma_j s (a_j - a_j^\dagger) \Biggr)
\end{eqnarray}
and $T_3$ is obtained from $T_2$ by replacing $\sin \psi$ by
$- \cos \psi$ and $\cos \psi$ by $\sin \psi$.
Thereby we get the site-diagonal contribution to the Hamiltonian as
\begin{eqnarray}
\label{DA1EQ}
\delta {\cal H} &=& 4 \Delta K S \sum_i a_i^\dagger a_i \ ,
\end{eqnarray}
where $\Delta K$ was defined in Eq. (\ref{DKEQ}).

The remaining contributions to the Hamiltonian are
found from Eq. (\ref{F4EQ}) to be
\begin{eqnarray}
\delta {\cal H} &=& \case 1/2 \sum_{ i \in II , j}
\Biggl\{ \case 1/4 K_1 S (1 + \sigma_i \sigma_j) 
(a_i^\dagger + a_i ) (a_j^\dagger + a_j ) \nonumber \\ && \
+ \case 1/2 K_2 S \Biggl[ \case 1/2 c^2 (1 - \sigma_i \sigma_j)
(a_i^\dagger + a_i) (a_j^\dagger + a_j)
- \sigma_i \sigma_j s^2 (a_i - a_i^\dagger) (a_j - a_j^\dagger)
\nonumber \\ && \ + ics \mu_i \mu_j (\sigma_j - \sigma_i) (a_i^+ + a_i)
(a_j- a_j^\dagger) \Biggl]  \ + \dots \Biggr\} \ ,
\end{eqnarray}
where $\dots$ indicates further terms in $K_3$ obtained from
those of $K_2$ by replacing $\cos \psi$ by $\sin \psi$ 
and $\sin \psi$ by $-\cos \psi$ and $j$ is summed over CuII
nearest neighbors in adjacent planes.
For $q_x=q_y=0$ the imaginary term
gives zero contribution to the dynamical matrices.
Then, the terms with $\sigma_i=\sigma_j$ give a contribution to the
Hamiltonian of
\begin{eqnarray}
\label{DA2EQ}
\delta  {\cal H} &=& {S \over 4} \sum_i \sum_{j:\sigma_j=\sigma_i}
\Biggl[ K_1 (a_i^\dagger + a_i) ( a_j^\dagger + a_j)
- K_2 s^2 (a_i -a_i^\dagger)(a_j-a_j^\dagger) \nonumber \\ && \ \
- K_3 c^2 (a_i -a_i^\dagger)(a_j-a_j^\dagger) \Biggr] \ .
\end{eqnarray}
The terms with $\sigma_i=-\sigma_j$ give a contribution to the
Hamiltonian of
\begin{eqnarray}
\label{DA3EQ}
\delta  {\cal H} &=& {S \over 4} \sum_i \sum_{j:\sigma_j=-\sigma_i}
\Biggl[
K_2 c^2 (a_i + a_i^\dagger)(a_j + a_j^\dagger) 
+ K_2 s^2 (a_i - a_i^\dagger)(a_j - a_j^\dagger) \nonumber \\ && \ \
+ K_3 s^2 (a_i + a_i^\dagger)(a_j + a_j^\dagger) 
+ K_3 c^2 (a_i - a_i^\dagger)(a_j - a_j^\dagger) \Biggr] \ . 
\end{eqnarray}
The term in Eq. (\ref{DA1EQ}) and the number conserving terms
in Eq. (\ref{DA2EQ}) reproduce Eq. (\ref{A55EQ}) and the other
terms in Eq. (\ref{DA2EQ}) reproduce Eq. (\ref{B55EQ}).
Equation (\ref{DA3EQ}) reproduces Eqs. (\ref{A56EQ}) and (\ref{B56EQ}).

\subsection{Dipolar Interactions}

For the dipolar interactions it is convenient to construct
the Hamiltonian explicitly rather than to identify it with the
pseudodipolar interaction.
We substitute Eq. (\ref{SEQ}) into the dipolar interaction to get
\begin{eqnarray}
{\cal H}_{ij} &=& g^2 \mu_B^2 r_{ij}^{-3} \left[ {\bf S}_i \cdot {\bf S}_j
- 3 ({\bf S}_i \cdot \hat {\bf r}_{ij})
({\bf S}_j \cdot \hat {\bf r}_{ij}) \right]
\rightarrow -3 g^2 \mu_B^2 r_{ij}^{-3}
({\bf S}_i \cdot \hat {\bf r}_{ij}) ({\bf S}_j \cdot \hat {\bf r}_{ij})
\nonumber \\ &=& - {3 g^2 \mu_B^2 \over r_{ij}^3} \Biggl[
- \sigma_i [S-a_i^\dagger a_i] (\hat \xi \cdot \hat {\bf r}_{ij})
+ \sqrt{S/2} (a_i + a_i^\dagger) (\hat \eta \cdot \hat {\bf r}_{ij} )
+ i \sigma_i \sqrt{S/2} (a_i - a_i^\dagger)(\hat z \cdot {\bf r}_{ij} )
\Biggr] \nonumber \\ && \ \ \times \Biggl[
- \sigma_j [S-a_j^\dagger a_j] (\hat \xi \cdot \hat {\bf r}_{ij})
+ \sqrt{S/2} (a_j + a_j^\dagger) (\hat \eta \cdot \hat {\bf r}_{ij} )
+ i \sigma_j \sqrt{S/2} (a_j - a_j^\dagger)(\hat z \cdot {\bf r}_{ij} )
\Biggr] \ .
\end{eqnarray}
Here we dropped the term in ${\bf S}_i \cdot {\bf S}_j$ which may be
included in the isotropic Heisenberg Hamiltonian.
At quadratic order this gives
\begin{eqnarray}
{\cal H} &=& \case 1/2 \sum_{i,j \in II} {\cal H}_{ij}
= \sum_{i,j \in II} {3 g^2 \mu_B^2 S \over 2r_{ij}^3 }
\Biggl[ \sigma_i \sigma_j ( a_j^\dagger a_j + a_i^\dagger a_i)
(\hat \xi \cdot \hat {\bf r}_{ij} )^2 - \case 1/2 (a_i+a_i^\dagger)
(a_j + a_j^\dagger) (\hat \eta \cdot {\bf r}_{ij})^2 \nonumber \\&&
+ \case 1/2 \sigma_i \sigma_j (a_i- a_i^\dagger)(a_j - a_j^\dagger)
(\hat z \cdot {\bf r}_{ij})^2
- i \sigma_j (a_i+a_i^\dagger ) (a_j-a_j^\dagger)
(\hat z \cdot \hat {\bf r}_{ij})
(\hat \eta \cdot \hat {\bf r}_{ij}) \Biggr] \ .
\end{eqnarray}
We now consider what contributions this gives to the dynamical matrix for
$q_x=q_y=0$.  Then the imaginary term can be dropped.
For simplicity we truncate the sums to include only
interactions between adjacent planes.  Then we have
\begin{mathletters}
\begin{eqnarray}
\delta a_{55} &=& \sum_{j \in e} {3 g^2 \mu_B^2 S \over r_{ij}^3} \Biggl[
(\hat \xi \cdot \hat r_{ij})^2
- \case 1/2 c_z ( \hat \eta \cdot \hat r_{ij})^2
- \case 1/2 c_z ( \hat z \cdot \hat r_{ij})^2 \Biggr]
+ \sum_{j \in f} {3 g^2 \mu_B^2 S \over r_{ij}^3}
\sigma_j (\hat \xi \cdot \hat r_{ij})^2 \\
\delta a_{56} &=& \sum_{j \in f} {3 g^2 \mu_B^2 S \over r_{ij}^3 }
\Biggl[ - \case 1/2 (\hat \eta \cdot \hat r_{ij})^2
+ \case 1/2 (\hat z \cdot \hat r_{ij})^2 \Biggr] c_z \\
\delta b_{55} &=& \sum_{j \in e} {3 g^2 \mu_B^2 S \over r_{ij}^3 }
\Biggl[ - \case 1/2 (\hat \eta \cdot \hat r_{ij})^2
+ \case 1/2 (\hat z \cdot \hat r_{ij})^2 \Biggr] c_z \\
\delta b_{56} &=& \sum_{j \in f} {3 g^2 \mu_B^2 S \over r_{ij}^3 }
\Biggl[ -\case 1/2 (\hat \eta \cdot \hat r_{ij})^2
- \case 1/2 (\hat z \cdot \hat r_{ij})^2 \Biggr] c_z \ ,
\end{eqnarray}
\end{mathletters}
where $c_z=\cos(q_zc/2)$, $i$ is a fixed site in the {\it e} sublattice,
and the sum over $j$ is restricted to the planes adjacent to site $i$.

This interaction is negligibly small except with respect to
the lowest in-plane mode.  So we only need the combination
\begin{eqnarray}
\delta ( a_{55} + b_{55} - a_{56} - b_{56}) &=&
\sum_{j \in II: z_{ij}=\pm c/2} {3 g^2 \mu_B^2 S \over r_{ij}^3 }
\Biggl[  \sigma_j (\hat \xi \cdot \hat r_{ij})^2
- \sigma_j (\hat \eta \cdot \hat r_{ij})^2 c_z \Biggr]
\nonumber \\
&=& \sum_{j \in II: z_{ij}=c/2} {3 g^2 \mu_B^2 S \over r_{ij}^5}
\Biggl[ \sigma_j (x_{ij} + y_{ij})^2
-  \sigma_j c_z (x_{ij} - y_{ij})^2 \Biggr] \ .
\end{eqnarray}
Note that the sum over sites $j$ in an adjacent plane from site $i$ vanishes:
\begin{eqnarray}
\sum_{j \in II: z_{ij}=c/2} {\sigma_j x_{ij}^2 \over r_{ij}^5 }
= \sum_{j \in II: z_{ij}=c/2} {\sigma_j y_{ij}^2 \over r_{ij}^5 } = 0 \ .
\end{eqnarray}
Thus
\begin{eqnarray}
\delta ( a_{55} + b_{55} - a_{56} - b_{56}) &=& 6 (1 + c_z)
g^2 \mu_B^2 S \sum_{j \in II: z_{ij} = c/2} {\sigma_j x_{ij} y_{ij} \over
r_{ij}^5} \ .
\end{eqnarray}

\section{Interplanar Anisotropic CuI - CuII Interaction}
\label{KI-II}

For the CuI sites we introduce further indicator variables
$\tau$ (which tells the direction of the moment) and $\rho$
(which discriminates between sublattices) such that
$\tau=\rho=1$ for an {\it a} site,
$-\tau=\rho=1$ for a {\it b} site,
$\tau=\rho=-1$ for a {\it c} site,
and $\tau=-\rho=1$ for a {\it d} site.
Then, from Fig. 5, we have the principal axes for the sites $i$
and $j$ where $i$ ($j$) is in the CuI (II) sublattice as
\begin{mathletters}
\begin{eqnarray}
\hat m_1^{(ij)} &=& - {\rho_i \sigma_j  \over \sqrt 2} \Biggl[
\hat \xi + \tau_i \hat \eta \Biggr] \\
\hat m_2^{(ij)} &=& \hat z \cos \phi 
+ { \rho_i \sigma_j \sin \phi \over \sqrt 2} \Biggl[
\hat \eta - \tau_i \hat \xi \Biggr] \\
\hat m_3^{(ij)} &=& - \hat z \sin \phi
+ {\rho_i \sigma_j \cos \phi \over \sqrt 2} \Biggl[
\hat \eta - \tau_i \hat \xi \Biggr] \ .
\end{eqnarray}
\end{mathletters}
In checking the above it is useful
to note that changing the sign of either $\rho_i$ or $\sigma_j$
induces a $180^{\rm o}$ rotation about the $z$-axis.

Also we use Eq. (\ref{SEQ}) for the CuII spins and
\begin{eqnarray}
{\bf S}_i = \tau_i (S - a_i^\dagger a_i) \hat \xi
+ \sqrt {S/2} (a_i + a_i^\dagger) \hat \eta + i \tau_i
\sqrt{S/2} (a_i^\dagger - a_i ) \hat z
\end{eqnarray}
for the CuI spins.  Thus if $i$ labels a CuI spin we have
\begin{mathletters}
\begin{eqnarray}
\hat m_1^{(ij)} \cdot {\bf S}_i &=& {\rho_i \tau_i \sigma_j \over \sqrt 2}
\Biggl[ - (S-a_i^\dagger a_i) - \sqrt {S/2} (a_i + a_i^\dagger) \Biggr] \\
\hat m_2^{(ij)} \cdot {\bf S}_i &=& {1 \over \sqrt 2} \Biggl[
- \rho_i \sigma_j s (S- a_i^\dagger a_i) + \rho_i \sigma_j s \sqrt {S/2}
(a_i + a_i^\dagger) + i \tau_i c \sqrt S (a_i^\dagger - a_i) \Biggr] \\
\hat m_3^{(ij)} \cdot {\bf S}_i &=& {1 \over \sqrt 2} \Biggl[
- \rho_i \sigma_j c (S- a_i^\dagger a_i) + \rho_i \sigma_j c \sqrt {S/2}
(a_i + a_i^\dagger) - i \tau_i s \sqrt S (a_i^\dagger - a_i) \Biggr]
\end{eqnarray}
\end{mathletters}
and if $j$ labels a CuII spin we have
\begin{mathletters}
\begin{eqnarray}
\hat m_1^{(ij)} \cdot {\bf S}_j &=& {\rho_i \over \sqrt 2} 
\Biggl[ (S-a_j^\dagger a_j)
- \tau_i \sigma_j \sqrt {S/2} (a_j + a_j^\dagger) \Biggr] \\
\hat m_2^{(ij)} \cdot {\bf S}_j &=& {1 \over \sqrt 2} \Biggl[
\rho_i \tau_i s (S- a_j^\dagger a_j) + \rho_i \sigma_j s \sqrt {S/2}
(a_j + a_j^\dagger) + i \sigma_j c \sqrt S (a_j - a_j^\dagger) \Biggr] \\
\hat m_3^{(ij)} \cdot {\bf S}_j &=& {1 \over \sqrt 2} \Biggl[
\rho_i \tau_i c (S- a_j^\dagger a_j) + \rho_i \sigma_j c \sqrt {S/2}
(a_j + a_j^\dagger) - i \sigma_j s \sqrt S (a_j - a_j^\dagger) \Biggr] \ ,
\end{eqnarray}
\end{mathletters}
where $c \equiv \cos \phi$ and $s \equiv \sin \phi$.  We now write
\begin{eqnarray}
{\cal H}_{ij} = \sum_{m=1}^3 [\hat m_m \cdot {\bf S}_i]
\sum_{m=1}^3 [\hat m_m \cdot {\bf S}_j] \equiv
S\sum_{m=1}^3 K_m' T_m \ ,
\end{eqnarray}
and at quadratic order we have
\begin{mathletters}
\begin{eqnarray}
T_1 &=& \case 1/2 \tau_i \sigma_j [a_i^\dagger a_i + a_j^\dagger a_j]
+ \case 1/4 [ a_i + a_i^\dagger ][ a_j + a_j^\dagger  ] \\
T_2  &=& \case 1/2 \tau_i \sigma_j s^2 [a_i^\dagger a_i + a_j^\dagger a_j]
\nonumber \\ && \ \ +
\Biggl[ \case 1/2 \rho_i \sigma_j s (a_i + a_i^\dagger)
+ { i \tau_i c \over \sqrt 2} (a_i^\dagger -a_i) \Biggr]
\Biggl[ \case 1/2 \rho_i \sigma_j s (a_j + a_j^\dagger)
+ { i \sigma_j c \over \sqrt 2} (a_j -a_j^\dagger ) \Biggr] \\
T_3  &=& \case 1/2 \tau_i \sigma_j c^2 [a_i^\dagger a_i + a_j^\dagger a_j]
\Biggl[ \case 1/2 \rho_i \sigma_j c (a_i + a_i^\dagger)
- { i \tau_i s \over \sqrt 2} (a_i^\dagger -a_i) \Biggr]
\nonumber \\ && \ \ +
\Biggl[ \case 1/2 \rho_i \sigma_j c (a_j + a_j^\dagger)
- { i \sigma_j s \over \sqrt 2} (a_j -a_j^\dagger ) \Biggr] \ .
\end{eqnarray}
\end{mathletters}
We drop terms which do not contribute to the dynamical matrix for
$q_x=q_y=0$ and thereby find that
\begin{eqnarray}
{\cal H} &=& \sum_{i \in II, j \in II} {\cal H}_{ij}
= S \sum_{i \in I, j \in II} \Biggl\{ \case 1/4
(a_i^\dagger + a_i) ( a_j^\dagger + a_j) \left( K_1' + K_2' s^2 
+ K_3' c^2 \right) \nonumber \\ && \ \ - \case 1/2
(a_i^\dagger - a_i) (a_j - a_j^\dagger) (K_2' c^2+K_3' s^2)\tau_i
\sigma_j \Biggr\}  \ .
\end{eqnarray}
This result reproduces that of Eq. (\ref{AB15EQ}).

\newpage

\newpage

\vspace{0.2 in}

\vspace{0.5 in}

\newpage
\vspace{0.5 in}

\vspace{0.5 in}

\vspace{0.2 in}
\begin{figure}
\centerline{\psfig{figure=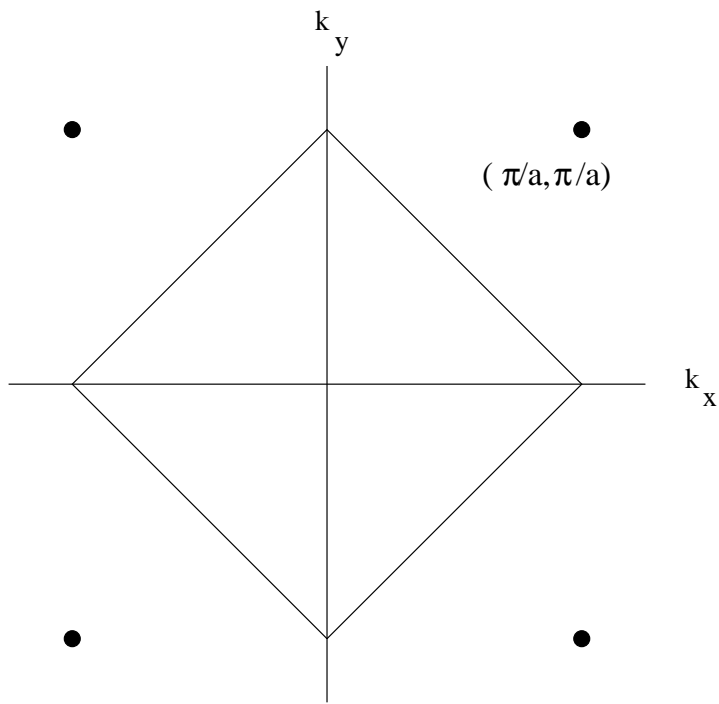}}

\vspace{0.2 in}
\caption{Brillouin Zone for the Square Lattice.}
\label{BZFIG}
\end{figure}

\newpage
\begin{table}
\caption{Definitions of Parameters.  Notation: $\delta J
\equiv (J_\perp - J_\parallel)/2.$}
\vspace{0.2 in}
\begin{tabular} { | c | c | c | c | c | c | c |}
$\alpha$ & $C_\alpha \approx 0.1686$ \ & $\tau$ &  $\zeta$ & $\xi$ &
$C_2 \approx 0.01^a$ & $J_{12}^{(n)}$ \ \\
$C_\alpha J_{12}^2 /J$ \ & \ Eq. (\ref{ALPHEQ}) \ & $(\delta J_1)^2 /J$ \ &
$(\delta J_{12})^2 /J_2$\ \ & \ $(\delta J_2)^2/J_2$ &
\ Eq. (\ref{C2EQ}) \ & Eqs. (\ref{VDMEQ}),
(\ref{VDM2EQ})\ \\
\hline \hline
$\Delta K$ & $\Delta K_{\rm eff}$ & $X=7\times 10^{-4} \AA^{-3}$ & & & & \\
Eq. (\ref{DKEQ}) & Eq. (\ref{KEFFEQ}) &  Eq. (\ref{XEQ}) & & & & \\
\end{tabular}
\label{PARAMS}
\end{table}
\noindent
a)  See Ref. \onlinecite{SOPR1}.

\newpage
\begin{table}
\caption{Estimated Values of Parameters from Experiment and Theory}
\vspace{0.2 in}
\begin{tabular} {| c ||c |c |c|c|}
Parameter & \multicolumn{4} {c|} {Values in meV} \\ \hline
& \multicolumn{2} {c|} {From Experiment}
& \multicolumn {2} {c|} {From Theory} \\ \hline
& Value & \ Reference$^a$ \ & Value & \ Reference$^a$ \ \\
\hline
$J$       & \ $130 \pm 5 $ \ & \ \onlinecite{RJB}
\  & \ 145 \ & \onlinecite{JCALC} \\
$J_3$     &\ $ 0.14 \pm 0.02$\  & \ \onlinecite{KIM,RJB} \ & & \\
$J_{12}$  & $-10 \pm 2$ & \onlinecite{RJB}, TW &  & \\
$J_2$     & $10.5 \pm 0.5$ & \ \onlinecite{KIM} \ & & \\
$\Delta J_1(T=0$K)     & $0.081 \pm 0.01$ & \ TW \ & 0.04 & 
\onlinecite{SOPR1,SOPR2} \\
$\Delta J_1(T=200$K)     & $0.068 \pm 0.011$ & \onlinecite{RJB} & & \\
$\Delta J_{12}$  & & & $1.3^d$ & \onlinecite{HAYN} \\
$\Delta J_2$     & $0.004\pm 0.004$  & \onlinecite{RIKEN2},
TW & 0.036 & \onlinecite{TORNOW} \\
$\delta J_1$     & $\pm 0.04$ & \onlinecite{RIKEN2}, TW
& $-0.02$ & \onlinecite{SOPR2} \\
$\delta J_{12}$  & $\pm 0.027$ & \onlinecite{KASTNER} & $-0.015$ & (b) \\
$\delta J_2$     & & & 0.4 & \\ 
$\Delta K_{\rm eff}$ & & & $2.73 \times 10^{-4}$ & TW,
Eq. (\ref{KEFFEQ}) \\ 
$\alpha$    & 0.135 & \onlinecite{KIM,RJB}, TW & 0.13 & TW, App. B \\
$\tau$ & $ 1.2 \times 10^{-5}$ & \onlinecite{RIKEN2}, TW
& ${\sim 10^{-5}}$ & (c) \\
$\zeta$ & $7 \times 10^{-5}$ & \onlinecite{KASTNER} & $2.2 \times 10^{-5}$
& (c) \\
$\xi$ & & & $10^{-6}$ & (c) \\
\end{tabular}
\label{VALUES}

\vspace{0.2 in} \noindent
a)  TW denotes this work.

\noindent
b)  This is the contribution to $\delta J_{12}$ from dipolar
interactions, which is much larger than that estimated from
$\delta J/J \sim 1.5 \times 10^{-4}$.

\noindent
c)  Evaluation based on the relevant $J$'s.

\noindent
d)  Evaluated for the similar compound Ba$_2$Cu$_3$O$_4$Cl$_2$.
\end{table}

\begin{table}
\vspace{0.2 in}
\caption{Renormalizations  $J \rightarrow ZJ$}

\vspace{0.2 in}
\begin{tabular} {| c || c | c | c |c|} 
Quantity & $J$ & $J_3^{\rm a}$ & $\sqrt{J \Delta J_I }$
& $\Delta K_{\rm eff}^{(\rm b)}$ \\
Renormalized to & $Z_cJ$ & $\tilde Z_3J_3$ & $Z_g \sqrt{J \Delta J_I}$
&\ \  $\Delta K_{\rm eff} \tilde Z_3^2$ \ \ \\ 
& $(1+ 0.085/S)J$ & $(1-0.2/S)J_3$ & $(1-0.2/S) \sqrt{J \Delta J_I}$ 
& $(1-0.2/S)K_{\rm eff}$ \\
Refer to: & Ref. \onlinecite{1/S2} & Eq. (\ref{Z3}) & Ref. \onlinecite{PETIT}  
& Eq. (\ref{J3EQ}) \\
\end{tabular}
\label{RENORM}
\end{table}
\vspace{0.1 in}
\noindent
a)  In the dynamics $JJ_3 \rightarrow Z_c \tilde Z_3JJ_3\equiv Z_3^2JJ_3$,
where we set $Z_3^2=0.77$.

\noindent
b)  In the dynamics $J_2\Delta K_{\rm eff} \rightarrow
Z_c \tilde Z_3 J_2 \Delta K_{\rm eff}\equiv Z_3^2 J_2 \Delta K_{\rm eff}$,
where we set $Z_3^2=0.77$.

\vspace{0.2 in}
\begin{table}
\caption{Experimental Values of Spin-wave Gaps at Zero Wave Vector}

\vspace{0.2 in}
\begin{tabular} {| c | c ||c |c|}
Mode & Temperature & Energy (meV)  & Ref \\ \hline
$\omega_+^>$ & $T=200$K & $5.5(3)^{\rm a}$ & \onlinecite{KIM} \\
$\omega_-^>$ & $T=200$K & 0.066(4)  & \onlinecite{RIKEN2} \\
$\omega_+^>$ & $T \rightarrow 0$K & 10.8(6) & \onlinecite{RJB} \\
$\omega_-^>$ & $T \rightarrow 0$K & 9.1(3) & \onlinecite{RJB} \\
$\omega_+^<$ & $T \rightarrow 0$K & 1.7473(4) & \onlinecite{RIKEN2} \\
$\omega_+^<$ & $T \rightarrow 0$K & 1.72(20) & \onlinecite{RJB} \\
$\omega_-^<$ & $T \rightarrow 0$K & 0.149(3) & \onlinecite{RIKEN2} \\
\end{tabular}
\label{EXPT}
\end{table}

\vspace{0.1 in} \noindent
a) Extrapolated to $T=0$.

\newpage
\begin{table}
\caption{Values in $(10^{-6}$meV) of the four-fold anisotropy constant $k$}

\vspace{0.2 in}
\begin{tabular} {| c || c |c|}
\hline
$k$ & $T=1.4$K & $T=100$K \\ \hline
Experimental: From statics \cite{KASTNER}&  25 & 2 \\
Experimental: Fitting Eq. (\ref{komeq}) to AFMR data\cite{RIKEN2} & 41 & 1 \\
Theoretical: See Eq. (\ref{kdeq}) & 56 & 1$^a$ \\
\hline
\end{tabular}
\label{ktab}
\end{table}

\vspace{0.1 in} \noindent
a)  $|J_\perp - J_\parallel|=0.041$ meV is fixed so that the dynamics
and theory agree. 

\begin{table}
\caption{Amplitude of the Dynamic Structure Factor}

\noindent
Results for wave vector 
$2 \pi (H \hat x/a + K \hat y/a + L \hat z /c ) + q_z$
for $H$ and $K$ half integral and $H+K+L$ an even integer.
Results are given only to leading order in $J$.
$y_3 = x_3 + 4\Delta J_1S + \case 1/2 \alpha =
2J_3 S [1-\cos (q_zc/2)] + 4\Delta J_1 S + \case 1/2 \alpha$.
The mode energies (without $1/S$ corrections) and intensities
[$I_r^{\alpha \beta}({\bf q})$]
are independent of the particular values of
$H$, $K$, and $L$ and are evaluated for $q_z=0$.

\vspace{0.2 in}
\begin{tabular}  {|c | c || c | c|}
Mode Energy & Energy$^{\rm a}$ (meV) & \multicolumn{2} {c|}
{Intensity} \\
& & Formula & Evaluation \\ \hline
{ \vspace{0.2 in} $ \omega_+^> =
\left[ 8JSy_3  \right]^{1/2}$ \vspace{-0.2 in} } 
& 10.8 & $I^{zz}_{>+}=0$  & $0$ \\
& & $I^{\eta \eta}_{>+}=0$ & $0$ \\
\hline
$\omega_+^< = \left\{ 8J_2S \left[ 4\Delta J_2S + {2 \alpha
(4\Delta J_1S+ x_3)\over y_3} \right]
\right\}^{1/2}$ & 1.72 & $I^{zz}_{<+}={16J_2S \over \omega_+^< }$
&  12 \\
& & $I^{\eta \eta}_{<+}=0$ & $0$ \\
\hline
$\omega_-^> = \left[ 8JS (x_3 + 2 \alpha )\right]^{1/2}$ &
9.1 & $I^{zz}_{>-}=0$ & $0$ \\
& & $I^{\eta \eta}_{>-}=0$ & $0$ \\
\hline
$\omega_-^< = \left[ 8J_2S{2 \alpha x_3 + 
64 \alpha ( 2 \tau - \xi) C_2 + 8 \alpha \zeta S
\over x_3 + \alpha } \right]^{1/2}$ &
$0.15$ &  $I^{zz}_{<-}=0$ & $0$ \\
& & $I^{\eta \eta}_{<-}={16J_2S \over \omega_-^< }$ & 140  \\
\end{tabular}
\label{IHALF}
\end{table}

\vspace{0.1 in} \noindent
a) See Table IV.

\begin{table}
\caption{Amplitude of the Dynamic Structure Factor}

\noindent
Results for wave vector 
$2 \pi (H \hat x/a + K \hat y/a + L \hat z /c ) + q_z$
for $H$ and $K$ integers and $H+K+L$ an even integer.
The notation is as in Table V.
Results are given only to leading order in $J$.
The intensities are evaluated for $q_z=0$ and
$H=L=1$ and $K=0$.

\vspace{0.2 in}
\begin{tabular}  {| c | c| c |}
Mode Energy$^{\rm a}$ (meV) & Formula for Intensity & Intensity \\
\hline $\omega_+^>=10.8$ & $I^{zz}_{>+} = {8JS \over \omega_+^>}
\left[ 1 - (-1)^L \right]^2$ & 50 \\
& $I^{\eta \eta}_{>+} = {8JS \over {\omega_+^>}^3}
\left| y_3[ 1 + (-1)^L ] + (-1)^H \alpha \right|^2$ & 0 \\
\hline
$\omega_+^<=1.72$ &
$I^{zz}_{<+} = \left[ 1 - (-1)^L \right]^2 {(8JS)^2 (4J_2S)
\alpha^2 \over {\omega_+^>}^4 \omega_+^<} $ & 26 \\
& $I^{\eta \eta}_{<+} = {\omega_+^< \over 4J_2 S}$ & 0 \\
\hline
$\omega_-^>=9.1$ &
$I^{zz}_{>-} = {8JS \over {\omega_-^>}^3} \left\{
\left[ 1 + (-1)^L \right][ x_3 + \case 1/2 \alpha ]
+ \alpha (-1)^H \right\}^2$ \ \ 
& 0 \\
& $I^{\eta \eta}_{>-} = {8JS \over \omega_-^>} \left[
1 - (-1)^L \right]^2$ & 59 \\
\hline
$\omega_-^<= 0.15$ meV
& $I^{zz}_{<-} = {\omega_-^< \over 4J_2S}$ & 0 \\
& $I^{\eta \eta}_{<-} =  { (4J_2S) (8JS)^2
\alpha^2 \over \omega_-^< {\omega_-^>}^4 } \left[ 1 - (-1)^L
\right]^2 $ & 570 \\
\end{tabular}

\vspace{0.2 in} \noindent
a) See Table IV.
\vspace{0.2 in} \noindent
\label{IINT}
\end{table}
\end{document}